\documentclass[a4paper,11pt]{article}
\usepackage{jheppub}
\usepackage[utf8]{inputenc}
\usepackage[T1]{fontenc}
\usepackage[english]{babel}
\usepackage{braket}
\usepackage{physics}
\usepackage{amsfonts}
\usepackage{mathtools}
\usepackage{bm}
\usepackage{enumitem}
\usepackage{array}
\usepackage{microtype}
\usepackage{float}
\usepackage{subcaption}
\usepackage{fmtcount}
\usepackage{romannum}
\usepackage{placeins}
\usepackage{tikz}
\usetikzlibrary{arrows.meta,calc,patterns,positioning}
\definecolor{UnifiedBraneWebBlue}{RGB}{50,85,210}
\usepackage[nameinlink]{cleveref}
\graphicspath{{figures/}}

\hypersetup{
 colorlinks=true,
 linkcolor=blue,
 filecolor=cyan,
 urlcolor=blue,
 citecolor=red,
 hypertexnames=false,
 pdftitle={Multipartite Entanglement Geometry in AdS/BCFT},
 pdfauthor={Debarshi Basu, Qiang Wen and Mingshuai Xu}
}
\numberwithin{equation}{section}
\makeatletter
\newcommand{\vast}{\bBigg@{3}}
\newcommand{\Vast}{\bBigg@{5}}
\makeatother

\newcommand{\zbar}{\raisebox{0.2ex}{--}\kern-0.6em Z}

\def\CC{{\cal C}}

\def\CF{{\cal F}}

\def\CL{{\cal L}}

\title{Multipartite Entanglement Geometry in AdS/BCFT}
\author[a,1]{Debarshi Basu\note{Corresponding author.}}
\author[a,2]{, Qiang Wen\note{Corresponding author.}}
\author[a,3]{and Mingshuai Xu\note{First author.}}

\affiliation[a]{School of Physics and Shing-Tung Yau Center, Southeast University, Nanjing, China}

\emailAdd{debarshi.128@gmail.com, wenqiang@seu.edu.cn, xumingshuai@seu.edu.cn}
\abstract{
		Multipartite entanglement measures probe how quantum correlations are organized beyond what can be inferred from pairwise entropies. We study how this structure is reshaped by a physical boundary in a holographic BCFT. The end-of-the-world brane opens new relative-homology channels for multipartite entanglement, allowing connected bulk webs to reorganize and factorize at scales that need not coincide with any ordinary Ryu--Takayanagi transition. In static AdS$_3$/BCFT$_2$, this produces a characteristic hierarchy of multipartite phases and a nontrivial dependence of genuine multi-entropy on the boundary. In the thermofield-double state, genuinely connected webs can instead be supported through the Einstein--Rosen bridge, leading to growth, reconnection and saturation scales distinct from the usual Hartman--Maldacena transitions. These results show that multipartite entanglement is sensitive not only to extremal-surface areas but also to how bulk entanglement structures are allowed to join, split and terminate in the presence of a boundary.
}

\begin{document}
\maketitle
\flushbottom
\section{Introduction}\label{sec:introduction}
The geometric encoding of entanglement is one of the most concrete lessons of the AdS/CFT correspondence \cite{Maldacena:1997re,Witten:1998qj,Gubser:1998bc,Aharony:1999ti}.  The broader quantum-field-theoretic study of entanglement originates in the area-law observations of \cite{Bombelli:1986rw,Srednicki:1993im} and, in two-dimensional conformal field theory, in the replica and conformal analyses of \cite{Holzhey:1994we,Calabrese:2004eu,Calabrese:2009qy,Casini:2009sr}; for a general quantum-information review see \cite{Horodecki:2009zz}.  For bipartite entropy, the Ryu--Takayanagi and Hubeny--Rangamani--Takayanagi prescriptions reduce a quantum-information quantity to the area of a codimension-two extremal surface \cite{Ryu:2006bv,Ryu:2006ef,Hubeny:2007xt,Headrick:2007km,Rangamani:2016dms}, while the gravitational replica trick explains the emergence of this surface from the bulk path integral \cite{Lewkowycz:2013nqa}.  A natural next question is whether genuinely multipartite correlations admit an equally sharp geometric organization.  Multi-entropy provides a particularly direct setting in which to address this question.  Its replica contraction involves $n^{q-1}$ copies for a $q$-partite state, and in holographic theories the $n\to1$ saddle is described by a multiway minimal cut, or ``brane web'', that partitions the bulk into chambers homologous to the boundary parties \cite{Gadde:2022cqi,Gadde:2023zzj}.  The resulting object is qualitatively different from a collection of pairwise RT surfaces: connected phases contain trivalent junctions satisfying a force-balance condition and therefore probe the global organization of multipartite entanglement.

The bipartite holographic dictionary itself has several complementary formulations that will serve as background below.  Useful reviews and structural results include the early overview of holographic entropy \cite{Nishioka:2009un}, the entanglement--geometry perspective \cite{VanRaamsdonk:2010pw}, causal constraints on holographic entropy \cite{Headrick:2014cta}, the cosmic-brane formulation of R\'enyi entropy \cite{Dong:2016fnf}, and the quantum-error-correcting interpretation of bulk reconstruction and the entanglement wedge \cite{Almheiri:2014lwa,Dong:2016eik,Harlow:2016vwg}.  The organization of holographic entropy inequalities into cones and arrangements has likewise been developed in~\cite{HernandezCuenca:2019wgh,Hubeny:2018ijt,He:2019ttu,He:2020xuo,Czech:2019lps}.

The holographic multi-entropy proposal has been investigated from several complementary directions.  The replica and multi-invariant framework has been developed through the classification of holographic multipartite measures, coarse-graining monotonicity, low-index CFT calculations, bulk replica-symmetry analysis, multipartite monotones and their classification, and random-tensor-network realizations, together with closely related multipartite edge-mode and triway-cut structures \cite{Penington:2022dhr,Gadde:2023zzj,Gadde:2023zni,Harper:2024ker,Gadde:2024taa,Gadde:2025csh,Gadde:2025ybn,Akella:2026xza,Hu:2026bhg,Akella:2026bci,Akers:2024wab,Akers:2024pgq}.  Genuine multi-entropy and closely related multipartite signals have in turn been developed as diagnostics that subtract or isolate lower-partite contributions \cite{Iizuka:2025ioc,Iizuka:2025bcc,Iizuka:2025caq,Balasubramanian:2024ysu,Balasubramanian:2025hxg,Gadde:2026msg,Ju:2026xfh,Balasubramanian:2026chr}.  Applications now include black holes, wormholes and real-time entanglement dynamics \cite{Iizuka:2024pzm,Anegawa:2025prn,Balasubramanian:2025jhq,Fujiki:2026qdt,Ju:2026lst,Anegawa:2026yee,Iizuka:2026kbo,Anegawa:2026npk}, as well as critical, topological, stabilizer, symmetry-resolved and Gaussian systems \cite{Harper:2025uui,Liu:2024ulq,Sheffer:2025jtc,Sheffer:2025zyr,Yuan:2025dgx,Chang:2026uds,Akella:2026rbe,Berthiere:2025toi,Iizuka:2025pqq,Camargo:2026xrd,DelZotto:2026fpw}.  Structural developments include holographic constraints and cones, junction-localization laws, IR-modified geometries and graph-encoded replica complexes \cite{Hernandez-Cuenca:2023iqh,Akella:2025owv,Ju:2025tgg,Ju:2026zbu,Iizuka:2026qqg,Iizuka:2026ahd,Ju:2025eyn,Chen:2026xtx,Iizuka:2026mlw}.  These developments make it important to understand how the multiway-cut prescription behaves in geometries in which extremal surfaces may end on additional gravitational boundaries.

Boundary conformal field theory is a minimal but nontrivial laboratory for this purpose.  In AdS/BCFT the bulk terminates on an end-of-the-world (EOW) brane $Q$, and ordinary RT/HRT surfaces are allowed to end on $Q$ subject to the appropriate relative-homology condition \cite{Takayanagi:2011zk,Fujita:2011fp,Nozaki:2012qd}. Recently, the replica-path-integral and large-$c$ boundary analyses have derived the relevant RT saddles and their phase transitions in AdS/BCFT and closely related Karch--Randall braneworld and multi-boundary settings  \cite{Geng:2024xpj,Bao:2025plr,Geng:2025efs}\footnote{See also \cite{Geng:2020qvw,Geng:2020fxl,Geng:2025rov,Geng:2025gns,Geng:2026wok} for related discussions of Karch--Randall braneworld holography and entanglement islands.}.  The AdS/BCFT setup has played a central role in holographic descriptions of black-hole entanglement and island-like transitions \cite{Rozali:2019day,Sully:2020pza,Chen:2020hmv,Deng:2020ent,Geng:2021iyq,Geng:2022dua}.  For a multiway cut, however, the EOW brane does more than provide a new endpoint condition.  It changes the topology of the admissible network: different connected components may terminate on $Q$, connected Steiner trees can partially factorize, and the transition of the multi-entropy need not coincide with the transition of any individual RT surface.  AdS/BCFT therefore exposes a genuinely multipartite version of relative homology.

The purpose of this paper is to develop this geometry in AdS$_3$/BCFT$_2$, test the static saddles directly against BCFT replica correlators, and compare the thermal saddles with the semiclassical channels suggested by the bulk topology.  We work on the semiclassical replica-selected branch near $n=1$ and, for the static problem, restrict to reduced planar, non-self-intersecting, equal-tension networks on a time-reflection-symmetric slice.  Within this sector we classify the allowed relative-homology webs for adjacent tripartitions and quadripartitions.  The tripartite problem has two reduced topologies, while the quadripartite problem has five.  Their competition already shows that multipartite factorization can occur at a scale that is invisible to the ordinary RT transition.  We also formulate the planar counting problem in a way that makes its Catalan structure manifest and illustrate the extension to higher numbers of adjacent parties.

A second goal is to make the bulk/BCFT comparison sufficiently sharp to test more than the endpoint dependence of the answer.  The doubling trick maps the BCFT twist correlators to reflection-symmetric chiral correlators, and at large central charge different bulk webs are naturally associated with different semiclassical fusion channels \cite{Fitzpatrick:2014vua,Sully:2020pza}.  For the connected quadripartite saddle, derivatives of the on-shell web action provide explicit generic accessory parameters.  We show that the Ward-identity and linearized-monodromy residuals factor through the same polynomial that determines the physical Steiner modulus in the bulk.  This identifies a nontrivial piece of the semiclassical block data geometrically.  The normalization of the block is then fixed by its OPE limit, so the only input that is not determined kinematically is the twist OPE coefficient.  This coefficient is genuine CFT data; the brane web predicts its leading holographic value.

We then turn to the thermofield-double (TFD) state.  In the left--right reflection-symmetric, principal non-winding sector we find five competing reduced saddles and determine their lower envelope.  The genuinely connected cross-boundary phase contains two trivalent vertices joined through the Einstein--Rosen bridge.  A direct extremization in Kruskal coordinates gives its complete Region-I/II continuation and, crucially, an additional Steiner admissibility condition.  In terms of the natural four-point cross-ratio,
\begin{equation}
 \eta_{\rm TFD}=
 \frac{\sinh^2\!\left(\frac{b_2-b_1}{2z_h}\right)}
 {\cosh^2(t/z_h)+\sinh^2\!\left(\frac{b_2-b_1}{2z_h}\right)},
\end{equation}
the separated two-vertex saddle exists only for $\eta_{\rm TFD}<3/4$; the equality point is the degenerate limit in which the internal edge shrinks to zero while the inherited angle is $2\pi/3$.  This existence line is distinct from the dominance transitions of the lower envelope.  The resulting equilibrium curve displays a multipartite reorganization scale that cannot be reconstructed from the two ordinary Hartman--Maldacena switches alone.  We call the corresponding growth-and-saturation pattern Page-like only in this geometric sense: the eternal black hole is an equilibrium TFD state, not an evaporating system.

It is also useful to distinguish the observable studied here from other multipartite mixed-state measures.  Entanglement of purification was introduced in~\cite{Terhal:2002riz}; its holographic and multipartite extensions are defined through optimized purifications and are related to entanglement-wedge cross sections \cite{Takayanagi:2017knl,Nguyen:2017yqw,Bao:2017nhh,Espindola:2018ozt,Bao:2018fso,Bao:2018gck,DeWolfe:2020vjp}.  The entanglement-wedge cross section is also closely tied to tripartite entanglement, negativity and balanced partial entanglement \cite{Akers:2019gcv,Kudler-Flam:2018qjo,Kusuki:2019zsp,Tamaoka:2018ned,Dong:2021clv,Wen:2021qgx}; genuinely tripartite information can also be isolated from multipartite entanglement of purification \cite{Bao:2025psl}.  Reflected entropy and multipartite reflected entropy instead use canonical purification \cite{Dutta:2019gen,Bao:2019zqc}.  Reflected multi-entropy applies the multi-entropy replica contraction to such a canonical purification and has a proposed multiway-cut dual \cite{Yuan:2024yfg}, while multipartite Markov gaps give a related mixed-state diagnostic \cite{Iizuka:2025elr}.  In brane and island settings, the earlier defect-extremal-surface construction for reflected entropy provides an important bipartite precursor \cite{Li:2021dmf}.  These constructions are closely related to the present geometry, but their replica sewings and bulk variational problems are not identical.

The present paper is restricted to the non-gravitating BCFT problem and its equilibrium thermal extension.  Two further developments are deferred to companion works: the island/generalized-entropy counterpart of the multiway-cut prescription (\emph{multipartite entropy islands}) \cite{BasuXu:2026MultipartiteIslands}, and \emph{reflected multi-entropy} in AdS/BCFT and in island phases \cite{BasuXu:2026ReflectedMultiEntropy}.  In particular, an endpoint on the fixed-background EOW brane studied here is a boundary condition and should not be conflated with a quantum extremal surface; we return to both directions in Sec.~\ref{sec:discussion}.

Finally, all completeness statements in this paper refer to the stated $n\to1$ sectors.  We do not assume that every unresolved multiway junction extends to a smooth dominant bulk filling at arbitrary integer R\'enyi index.  Replica-symmetry analyses, random tensor networks and graph-encoded replica complexes show that the finite-$n$ problem can contain additional saddles or obstructions \cite{Penington:2022dhr,Gadde:2024taa,Hu:2026bhg,Akella:2026bci,Iizuka:2026mlw}.  Similarly, the thermal classification is restricted to left--right reflection symmetry and the principal non-winding BTZ homotopy class.

The rest of the paper is organized as follows.  Section~\ref{sec:review} reviews multi-entropy and the AdS/BCFT ingredients.  Section~\ref{sec:brane-web} develops the static brane web geometry, while Sec.~\ref{sec:field-theory} performs the BCFT replica and semiclassical-block comparison.  Section~\ref{sec:eternal-black-hole} analyzes the thermal TFD saddles, their CFT channels and the resulting equilibrium multi-entropy curve.  Section~\ref{sec:discussion} summarizes the physical picture and discusses extensions.  Appendix~\ref{app:hyperbolic} gives complementary derivations based on hyperbolic geometry, and Appendix~\ref{app:web-counting} develops the planar topology-counting argument and its Catalan interpretation.

\section{Review}\label{sec:review}

\subsection{Multi-entropy}\label{sec:review-multi}
We collect only the ingredients of multi-entropy needed below; detailed discussions of the original construction, its holographic generalizations, coarse-graining monotonicity, low-R\'enyi CFT realization and multi-invariant formulation can be found in~\cite{Gadde:2022cqi,Gadde:2023zzj,Gadde:2023zni,Harper:2024ker,Gadde:2024taa}.

\subsubsection{Definition}

Consider a pure state on $q$ subsystems,
\begin{equation}
	\ket{\Psi} = \sum_{\alpha_1}^{d_1} \cdots \sum_{\alpha_q}^{d_q} \psi_{\alpha_1 \cdots \alpha_q} |\alpha_1\rangle \otimes \cdots \otimes |\alpha_q\rangle \in \otimes_{a=1}^{q} \mathcal{H}_a,
\end{equation}
where $\ket{\alpha_a}$ is a basis for the $d_a$-dimensional Hilbert space $\mathcal H_a$ associated with subsystem $A_a$.  Tracing over $A_1$ gives
\begin{equation}
	\rho=\rho_{A_2\cdots A_q}=\operatorname{Tr}_{A_1}\ket{\Psi}\bra{\Psi},\quad \rho_{k_2\cdots k_q}^{j_2\cdots j_q}=\psi_{k_1k_2\cdots k_q}\bar{\psi}^{k_1j_2\cdots j_q}.
\end{equation} 
To define the $q$-partite replica quantity, take $n^{q-1}$ copies of $\rho$ and label the sheets by
\begin{equation}
	\bm{i}=\left(i_2,\ldots,i_q\right)\in \mathbb Z_{n}^{q-1}.
\end{equation} 
\begin{figure}[H]
 \centering
 \begin{subfigure}[c]{0.3\linewidth}
  \centering
  \includegraphics[width=\linewidth]{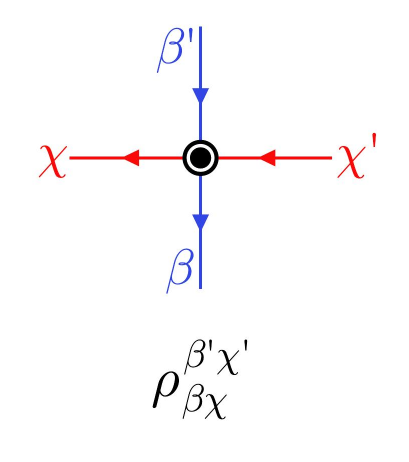}
 
 \end{subfigure}
 \begin{subfigure}[c]{0.41\linewidth}
  \centering
  \includegraphics[width=\linewidth]{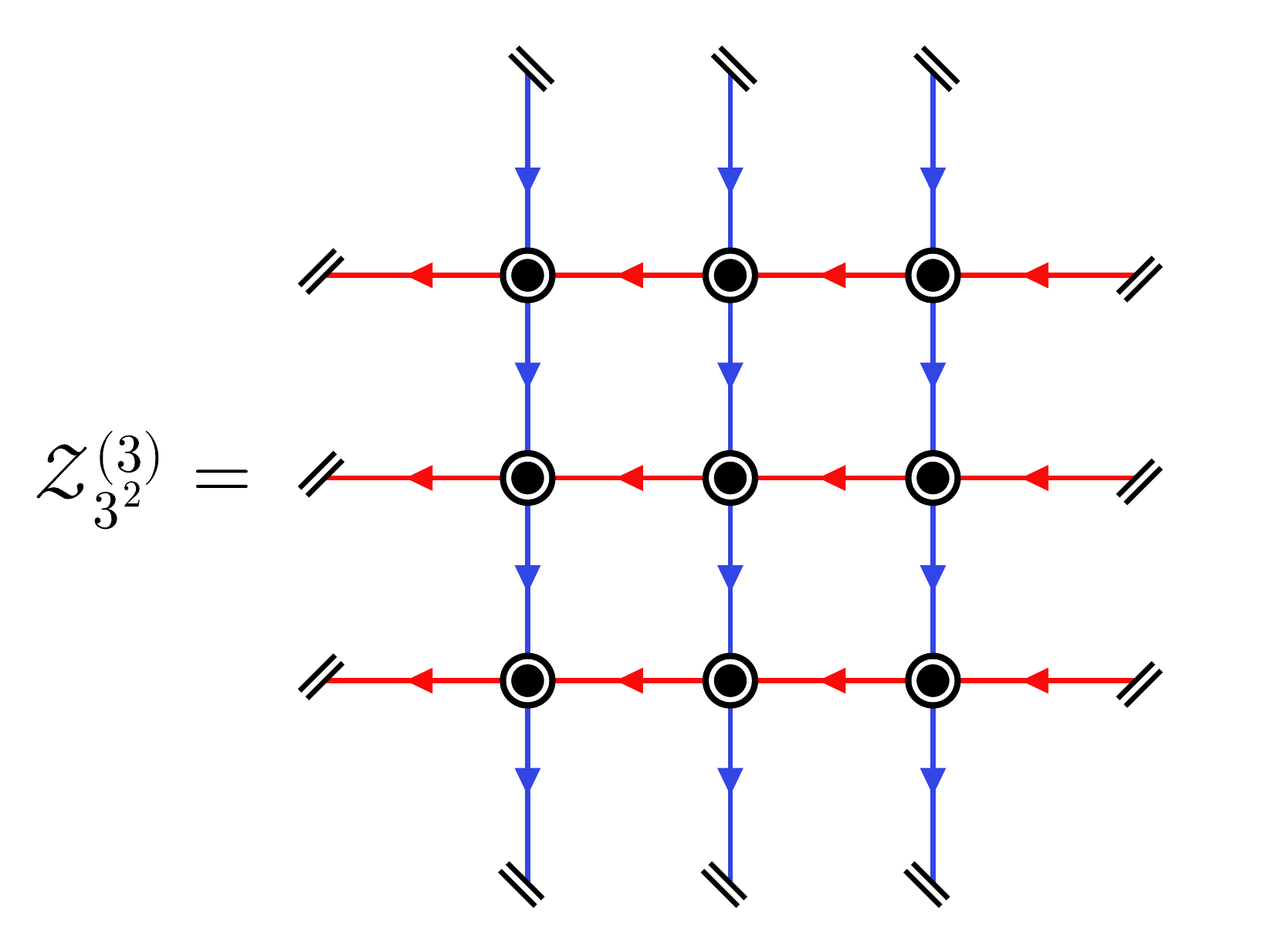}
 \end{subfigure}
 \caption{Graphical notation for the reduced density matrix and the replica construction of $Z_{n^{q-1}}^{\left(q\right)}$ for the case $q=3,n=3$.  In the left panel the lower and upper indices of $\rho=\rho_{A_2A_3}$ associated with $A_2$ and $A_3$ are denoted $(\beta,\beta')$ and $(\chi,\chi')$, respectively; in the right panel, each node represents one copy of $\rho$, and the lines of the two colors represent cyclic contractions in the two replica directions. The figure is adapted from \cite{Yuan:2025dgx}.}
 \label{fig:review-replica-grid}
\end{figure}
The analogue of $\operatorname{Tr}\rho^n$ is obtained by contracting these copies with cyclic shifts in the replica directions\footnote{We follow the replica conventions of \cite{Yuan:2025dgx,Yuan:2024yfg}.},
\begin{equation}
	\begin{aligned}
		Z_{n^{q-1}}^{(q)} =\prod_{\bm{i}\in \mathbb Z_{n}^{q-1}}\rho_{\alpha_2^{\left(\bm{i}\right)}\cdots \alpha_q^{\left(\bm{i}\right)}}^{\alpha_2^{\left(\sigma_{2}\cdot \bm{i}\right)}\cdots \alpha_q^{\left(\sigma_q\cdot \bm{i}\right)}}
		=&\prod_{\bm{i}\in \mathbb Z_{n}^{q-1}}\left(\psi_{\alpha_1^{(\bm{i})} \alpha_2^{(\bm{i})}\cdots \alpha_q^{(\bm{i})}} \bar{\psi}^{\alpha_1^{(\bm{i})} \alpha_2^{(\sigma_{2}\cdot \bm{i})}\cdots \alpha_q^{(\sigma_{q}\cdot \bm{i})}}\right),
	\end{aligned}
\end{equation}
where $\sigma_a$ is the cyclic shift in the replica direction associated with party $A_a$,
\begin{equation}
	\sigma_a\cdot\left(\ldots,i_a,\ldots\right)=\left(\ldots,i_a+1,\ldots\right),\qquad a=2,\ldots,q.
\end{equation}
Figure~\ref{fig:review-replica-grid} illustrates the construction for $q=3$ and $n=3$.  The R\'{e}nyi multi-entropy and its $n\to1$ limit are
\begin{equation}\label{definition multi-entropy}
	S_n^{(q)}\left(A_1,\cdots,A_q\right)=\frac{1}{1-n}\frac{1}{n^{q-2}}
	\log\left[
	Z_{n^{q-1}}^{\left(q\right)}/\left(Z_1^{\left(q\right)}\right)^{n^{q-1}}\right],\qquad
	S^{(q)}=\lim_{n\to1}S_n^{(q)}.
\end{equation}
The construction is symmetric under permutations of the parties and reduces to the ordinary R\'{e}nyi entropy for $q=2$.

In a two-dimensional CFT the replicated partition function can be represented as a twist-operator correlator at the interfaces $x_i$ \cite{Gadde:2022cqi,Gadde:2023zzj},
\begin{equation}\label{part function replica}
	Z_{n^{q-1}}^{\left(q\right)}=\left\langle\sigma_{g_1}\left(x_1\right)\cdots \sigma_{g_q}\left(x_q\right)\right\rangle_{\text{CFT}^{\otimes n^{q-1}}},\qquad \prod_i g_i =\mathrm{id}.
\end{equation}
The corresponding twist dimension is \cite{Gadde:2022cqi,Gadde:2023zzj}
\begin{equation}\label{eq:multi-twist-dim}
\Delta_n^{(q)}=\frac{c}{12}n^{q-2}\left(n-\frac1n\right).
\end{equation}

\subsubsection{Genuine multi-entropy}

Multi-entropy also contains contributions inherited from lower-partite correlations \cite{Iizuka:2025ioc,Iizuka:2025caq}.  For example, consider the triangle state
\begin{equation}\label{tri state}
	\ket{\Delta}_{ABC}=\ket{\psi_1}_{A_1B_1}\otimes\ket{\psi_2}_{A_2C_1}\otimes \ket{\psi_3}_{B_2C_2},
\end{equation}
for which the tripartite multi-entropy does not vanish:
\begin{equation}
	S^{\left(3\right)}\left(A:B:C\right)=\frac{1}{2}\left(S_A+S_B+S_C\right).
\end{equation}
A ``genuine'' multi-entropy subtracts the corresponding lower-partite contributions.  For three parties,
\begin{equation}\label{eq:genuine-multi}
	\mathrm{GM}^{(3)}(A:B:C)=S^{(3)}(A:B:C)-\frac12\left[S_A+S_B+S_C\right].
\end{equation}
Equation~\eqref{eq:genuine-multi} vanishes on the triangle state.  The three-party quantity in Eq.~\eqref{eq:genuine-multi} is the only genuine multi-entropy subtraction needed in the calculations below; higher-partite generalizations, multipartite-signal constructions and their remaining ambiguities are discussed in~\cite{Iizuka:2025ioc,Iizuka:2025bcc,Iizuka:2025caq,Gadde:2026msg,Ju:2026xfh,Balasubramanian:2026chr}.  Genuine multi-entropy has also been tested in stabilizer, topological, Chern--Simons, symmetry-resolved, Lifshitz and Gaussian states \cite{Akella:2026xza,Akella:2026rbe,Yuan:2025dgx,Chang:2026uds,Iizuka:2025pqq,Berthiere:2025toi,Camargo:2026xrd}.

\subsubsection{Holographic dual: minimal brane web}\label{sec:review-web}

\begin{figure}[t]
	\centering
	\includegraphics[width=0.86\linewidth]{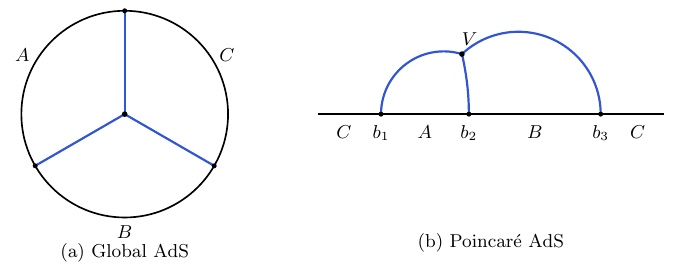}
	\caption{Basic tripartite multi-entropy webs in global and Poincar\'e AdS.  The blue geodesic walls meet at a trivalent Steiner vertex and partition the bulk into the three boundary-color chambers.}
	\label{fig:review-basic-brane-webs}
\end{figure}

The proposed classical holographic dual is a minimal brane web, or ``soap film'' \cite{Gadde:2022cqi,Gadde:2023zzj}.  Let $\mathcal M$ be a bulk Cauchy slice whose asymptotic boundary is partitioned into the regions $A_a$. One can partition $\mathcal M$ into chambers $\{\mathcal C_a\}$ such that each chamber is homologous to the corresponding boundary region. The interfaces $\mathcal W_{ab}$ between chambers form the so-called brane web $\mathcal W$, and at leading order
\begin{equation}\label{ME-holo}
	S^{(q)}=\min_{\{\mathcal C_a\}}\frac{{\rm Area}(\mathcal W)}{4G}.
\end{equation}
The minimization is over topology as well as vertex positions.  Consequently, the physically meaningful observable is the lower envelope of all admissible homology-compatible webs; a stationary web that exists locally may not dominate globally. This distinction between existence, extremality and dominance will recur throughout the analyses below. The elementary brane-web geometry in AdS/CFT is shown in Fig.~\ref{fig:review-basic-brane-webs}.

In a static AdS$_3$ slice the problem reduces to a hyperbolic Steiner tree.  Equal-tension trivalent edges meet at $2\pi/3$, while changes of topology occur when a Steiner vertex degenerates or a competing disconnected web becomes shorter.  This network perspective is closely related to graph and cut-and-paste descriptions of holographic entropy inequalities, as well as to bit-thread and tensor-network constructions \cite{Hayden:2011ag,Bao:2015bfa,Cui:2018dyq,Hubeny:2018trv,Pastawski:2015qua,Penington:2022dhr}.  Recent studies of multipartite constraints, multi-entropy cones and junction laws further emphasize that the connectivity and junction data of the web contain information not captured by pairwise entropies alone \cite{Balasubramanian:2024ysu,Ju:2025tgg,Ju:2026zbu,Iizuka:2026qqg,Iizuka:2026ahd}.

Let us consider a concrete example (the right panel of Fig.~\ref{fig:review-basic-brane-webs}) to show how the holographic prescription \eqref{ME-holo} works in Poincar\'{e} AdS$_3$. Consider the three parties
\begin{equation}
	A=[b_1,b_2],\quad B=[b_2,b_3],\quad C=(-\infty,b_1)\cup(b_3,\infty),
\end{equation}
with $b_1<b_2<b_3$.  The total length of the brane web is
\begin{equation}\label{tri multi}
	\begin{aligned}
		\mathcal{L}_{\text{tot}}=&\CL\left(V,b_1\right)+\CL\left(V,b_2\right)+\CL\left(V,b_3\right)\\
		=&\log\left[\frac{(x-b_1)^2+z^2}{\epsilon z}\right]
 			+\log\left[\frac{(x-b_2)^2+z^2}{\epsilon z}\right]+\log\left[\frac{(x-b_3)^2+z^2}{\epsilon z}\right],
	\end{aligned}
\end{equation}
where $V=\left(x,z\right)$ denotes the vertex. The extremality conditions are given by
\begin{equation}
	\partial_{x}\mathcal{L}_{\text{tot}}=\partial_{z}\mathcal{L}_{\text{tot}}=0.
\end{equation}
The solutions of these conditions read
\begin{equation}\label{location vertex}
	\begin{aligned}
		x=&\:\frac{(b_1^2+b_2 b_3)(b_2+b_3)+b_1(b_2^2-6 b_2 b_3+b_3^2)}
		{2(b_1^2+b_2^2+b_3^2-b_1b_2-b_2b_3-b_1b_3)},\\
		z=&\:\frac{\sqrt{3}(b_2-b_1)(b_3-b_2)(b_3-b_1)}
		{2(b_1^2+b_2^2+b_3^2-b_1b_2-b_2b_3-b_1b_3)}.
	\end{aligned}
\end{equation}
Substituting these into \eqref{tri multi}, we obtain the extremal (minimal) length of the brane web,
\begin{equation}
	\mathcal{L}_{\text{ext}}=\log\left[\frac{\left(b_2-b_1\right)\left(b_3-b_2\right)\left(b_3-b_1\right)}{\epsilon^3}\right]+3\log\frac{2}{\sqrt{3}}.
\end{equation}
Therefore, according to \eqref{ME-holo}, the tripartite multi-entropy is given by
\begin{equation}
	S^{\left(3\right)}\left(A:B:C\right)=\frac{1}{4G}\log\left[\frac{8\left(b_2-b_1\right)\left(b_3-b_2\right)\left(b_3-b_1\right)}{3\sqrt{3}\epsilon^3}\right],
\end{equation}
which agrees with the field-theoretic calculations \cite{Gadde:2022cqi,Gadde:2023zzj}.

\subsection{AdS/BCFT}

Boundary conformal field theory and its bulk/defect operator expansions were developed in~\cite{Cardy:1984bb,Cardy:1989ir,Cardy:2004hm,McAvity:1993ue}, with holographic boundary/defect realizations discussed more broadly in~\cite{Karch:2000ct,DeWolfe:2001pq,Jensen:2013lxa}.  The AdS/BCFT construction associates a BCFT on a manifold $M$ with physical boundary $P=\partial M$ to a bulk region $N$ bounded by the asymptotic AdS boundary $M$ and an end-of-the-world (EOW) brane $Q$, with $\partial N=M\cup Q$ and $\partial Q=P$ \cite{Takayanagi:2011zk,Fujita:2011fp}.  For three-dimensional Einstein gravity we use
\begin{equation}
 I=\frac{1}{16\pi G}\int_N\!\sqrt{-g}\,(R-2\Lambda)
 +\frac{1}{8\pi G}\int_Q\!\sqrt{-h}\,(K-T),
 \label{eq:ads-bcft-action}
\end{equation}
where $h_{ab}$ is the induced metric on $Q$, $K_{ab}$ is its extrinsic curvature, and $T$ is the brane tension.  The Neumann condition on $Q$ gives \cite{Takayanagi:2011zk,Fujita:2011fp}
\begin{equation}
 K_{ab}-(K-T)h_{ab}=0.
 \label{eq:neumann-bcft}
\end{equation}

\subsubsection{Vacuum state}

For the vacuum BCFT$_2$ on the half-line $x\geq0$, the dual geometry is Poincar\'e AdS$_3$\footnote{We set the AdS radius $\ell=1$ throughout.},
\begin{equation}\label{eq:poincare-ads3}
 ds^2=\frac{1}{z^2}\left(-dt^2+dx^2+dz^2\right).
\end{equation}
Poincar\'e AdS$_3$ also admits the AdS$_2$ foliation
\begin{equation}
 ds^2=d\rho^2+\cosh^2\rho
 \frac{-d t^2+dy^2}{y^2},
\end{equation}
with coordinate transformation
\begin{equation}
 x=-y\tanh\rho,\qquad
 z=y\operatorname{sech}\rho.
\end{equation}
The brane is the constant-$\rho$ surface $\rho=\rho_0$, with $T=\tanh\rho_0$. Throughout the paper we focus on non-negative tension, $T\geq0$, so that $\rho_0\ge0$. We also use the angle $\theta$, defined by $\cos\theta=\operatorname{sech}\rho_0$, so that on the static Poincar\'{e} slice $Q$ is the line $x=-z\tan\theta$. The bulk region $N$ is then
\begin{equation}\label{eq:static-physical-wedge}
	N=\{(x,z):z>0,\ x+z\tan\theta> 0\}.
\end{equation}

Since the EOW brane $Q$ is part of the bulk boundary, RT surfaces may end on it, subject to the usual extremality condition \cite{Takayanagi:2011zk,Fujita:2011fp}.

\begin{figure}[ht]
	\centering
	\def\FigThreeOneTheta{30}
	\def\FigThreeOneA{3}
	\begin{subfigure}[t]{0.46\linewidth}
		\raggedleft
		\begin{tikzpicture}[x=0.99cm,y=0.99cm,>=stealth]
			\pgfmathsetmacro{\FigThreeOneBraneAngle}{90+\FigThreeOneTheta}
			\draw[->,line width=0.6pt] (0,0) -- ({\FigThreeOneA+1.3},0)
			node[right] {$x$};
			\draw[->,densely dashed,gray,line width=0.45pt]
			(0,0) -- (0,{\FigThreeOneA+0.9})
			node[above,text=black] {$z$};
			\draw[red,line width=0.8pt]
			(0,0) -- (\FigThreeOneBraneAngle:{\FigThreeOneA+1.3})
			node[above left,red] {$Q$};
			\draw[line width=0.5pt] (90:0.65) arc[start angle=90,
			end angle=\FigThreeOneBraneAngle,radius=0.65];
			\node at ({90+0.5*\FigThreeOneTheta}:0.9) {$\theta$};
			\draw[UnifiedBraneWebBlue,line width=1pt] (\FigThreeOneA,0)
			arc[start angle=0,end angle=\FigThreeOneBraneAngle,
			radius=\FigThreeOneA]
			node[pos=0.44,above right] {$\gamma_A$};
			\fill (0,0) circle[radius=1.3pt]
			node[below left=2pt] {$O$};
			\fill (\FigThreeOneA,0) circle[radius=1.6pt]
			node[below=2pt] {$a$};
			\node[below=2pt] at ({0.5*\FigThreeOneA},0) {$A$};
			\fill (\FigThreeOneBraneAngle:\FigThreeOneA)
			circle[radius=1.6pt]
			node[left=2pt,xshift=5pt,yshift=-4pt] {$a$};
		\end{tikzpicture}
	\end{subfigure}%
	\hspace{0.03\linewidth}%
	\begin{subfigure}[t]{0.46\linewidth}
		\raggedright
		\begingroup
		\definecolor{TwoSidedRTGreen}{RGB}{25,155,85}
		\begin{tikzpicture}[x=0.92cm,y=0.92cm,>=stealth,
			every node/.style={font=\small}]
			\def\TwoSidedRTTheta{30}
			\def\TwoSidedRTA{2}
			\def\TwoSidedRTComplement{1.10}
			\def\TwoSidedRTHalfGap{2.8}
			\def\TwoSidedRTBraneLength{3.0}
			\pgfmathsetmacro{\TwoSidedRTBraneAngle}{90+\TwoSidedRTTheta}
			\pgfmathsetmacro{\TwoSidedRTUpperBraneAngle}{270-\TwoSidedRTTheta}
			
			\coordinate (UpperO) at (0,\TwoSidedRTHalfGap);
			\coordinate (LowerO) at (0,-\TwoSidedRTHalfGap);
			\coordinate (UpperA) at (\TwoSidedRTA,\TwoSidedRTHalfGap);
			\coordinate (LowerA) at (\TwoSidedRTA,-\TwoSidedRTHalfGap);
			
			\draw[line width=0.65pt] (UpperO) --
			({\TwoSidedRTA+\TwoSidedRTComplement},\TwoSidedRTHalfGap)
			node[right=2pt,font=\scriptsize] {$\mathrm{BCFT}_{L}$};
			\draw[line width=0.65pt] (LowerO) --
			({\TwoSidedRTA+\TwoSidedRTComplement},-\TwoSidedRTHalfGap)
			node[right=2pt,font=\scriptsize] {$\mathrm{BCFT}_{R}$};
			\draw[red,line width=0.9pt] (UpperO) -- ++
			(\TwoSidedRTUpperBraneAngle:\TwoSidedRTBraneLength)
			node[above left,text=red] {$Q$};
			\draw[red,line width=0.9pt] (LowerO) -- ++
			(\TwoSidedRTBraneAngle:\TwoSidedRTBraneLength)
			node[below left,text=red] {$Q$};
			
			\begin{scope}[shift={(0,\TwoSidedRTHalfGap)},yscale=-1]
				\coordinate (UpperQRT) at
				(\TwoSidedRTBraneAngle:\TwoSidedRTA);
				\draw[UnifiedBraneWebBlue,line width=1.2pt]
				(\TwoSidedRTA,0) arc[start angle=0,
				end angle=\TwoSidedRTBraneAngle,
				radius=\TwoSidedRTA];
			\end{scope}
			\begin{scope}[shift={(0,-\TwoSidedRTHalfGap)}]
				\coordinate (LowerQRT) at
				(\TwoSidedRTBraneAngle:\TwoSidedRTA);
				\draw[UnifiedBraneWebBlue,line width=1.2pt]
				(\TwoSidedRTA,0) arc[start angle=0,
				end angle=\TwoSidedRTBraneAngle,
				radius=\TwoSidedRTA];
			\end{scope}
			
			\draw[TwoSidedRTGreen,line width=1.3pt]
			(UpperA) -- (LowerA);
			\node[above=3pt]
			at ({0.5*\TwoSidedRTA},\TwoSidedRTHalfGap) {$A_L$};
			\node[below=3pt]
			at ({0.5*\TwoSidedRTA},-\TwoSidedRTHalfGap) {$A_R$};
			\node[text=UnifiedBraneWebBlue,above right=1pt]
			at (-0.10,-0.70) {$\gamma_A^{\mathrm{island}}$};
			\node[text=TwoSidedRTGreen,right=5pt,fill=white,inner sep=1pt]
			at (\TwoSidedRTA,0) {$\gamma_A^{\mathrm{HM}}$};
			
			\fill (UpperO) circle[radius=1.3pt]
			node[above left=1pt] {$O$};
			\fill (LowerO) circle[radius=1.3pt]
			node[below left=1pt] {$O$};
			\fill (UpperA) circle[radius=1.5pt]
			node[above=2pt] {$a$};
			\fill (LowerA) circle[radius=1.5pt]
			node[below=2pt] {$a$};
			\fill (UpperQRT) circle[radius=1.3pt]
			node[left=1pt] {$a$};
			\fill (LowerQRT) circle[radius=1.3pt]
			node[left=1pt] {$a$};
		\end{tikzpicture}
		\endgroup
	\end{subfigure}
	\caption{Holographic entanglement entropy in the AdS/BCFT setup. Left panel: vacuum state. Right panel: TFD state.}
	\label{fig:ads-bcft-review}
	\label{fig:two-sided-interval-RT-candidates}
\end{figure}
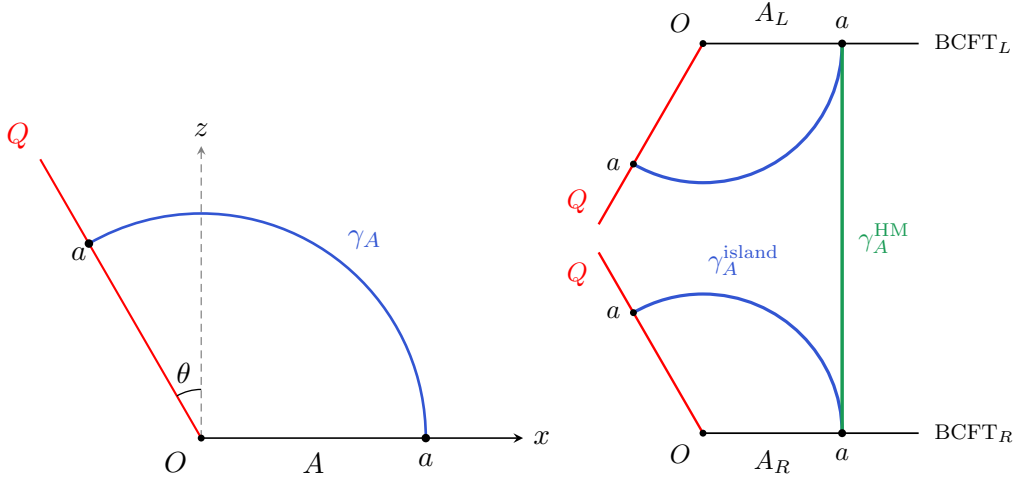

For an interval $A=[0,a]$, the RT geodesic ends orthogonally on $Q$ and gives
\begin{equation}\label{HEE AdS/BCFT}
  S_A=\frac{1}{4G}\log\left(\frac{2a}{\epsilon}\right)+\frac{1}{4G}\operatorname{arcsech}(\cos\theta)=\frac{c}{6}\log\left(\frac{2a}{\epsilon}\right)+\log g_b,
\end{equation}
where we used the Brown--Henneaux relation $c=3/(2G)$ \cite{Brown:1986nw} and introduced the boundary entropy \cite{Affleck:1991tk,Azeyanagi:2007qj,Takayanagi:2011zk}, 
\begin{equation}
	\log g_b=\frac{c}{6}\operatorname{arcsech}(\cos\theta)=\frac{c}{6}\rho_0.
\end{equation}

For an interval $A=[a,b]$ away from the BCFT boundary, the connected geodesic competes with two disconnected geodesics ending on $Q$ \cite{Takayanagi:2011zk,Fujita:2011fp,Sully:2020pza}.  The entropy is given by the smaller of the two candidates,
\begin{equation}\label{HEE two phases}
  S_A=\begin{cases}
	\frac{c}{3}\log\frac{b-a}{\epsilon},\quad &b/a< \eta_{c,1}\\
    \frac{c}{6}\log\frac{4ab}{\epsilon^2}+\frac{c}{3}\rho_0,\quad &b/a\geq \eta_{c,1},
  \end{cases}
\end{equation} 
where the transition occurs at
\begin{equation}\label{eta c 1}
  \eta_{c,1}=1+2e^{\rho_0}\left(e^{\rho_0}+\sqrt{1+e^{2\rho_0}}\right).
\end{equation}
The geometric transition is mirrored by a change of dominant bulk or boundary OPE channel in the large-$c$ limit of the BCFT \cite{Sully:2020pza}.  

\subsubsection{TFD state}

The normalized thermofield-double state of two identical BCFTs is
\begin{equation}
 \ket{\mathrm{TFD}(\beta)}=\frac{1}{\sqrt{Z(\beta)}}
 \sum_n e^{-\beta E_n/2}\ket{n}_L\otimes\ket{n}_R .
\end{equation}
Here $\beta$ is the inverse temperature and $\ket n_{L,R}$ are energy eigenstates.  Its bulk dual is an eternal two-sided black hole with one EOW brane associated with each exterior.  In each exterior the metric is the planar BTZ black string
\begin{equation}\label{BTZ BH}
 ds^2=\frac{1}{z^2}\left[-h(z)dt^2+\frac{dz^2}{h(z)}+dx^2\right],\qquad
 h(z)=1-\frac{z^2}{z_h^2},
\end{equation}
with $z_h=\beta/(2\pi)$.  The EOW-brane profile is \cite{Rozali:2019day,Sully:2020pza}
\begin{equation}\label{BTZ EOW}
 z=-\kappa z_h\sinh\!\left(\frac{x}{z_h}\right),\qquad
 \kappa=\frac{\sqrt{1-T^2}}{T}=\frac{1}{\sinh\rho_0},
\end{equation}
where we used $T=\tanh\rho_0$; in particular $\operatorname{arccsch}\kappa=\rho_0$, so the thermal brane is characterized by the same parameter $\rho_0$ as the vacuum one. For the two-sided interval $A=\left(0,a\right)_L\cup\left(0,a\right)_R$, there are two types of RT surfaces:
the island surface, which connects each endpoint of the interval to the nearby brane, and the Hartman--Maldacena (HM) surface, which passes through the interior of the black string
to connect the two endpoints of the interval. The island and HM surfaces are shown as the blue and green curves, respectively, in the right panel of Fig.~\ref{fig:two-sided-interval-RT-candidates}. In the TFD state the time coordinate of the right patch is chosen to be $t_R=t$, while that of the left patch is $t_L=-t+i\beta/2$ \cite{Maldacena:2001kr,Hartman:2013qma}. The entanglement entropy of the two-sided interval $A$ is then
\begin{equation}\label{HEE A TFD state}
	\begin{aligned}
		S_A=&\min\left\{S_A^{\text{island}},S_A^{\text{HM}}\right\}\\
		=&\min\left\{\frac{c}{3}\log\left[\frac{\beta}{\pi \epsilon}\sinh\left(\frac{2\pi a}{\beta}\right)\right]+2\log g_b,\: \frac{c}{3}\log\left[\frac{\beta}{\pi \epsilon}\cosh\left(\frac{2\pi t}{\beta}\right)\right]\right\}.
	\end{aligned}
\end{equation}
The thermal boundary entropy is the same as in the vacuum,
\begin{equation}\label{boundary entropy TFD state}
 \log g_b=\frac{c}{6}\operatorname{arccsch}\kappa=\frac{c}{6}\rho_0.
\end{equation}

\section{Holographic multi-entropy in AdS/BCFT}\label{sec:brane-web}
We first classify the allowed relative-homology partitions and then extremize their lengths. The static solutions supply the local building blocks used later in the thermal problem.

\subsection{Relative homology and admissible brane web topologies}\label{par:extended-homology-review}
The holographic multi-entropy is computed, at leading order in the classical bulk limit, by a minimal brane web (or ``soap film'') whose interfaces divide a bulk Cauchy slice into regions associated with the boundary parties \cite{Gadde:2022cqi,Gadde:2023zzj,Penington:2022dhr,Harper:2024ker}.  In the present AdS/BCFT geometry the end-of-the-world brane $Q$ is itself part of the boundary of the bulk slice.  The appropriate homology condition is therefore a relative one: the bulk region associated with a boundary party may close partly on $Q$, exactly as an RT surface in AdS/BCFT may end on the EOW brane \cite{Takayanagi:2011zk,Fujita:2011fp,Nozaki:2012qd,Sully:2020pza}.

More explicitly, let $A_a$ be one of the boundary parties, let $\mathcal C_a$ denote the corresponding bulk region, and let $\mathcal Q_a\subset Q$ be the portion of the EOW brane assigned to that region.  Suppressing orientations, the relative-homology condition may be written as
\begin{equation}\label{eq:extended-homology-chamber}
 \partial\mathcal C_a=A_a\cup\mathcal Q_a
 \cup\bigcup_{b\ne a}\mathcal W_{ab},\qquad
 \mathcal W=\bigcup_{a<b}\mathcal W_{ab},
\end{equation}
where $\mathcal W_{ab}$ is the part of the brane web separating $\mathcal C_a$ from $\mathcal C_b$.  The nonempty portions $\mathcal Q_a$ cover the accessible part of $Q$.  Whenever the assignment of $Q$ changes from one party to another, a component of $\mathcal W$ ends on the brane.  We denote the number of such EOW-brane endpoints by
\begin{equation}
 N_Q\equiv \#(\mathcal W\cap Q).
\end{equation}
For the ordinary $q$-party multi-entropy all web segments have the same tension.  Consequently a non-degenerate bulk junction is trivalent and obeys the usual equal-tension force-balance condition, so the three incident geodesic segments meet pairwise at $2\pi/3$ \cite{Gadde:2022cqi,Harper:2024ker,Iizuka:2026qqg,Iizuka:2026ahd}.

We restrict the static analysis to planar, non-self-intersecting brane webs on the disk-like time-reflection-symmetric slice.  We also work with reduced configurations: closed web components, degree-two bulk points and bulk regions that do not touch any asymptotic boundary party are discarded, since they cannot lower the web length.  For $q$ adjacent parties
\begin{equation}
 A_1|A_2|\cdots|A_q
\end{equation}
the web is anchored at the $q-1$ boundary entangling points $b_1,\ldots,b_{q-1}$.  Planarity and the relative-homology condition imply that the nonempty portions of $Q$ appear in the same order as the boundary parties, although some intermediate parties need not reach the brane.  Equivalently, the $N_Q$ brane endpoints divide $Q$ into $N_Q+1$ portions whose labels form an ordered subsequence beginning with $A_1$ and ending with $A_q$.  This immediately gives
\begin{equation}\label{eq:static-terminal-bound}
 1\le N_Q\le q-1.
\end{equation}
The detailed proof, together with the enumeration of the allowed web connectivities, is given in Appendix~\ref{app:web-counting}.  It is useful already here to contrast this result with the ordinary AdS/CFT problem.  For a pure partition of a single connected asymptotic boundary into $q$ adjacent connected regions, the holographic multi-entropy is represented by a connected planar $q$-way cut.  Such a cut has $q-2$ equiangular trivalent junctions \cite{Gadde:2022cqi,Gadde:2023zzj,Iizuka:2025caq,Iizuka:2025bcc,Iizuka:2026qqg}; with the cyclic ordering of the boundary regions held fixed, its planar connectivity classes are counted by the Catalan number $C_{q-2}$.  The fully connected AdS/BCFT sector $N_Q=1$ has the same counting: the $q-1$ asymptotic entangling points together with the single endpoint on $Q$ form the same $q$ external legs of a planar trivalent tree.  The EOW brane, however, permits additional relative-homology sectors with $N_Q>1$, in which the web separates into several brane-ending components.  Summing over these sectors enlarges the counting by one Catalan level,
\begin{equation}\label{eq:main-catalan-comparison}
 \mathcal N_q^{\rm AdS/CFT}=C_{q-2},\qquad
 \mathcal N_{q,N_Q=1}^{\rm AdS/BCFT}=C_{q-2},\qquad
 \mathcal N_q^{\rm AdS/BCFT}=C_{q-1}.
\end{equation}
Here $C_n=(n+1)^{-1}\binom{2n}{n}$.  Thus the ordinary AdS/CFT sequence $1,2,5,14,\ldots$ for $q=3,4,5,6,\ldots$ becomes $2,5,14,42,\ldots$ in AdS/BCFT once all allowed brane-ending sectors are included.  Appendix~\ref{app:web-counting} derives both statements and makes the relation between them precise.  The only information needed for the calculations below is the resulting list of $q=3$ and $q=4$ topologies.

For three adjacent parties, $N_Q=1$ gives the connected $Y$ web and $N_Q=2$ gives two independent boundary-to-brane geodesics.  For four adjacent parties, $N_Q=1$ gives the two fully connected brane webs (phases~I and II), $N_Q=2$ gives the two partially factorized configurations (phases~III and IV), and $N_Q=3$ gives the completely factorized phase~V.  The corresponding assignments of the EOW brane are summarized in Table~\ref{tab:allowed-Q-partitions}.  Here a vertical bar denotes a brane endpoint at which the relative-homology assignment changes.
\begin{table}[H]
 \centering
 \begin{tabular}{c|c|c|c}
 configuration & $N_Q$ & brane web topology & assignment along $Q$\\\hline
 $q=3$, connected & 1 & one connected $Y$ web & $\mathcal Q_A|\mathcal Q_C$\\
 $q=3$, disconnected & 2 & two brane-ending geodesics & $\mathcal Q_A|\mathcal Q_B|\mathcal Q_C$\\\hline
 $q=4$, phases I/II & 1 & one connected four-leg web & $\mathcal Q_A|\mathcal Q_D$\\
 $q=4$, phase III & 2 & $Y$ web $+$ geodesic & $\mathcal Q_A|\mathcal Q_C|\mathcal Q_D$\\
 $q=4$, phase IV & 2 & geodesic $+$ $Y$ web & $\mathcal Q_A|\mathcal Q_B|\mathcal Q_D$\\
 $q=4$, phase V & 3 & three brane-ending geodesics & $\mathcal Q_A|\mathcal Q_B|\mathcal Q_C|\mathcal Q_D$
 \end{tabular}
 \caption{Admissible relative-homology assignments and planar brane web topologies for the adjacent tripartite and quadripartite partitions considered below.  The table classifies the topology only.  Each candidate must still be extremized, checked for geometric admissibility, and compared with the other on-shell saddles.}
 \label{tab:allowed-Q-partitions}
\end{table}

Thus the topology problem and the geometric extremization are logically separate.  The present section supplies the finite set of admissible connectivity classes; Secs.~\ref{three-party} and~\ref{four-party} determine the stationary representative and on-shell length in each class.  Appendix~\ref{app:web-counting} gives the detailed counting, first recalling the ordinary AdS/CFT result $C_{q-2}$ and then showing how relative homology on the EOW brane extends it to $C_{q-1}$ in AdS/BCFT.  The five-party example is worked out explicitly.
\subsection{Tripartite case}\label{three-party}

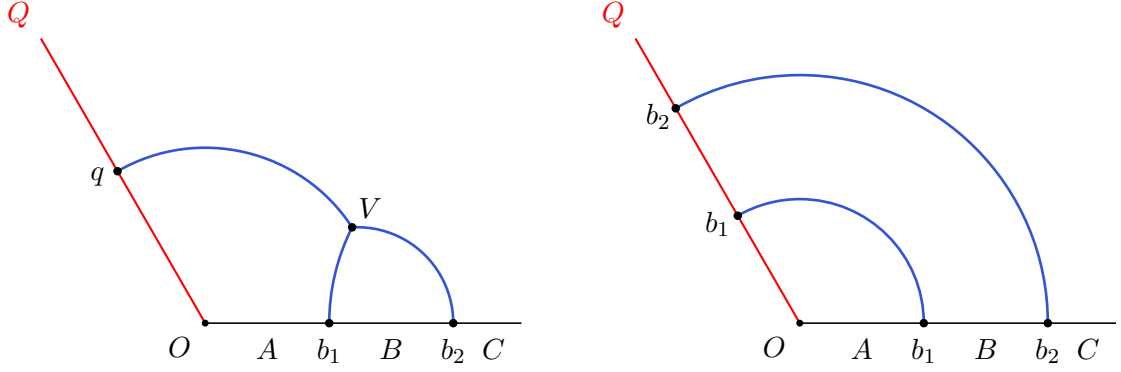
\begin{figure}[ht]
	\centering
	\def\TripartiteXAxisLength{5.10}
	\def\TripartiteYAxisLength{5.30}
	\begin{subfigure}[t]{0.48\linewidth}
	\centering
	\def\FigTwoQTheta{30}
	\def\FigTwoQA{2}
	\def\FigTwoQB{4}
	\begin{tikzpicture}[x=0.82cm,y=0.82cm,>=stealth]
		\pgfmathsetmacro{\FigTwoQBraneAngle}{90+\FigTwoQTheta}
		\pgfmathsetmacro{\FigTwoQW}{sqrt(\FigTwoQA*\FigTwoQB)}
		\pgfmathsetmacro{\FigTwoQQ}{\FigTwoQW}
		\pgfmathsetmacro{\FigTwoQVX}{
			-\FigTwoQW*(\FigTwoQA+\FigTwoQB-4*\FigTwoQW)
			/(2*(\FigTwoQA+\FigTwoQB-\FigTwoQW))}
		\pgfmathsetmacro{\FigTwoQVZ}{sqrt(\FigTwoQW*\FigTwoQW-\FigTwoQVX*\FigTwoQVX)}

		\pgfmathsetmacro{\FigTwoQLeftCenter}{
			(\FigTwoQVX*\FigTwoQVX+\FigTwoQVZ*\FigTwoQVZ-\FigTwoQA*\FigTwoQA)
			/(2*(\FigTwoQVX-\FigTwoQA))}
		\pgfmathsetmacro{\FigTwoQLeftRadius}{abs(\FigTwoQA-\FigTwoQLeftCenter)}
		\pgfmathsetmacro{\FigTwoQLeftVAngle}{
			atan2(\FigTwoQVZ,\FigTwoQVX-\FigTwoQLeftCenter)}
		\pgfmathsetmacro{\FigTwoQRightCenter}{
			(\FigTwoQVX*\FigTwoQVX+\FigTwoQVZ*\FigTwoQVZ-\FigTwoQB*\FigTwoQB)
			/(2*(\FigTwoQVX-\FigTwoQB))}
		\pgfmathsetmacro{\FigTwoQRightRadius}{abs(\FigTwoQB-\FigTwoQRightCenter)}
		\pgfmathsetmacro{\FigTwoQRightVAngle}{
			atan2(\FigTwoQVZ,\FigTwoQVX-\FigTwoQRightCenter)}
		\pgfmathsetmacro{\FigTwoQVertexAngle}{atan2(\FigTwoQVZ,\FigTwoQVX)}

		\coordinate (O) at (0,0);
		\coordinate (A) at (\FigTwoQA,0);
		\coordinate (B) at (\FigTwoQB,0);
		\coordinate (V) at (\FigTwoQVX,\FigTwoQVZ);
		\coordinate (Q) at (\FigTwoQBraneAngle:\FigTwoQQ);

		\draw[line width=0.6pt] (O) -- (\TripartiteXAxisLength,0);
		\draw[red,line width=0.8pt]
			(O) -- (\FigTwoQBraneAngle:\TripartiteYAxisLength)
			node[above left,text=red] {$Q$};

		\draw[UnifiedBraneWebBlue,line width=1pt] (A)
			arc[start angle=180,end angle=\FigTwoQLeftVAngle,
				radius=\FigTwoQLeftRadius];
		\draw[UnifiedBraneWebBlue,line width=1pt] (B)
			arc[start angle=0,end angle=\FigTwoQRightVAngle,
				radius=\FigTwoQRightRadius];
		\draw[UnifiedBraneWebBlue,line width=1pt] (Q)
			arc[start angle=\FigTwoQBraneAngle,end angle=\FigTwoQVertexAngle,
				radius=\FigTwoQQ];

		\fill (O) circle[radius=1.3pt] node[below left=2pt] {$O$};
		\fill (A) circle[radius=1.6pt] node[below=2pt] {$b_1$};
		\fill (B) circle[radius=1.6pt] node[below=2pt] {$b_2$};
		\node[below=2pt] at ({0.5*\FigTwoQA},0) {$A$};
		\node[below=2pt] at ({0.5*(\FigTwoQA+\FigTwoQB)},0) {$B$};
		\node[below=2pt]
			at ({0.5*(\FigTwoQB+\TripartiteXAxisLength)+0.1},0) {$C$};
		\fill (Q) circle[radius=1.6pt] node[left=1pt,yshift=-2pt] {$q$};
		\fill (V) circle[radius=1.6pt] node[above right=-2pt,yshift=1pt] {$V$};
	\end{tikzpicture}
	\label{fig:phase-A-Q-brane}
	\end{subfigure}
	\hfill
	\begin{subfigure}[t]{0.48\linewidth}
	\centering
	\def\FigThreeQTheta{30}
	\def\FigThreeQA{2}
	\def\FigThreeQB{4}
	\begin{tikzpicture}[x=0.82cm,y=0.82cm,>=stealth]
		\pgfmathsetmacro{\FigThreeQBraneAngle}{90+\FigThreeQTheta}

		\coordinate (O) at (0,0);
		\coordinate (A) at (\FigThreeQA,0);
		\coordinate (B) at (\FigThreeQB,0);
		\coordinate (QOne) at (\FigThreeQBraneAngle:\FigThreeQA);
		\coordinate (QTwo) at (\FigThreeQBraneAngle:\FigThreeQB);

		\draw[line width=0.6pt] (O) -- (\TripartiteXAxisLength,0);
		\draw[red,line width=0.8pt]
			(O) -- (\FigThreeQBraneAngle:\TripartiteYAxisLength)
			node[above left,text=red] {$Q$};

		\draw[UnifiedBraneWebBlue,line width=1pt] (A)
			arc[start angle=0,end angle=\FigThreeQBraneAngle,
				radius=\FigThreeQA];
		\draw[UnifiedBraneWebBlue,line width=1pt] (B)
			arc[start angle=0,end angle=\FigThreeQBraneAngle,
				radius=\FigThreeQB];

		\fill (O) circle[radius=1.3pt] node[below left=2pt] {$O$};
		\fill (A) circle[radius=1.6pt] node[below=2pt] {$b_1$};
		\fill (B) circle[radius=1.6pt] node[below=2pt] {$b_2$};
		\node[below=2pt] at ({0.5*\FigThreeQA},0) {$A$};
		\node[below=2pt] at ({0.5*(\FigThreeQA+\FigThreeQB)},0) {$B$};
		\node[below=2pt]
			at ({0.5*(\FigThreeQB+\TripartiteXAxisLength)+0.1},0) {$C$};
		\fill (QOne) circle[radius=1.6pt]
			node[left=-1pt,yshift=-3pt] {$b_1$};
		\fill (QTwo) circle[radius=1.6pt]
			node[left=-2pt,yshift=-3pt] {$b_2$};
	\end{tikzpicture}
	\label{fig:q_3_trivial_phase-Q-brane}
	\end{subfigure}
	\caption{The connected (left) and disconnected (right) phases of the brane web in the tripartite case.}
	\label{fig:phase-A}
	\label{fig:tripartite-phases-Q-brane}
\end{figure}

We begin with three adjacent boundary regions
\begin{equation}\label{eq:q3-partition-b}
 A=[0,b_1],\qquad B=[b_1,b_2],\qquad C=[b_2,\infty).
\end{equation}
Following the relative-homology classification above, there are two reduced planar topologies for the brane web: a connected $Y$ web with one trivalent vertex $V$, and a disconnected configuration consisting of two independent boundary-to-brane geodesics.  This is the AdS/BCFT analogue of the minimal multiway-cut geometry of holographic multi-entropy \cite{Gadde:2022cqi,Gadde:2023zzj,Penington:2022dhr}. They are shown in the left and right panels of Fig.~\ref{fig:tripartite-phases-Q-brane}, respectively.

For the connected phase, the total web length is
\begin{equation}\label{tri-total-length-1}
	\begin{aligned}
		\CL_{\text{tot}}&=\CL\left(V,b_1\right)+\CL\left(V,b_2\right)+\CL\left(V,q\right)\\
			&=\log\left[\frac{(x-b_1)^2+z^2}{\epsilon z}\right]
 			+\log\left[\frac{(x-b_2)^2+z^2}{\epsilon z}\right]\\
			&\quad+\operatorname{arccosh}\left[\frac{(x+q\sin\theta)^2+z^2+(q\cos\theta)^2}{2z\,q\cos\theta}\right],
	\end{aligned}
\end{equation}
where $q$ parametrizes the endpoint on the EOW brane and $V=(x,z)$ is the trivalent vertex; see Fig.~\ref{fig:tripartite-phases-Q-brane}.  We extremize the length within this topology,
\begin{equation}
	\partial_{q}\CL_{\text{tot}}=\partial_{x}\CL_{\text{tot}}=\partial_{z}\CL_{\text{tot}}=0.
\end{equation}
The $q$ variation first gives
\begin{equation}
	q=\sqrt{x^2+z^2}\,,
\end{equation}
so Eq.~\eqref{tri-total-length-1} becomes 
\begin{equation}\label{tri-total-length-phase-1-2}
	\CL_{\text{tot}}=\log\left[\frac{(w+x) \left(b_1^2-2 b_1 x+w^2\right) \left(b_2^2-2 b_2 x+w^2\right)}{\epsilon ^2 \left(w^2-x^2\right)^{3/2}}\right]+\operatorname{arccosh}(\sec\theta),
\end{equation}
where $w=\sqrt{x^2+z^2}$.  Extremizing with respect to $w$ and $x$ then gives
\begin{equation}\label{tri vertex coordinate}
	w=\sqrt{b_1\,b_2},\qquad x=-\frac{\sqrt{b_1\,b_2}\left(b_1+b_2-4 \sqrt{b_1\,b_2}\right)}{2 \left(b_1+b_2-\sqrt{b_1\,b_2}\right)}.
\end{equation}
Substituting this stationary point into Eq.~\eqref{tri-total-length-phase-1-2} gives the extremized connected-web length,
\begin{equation}\label{tri-brane-web-length-phase-1}
	\CL_{\rm ext}=\log \left[\frac{4(b_2-b_1)^2}{3\sqrt{3}\epsilon^2}\frac{\sqrt{b_2}+\sqrt{b_1}}{\sqrt{b_2}-\sqrt{b_1}}\right]+\operatorname{arccosh}(\sec\theta)\,.
\end{equation}
The holographic prescription \eqref{ME-holo} then gives
\begin{equation}\label{tri multi ee phase 1}
	S^{\left(3\right)}_{\text{con}}\left(A:B:C\right)=\frac{c}{6}\log \left[\frac{4(b_2-b_1)^2}{3\sqrt{3}\epsilon^2}\frac{\sqrt{b_2}+\sqrt{b_1}}{\sqrt{b_2}-\sqrt{b_1}}\right]+\frac{c}{6}\rho_0,
\end{equation}
where the subscript ``con'' denotes the connected topology and we used the Brown--Henneaux relation $c=3/(2G)$ \cite{Brown:1986nw}.

Before turning to the disconnected phase, it is worth checking the geometric admissibility of the brane web, i.e.\ that the web lies entirely inside the bulk region \eqref{eq:static-physical-wedge}. In particular, the vertex $V$ must lie inside this region. Using
 Eq.~\eqref{tri vertex coordinate}, this requires
\begin{equation}\label{eq:q3-static-admissibility}
 4\sqrt{b_2/b_1}-1-b_2/b_1+\sqrt3\,(b_2/b_1-1)\sinh\rho_0> 0.
\end{equation}
For $\rho_0\geq \log\sqrt{3}$ the inequality is automatically satisfied, so the ratio $b_2/b_1$ can take any value. For $\rho_0<\log\sqrt3$ the inequality imposes an upper bound $\eta_{\rm geom}$ on $b_2/b_1$,
\begin{equation}
 b_2/b_1<\eta_{\rm geom}= \left(\frac{2+\sqrt3\cosh\rho_0}{1-\sqrt3\sinh\rho_0}\right)^2.
\end{equation}
When $b_2/b_1$ reaches or exceeds $\eta_{\rm geom}$, the expression in Eq.~\eqref{tri multi ee phase 1} is no longer geometrically admissible and should be regarded only as a formal analytic continuation rather than as a physical saddle.

In the disconnected phase, the brane web factorizes into two independent boundary-to-brane geodesics. Each component is extremized by meeting the EOW brane orthogonally, exactly as for the disconnected RT surface in AdS/BCFT \cite{Takayanagi:2011zk,Fujita:2011fp,Sully:2020pza}; hence 
\begin{equation}\label{tri multi ee phase 2}
	S^{\left(3\right)}_{\text{dis}}\left(A:B:C\right)=S_A+S_C=\frac{c}{6}\log\left(\frac{4b_1 b_2}{\epsilon^2}\right)+\frac{c}{3}\rho_0,
\end{equation}
where we used Eq.~\eqref{HEE AdS/BCFT}.

\begin{figure}[H]
	\centering
	\includegraphics[width=0.49\linewidth]{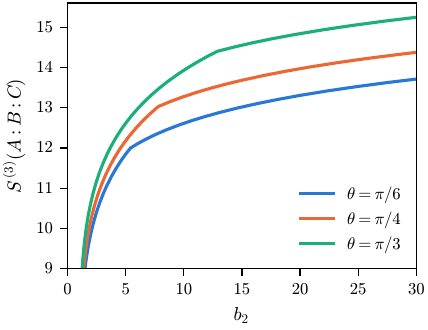}\hfill
	\includegraphics[width=0.49\linewidth]{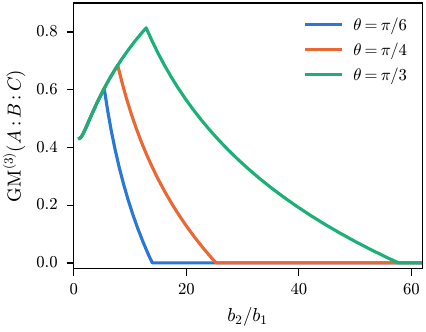}
	\caption{$S^{(3)}$ (left) and $\mathrm{GM}^{(3)}$ (right) for $c=6$, $\epsilon=0.02$ and three brane angles $\theta$.  In the left panel $b_1=1$ is fixed and $S^{(3)}$ is plotted against $b_2$; the right panel uses the ratio $b_2/b_1$.  The kink of each curve in the left panel is the web transition at $b_2/b_1=\eta_{c,2}$.  In the right panel $\mathrm{GM}^{(3)}$ peaks at $\eta_{c,2}$ and vanishes beyond the ordinary RT transition at $\eta_{c,1}$.}
	\label{fig:tri-multi}
\end{figure}

To obtain the physical tripartite multi-entropy, we compare the two phases, Eqs.~\eqref{tri multi ee phase 1} and \eqref{tri multi ee phase 2}:
\begin{equation}
	S_{\text{con}}^{\left(3\right)}\leq S_{\text{dis}}^{\left(3\right)},\quad \text{for}\quad  b_2/b_1\leq \eta_{c,2},\quad  S_{\text{con}}^{\left(3\right)}> S_{\text{dis}}^{\left(3\right)},\quad \text{for}\quad b_2/b_1>\eta_{c,2}.
\end{equation}
The transition point $\eta_{c,2}$ is fixed by equality of the two on-shell lengths,
\begin{equation}
	\log \left(\frac{(\eta_{c,2}-1)^2}{4 \eta_{c,2}}\right)+\log \left(\frac{\sqrt{\eta_{c,2}}+1}{2 \left(\sqrt{\eta_{c,2}}-1\right)}\right)+3 \log \left(\frac{2}{\sqrt{3}}\right)-\rho_0=0.
\end{equation}
This equation does not admit a useful closed-form solution. Furthermore, one can easily check that $\eta_{\rm geom}>\eta_{c,2}$ whenever $\rho_0<\log\sqrt{3}$. In other words, before the connected phase becomes geometrically inadmissible, the tripartite multi-entropy has already entered the disconnected phase, so that it is simply the lower envelope of Eqs.~\eqref{tri multi ee phase 1} and \eqref{tri multi ee phase 2}:
\begin{equation}\label{S3 A B C}
	S^{\left(3\right)}\left(A:B:C\right)=\begin{cases}
		\frac{c}{6}\log\left(\frac{4b_1b_2}{\epsilon^2}\right)+\frac{c}{3}\rho_0, & b_2/b_1>\eta_{c,2}, \\
		\frac{c}{6}\log \left[\frac{4(b_2-b_1)^2}{3\sqrt{3}\epsilon^2}\frac{\sqrt{b_2}+\sqrt{b_1}}{\sqrt{b_2}-\sqrt{b_1}}\right]+\frac{c}{6}\rho_0, & b_2/b_1<\eta_{c,2}.
	\end{cases}
\end{equation}
The dependence of $S^{(3)}$ on $b_2$ is shown in Fig.~\ref{fig:tri-multi}.

\begin{figure}[ht]
  \centering
  \includegraphics[width=0.92\linewidth]{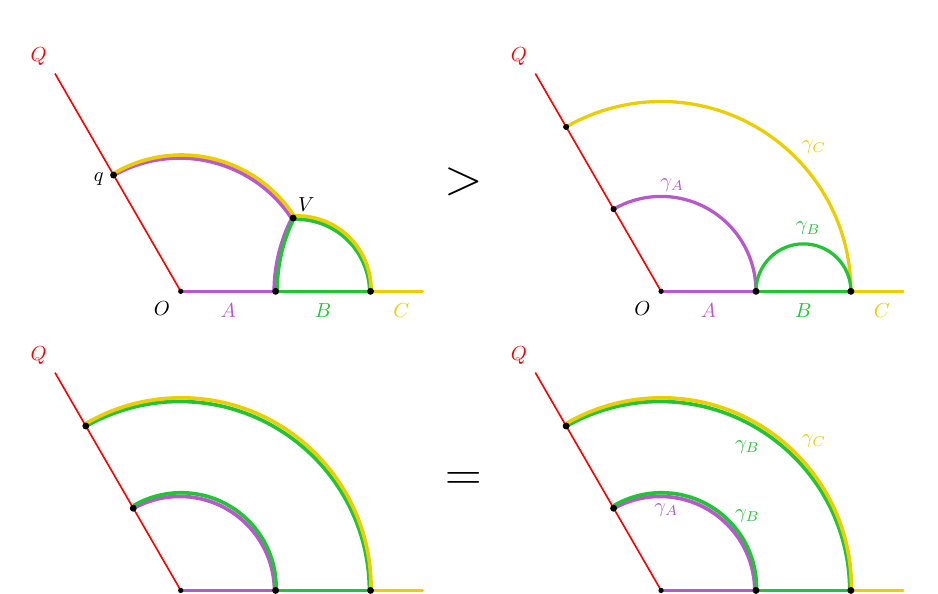}
  \caption{Geometric decomposition underlying the non-negativity of $\mathrm{GM}^{(3)}$.  $\gamma_A$, $\gamma_B$ and $\gamma_C$ represent the RT surfaces of $A$, $B$ and $C$, respectively.  $2S^{(3)}$ and $S_A+S_B+S_C$ correspond to the minimal values of the two left and the two right panels, respectively; hence $\mathrm{GM}^{(3)}\geq0$.  Top row: connected web, where the inequality is strict.  Bottom row: fully factorized configuration, where it is saturated.}
  \label{fig:pairwise-coloured-web-comparisons}
\end{figure}

We next consider the genuine tripartite multi-entropy $\mathrm{GM}^{(3)}$ of Eq.~\eqref{eq:genuine-multi} \cite{Iizuka:2025ioc,Iizuka:2025caq}.  For time-reflection-symmetric holographic states, $\mathrm{GM}^{(3)}$ is non-negative as a consequence of geometric inequalities \cite{Iizuka:2025ioc,Iizuka:2025caq}\footnote{This non-negativity is not a universal property of arbitrary quantum states.}.  The same cut-and-paste argument applies in the present AdS/BCFT geometry: two copies of the minimal brane web can be repartitioned into candidate homologous surfaces for $A$, $B$, and $C$, giving
\begin{equation}
	2S^{\left(3\right)}\left(A:B:C\right)-\left(S_A+S_B+S_C\right)\geq 0 \quad \Rightarrow \quad \mathrm{GM}^{\left(3\right)}\left(A:B:C\right)\geq 0,
\end{equation}
because each entanglement entropy is bounded above by the corresponding candidate homologous surface; see Fig.~\ref{fig:pairwise-coloured-web-comparisons}.

To obtain an explicit expression for $\mathrm{GM}^{\left(3\right)}\left(A:B:C\right)$, we substitute Eqs.~\eqref{HEE AdS/BCFT}, \eqref{HEE two phases}, and \eqref{S3 A B C} into Eq.~\eqref{eq:genuine-multi}, which gives
\begin{equation}
		\mathrm{GM}^{(3)}(A : B : C)=\begin{cases}
		0, & b_2/b_1>\eta_{c,1}, \\
		-\frac{c}{12}\log\left[\frac{\left(b_2-b_1\right)^2}{4b_1b_2}\right]+\frac{c}{6}\rho_0, & \eta_{c,2}<b_2/b_1<\eta_{c,1},\\
		\frac{c}{6}\log\left[\frac{\left(\sqrt{b_1}+\sqrt{b_2}\right)^2}{4\sqrt{b_1b_2}}\right]+\frac{c}{2}\log\left(\frac{2}{\sqrt{3}}\right), & b_2/b_1<\eta_{c,2}.
	\end{cases}
\end{equation}
The dependence of $\mathrm{GM}^{(3)}$ on $b_2/b_1$ is shown in Fig.~\ref{fig:tri-multi}.  The three physical branches join continuously and remain non-negative.  The connected branch increases with $b_2/b_1$, whereas the intermediate branch decreases; their meeting point at $\eta_{c,2}$ is therefore the maximum of $\mathrm{GM}^{(3)}$:
\begin{equation}
	\begin{aligned}
		\max\left[\mathrm{GM}^{(3)}(A : B : C)\right]=&\:\mathrm{GM}^{(3)}(A : B : C)|_{b_2/b_1\rightarrow \eta_{c,2}}\\
		=&\:\frac{c}{6}\log\left[\frac{\left(1+\sqrt{\eta_{c,2}}\right)^2}{4\sqrt{\eta_{c,2}}}\right]+\frac{c}{2}\log\left(\frac{2}{\sqrt{3}}\right)>\frac{c}{2}\log\left(\frac{2}{\sqrt{3}}\right).
	\end{aligned}
\end{equation}
Note that the web transition at $\eta_{c,2}$ precedes the ordinary RT transition, $\eta_{c,2}<\eta_{c,1}$: in the window $\eta_{c,2}<b_2/b_1<\eta_{c,1}$, $S^{(3)}$ has already selected the disconnected phase while $S_B$ still probes a connected RT surface. This intermediate window makes explicit that multipartite reorganization is not tied to the ordinary RT transition.  Eventually, when $S^{\left(3\right)}$ and $S_B$ both enter the disconnected phase, $\mathrm{GM}^{\left(3\right)}$ vanishes. In contrast to the corresponding AdS/CFT setting, where $\mathrm{GM}^{(3)}$ takes the constant value $\frac{c}{2}\log(2/\sqrt{3})$ \cite{Iizuka:2025ioc,Iizuka:2025caq}, the AdS/BCFT result depends nontrivially on the boundary parameters. This behavior shows explicitly that the EOW brane can reorganize multipartite entanglement by opening relative-homology channels that have no analogue for a single connected asymptotic boundary; this complements recent work emphasizing the role of multiway-cut junctions and genuine multipartite structure in holography \cite{Iizuka:2025bcc,Iizuka:2026qqg,Iizuka:2026ahd}. These observations motivate further investigations into how the geometry and dynamics of the EOW brane control multipartite entanglement in AdS/BCFT.

\subsection{Quadripartite case}\label{four-party}

We now consider four adjacent boundary intervals,
\begin{equation}\label{eq:q4-partition-b}
 A=[0,b_1],\qquad B=[b_1,b_2],\qquad C=[b_2,b_3],\qquad D=[b_3,\infty).
\end{equation}
Following the classification in Sec.~\ref{par:extended-homology-review}, there are five reduced quadripartite topologies, shown in Fig.~\ref{fig:quadripartite-phases}. 

\paragraph{Phases \uppercase\expandafter{\romannumeral 1} and \uppercase\expandafter{\romannumeral 2}.}
Phases~I and II are fully connected and contain two trivalent vertices, $U$ and $V$.  Instead of solving the coupled coordinate extremization directly, we impose the equivalent equal-tension force-balance condition at each vertex.  The three geodesic segments incident on a trivalent vertex therefore meet pairwise at $2\pi/3$\footnote{For an equal-tension multiway cut this is precisely the vanishing of the first variation with respect to the vertex position \cite{Gadde:2022cqi,Gadde:2023zzj,Harper:2024ker,Iizuka:2026qqg}.}.  To implement this condition, we extend the geodesic containing the internal edge $UV$ until it intersects the asymptotic boundary at two auxiliary points $p$ and $m$; see Fig.~\ref{fig:me4}.  The sub-web $(b_1,b_2)\to V\to m$ can then be treated as the standard tripartite Steiner geometry in AdS/CFT \cite{Gadde:2022cqi,Gadde:2023zzj, Harper:2024ker}. The vertex $V$ is therefore given by Eq.~\eqref{location vertex} with $b_3$ replaced by $m$,
\begin{equation}\label{coordinate vertex-V}
	\begin{aligned}
		x_{V}=\:&\frac{(b_1^2+b_2 m)(b_2+m)+b_1(b_2^2-6 b_2 m+m^2)}
		{2(b_1^2+b_2^2+m^2-b_1b_2-b_2m-b_1m)},\\
		z_{V}=\:&\frac{\sqrt{3}(b_2-b_1)(m-b_2)(m-b_1)}
		{2(b_1^2+b_2^2+m^2-b_1b_2-b_2m-b_1m)},
	\end{aligned}
\end{equation}
\def\QuadripartiteXAxisLength{7.35}

\def\QuadripartiteShortXAxisLength{5.15}

\def\QuadripartiteYAxisLength{5.30}

\def\QuadripartiteFigureSevenShortXAxisLength{5.15}

\def\QuadripartiteFigureSevenPhaseAYAxisLength{4.35}

\def\QuadripartiteFigureSevenPhaseBYAxisLength{4.05}

\def\QuadripartiteFigureSevenPhaseCYAxisLength{5.30}

\def\QuadripartiteFigureSevenPhaseDYAxisLength{4.55}

\def\QuadripartiteFigureSevenPhaseEYAxisLength{5.30}

\def\QuadripartitePhaseAYAxisLength{4.35}

\def\QuadripartitePhaseBYAxisLength{\QuadripartitePhaseAYAxisLength}

\def\QuadripartitePhaseDYAxisLength{\QuadripartiteYAxisLength}

\def\QuadripartiteRegionLabelOffset{4pt}

\newif\ifQuadripartiteIslandView

\QuadripartiteIslandViewfalse

\definecolor{QuadripartiteAColor}{RGB}{50,85,210}

\definecolor{QuadripartiteBColor}{RGB}{35,195,55}

\definecolor{QuadripartiteCColor}{RGB}{235,205,0}

\definecolor{QuadripartiteDColor}{RGB}{215,55,70}

\colorlet{QuadripartiteAFillColor}{QuadripartiteAColor!18}

\colorlet{QuadripartiteBFillColor}{QuadripartiteBColor!20}

\colorlet{QuadripartiteCFillColor}{QuadripartiteCColor!16}

\colorlet{QuadripartiteDFillColor}{QuadripartiteDColor!14}

\tikzset{
	QuadripartiteQLabel/.style={left=0pt,yshift=-3pt},
	QuadripartiteIndexedQLabel/.style={left=0pt,xshift=1.5pt,
		yshift=-3pt},
	QuadripartiteIslandLabel/.style={left=0pt,yshift=-3pt},
	QuadripartiteARegion/.style={shade,bottom color=QuadripartiteAColor!24,
		top color=white},
	QuadripartiteBRegion/.style={shade,bottom color=QuadripartiteBColor!22,
		top color=white},
	QuadripartiteCRegion/.style={shade,left color=QuadripartiteCColor!30,
		right color=QuadripartiteCColor!5,shading angle=22},
	QuadripartiteDRegion/.style={shade,
		bottom color=QuadripartiteDColor!38,
		middle color=QuadripartiteDColor!22,
		top color=QuadripartiteDColor!10}
}

\newcommand{\DrawQuadripartitePhaseA}{%
	\def\FigFiveTheta{30}
	\ifQuadripartiteIslandView
		\def\FigFiveA{2}
		\def\FigFiveB{3}
		\def\FigFiveC{4}
		\def\FigFiveP{2.5643093177}
	\else
		\def\FigFiveA{2}
		\def\FigFiveB{3}
		\def\FigFiveC{4}
		\def\FigFiveP{2.5643093177}
	\fi
	\begin{tikzpicture}[x=0.68cm,y=0.68cm,>=stealth]
		\pgfmathsetmacro{\FigFiveBraneAngle}{90+\FigFiveTheta}
		\ifQuadripartiteIslandView
			\def\FigFiveXAxisLength{\QuadripartiteShortXAxisLength}
			\def\FigFiveYAxisLength{\QuadripartitePhaseAYAxisLength}
		\else
			\def\FigFiveXAxisLength{\QuadripartiteXAxisLength}
			\def\FigFiveYAxisLength{\QuadripartiteFigureSevenPhaseAYAxisLength}
		\fi
		\pgfmathsetmacro{\FigFiveIslandZCut}{
			\FigFiveYAxisLength*sin(\FigFiveBraneAngle)}
		\pgfmathsetmacro{\FigFiveCutLeftX}{
			\FigFiveYAxisLength*cos(\FigFiveBraneAngle)}
		\pgfmathsetmacro{\FigFiveM}{
			(2*\FigFiveA*\FigFiveB-(\FigFiveA+\FigFiveB)*\FigFiveP)
			/(\FigFiveA+\FigFiveB-2*\FigFiveP)}
		\pgfmathsetmacro{\FigFiveQ}{sqrt(\FigFiveP*\FigFiveC)}

		\pgfmathsetmacro{\FigFiveUX}{
			-\FigFiveQ*(\FigFiveP+\FigFiveC-4*\FigFiveQ)
			/(2*(\FigFiveP+\FigFiveC-\FigFiveQ))}
		\pgfmathsetmacro{\FigFiveUZ}{
			sqrt(\FigFiveP*\FigFiveC-\FigFiveUX*\FigFiveUX)}
		\pgfmathsetmacro{\FigFiveVDen}{
			2*(\FigFiveA*\FigFiveA+\FigFiveB*\FigFiveB
			+\FigFiveM*\FigFiveM-\FigFiveA*\FigFiveB
			-\FigFiveB*\FigFiveM-\FigFiveA*\FigFiveM)}
		\pgfmathsetmacro{\FigFiveVX}{
			((\FigFiveA*\FigFiveA+\FigFiveB*\FigFiveM)
			*(\FigFiveB+\FigFiveM)
			+\FigFiveA*(\FigFiveB*\FigFiveB
			-6*\FigFiveB*\FigFiveM+\FigFiveM*\FigFiveM))
			/\FigFiveVDen}
		\pgfmathsetmacro{\FigFiveVZ}{
			sqrt(3)*(\FigFiveB-\FigFiveA)*(\FigFiveM-\FigFiveB)
			*(\FigFiveM-\FigFiveA)/\FigFiveVDen}

		\pgfmathsetmacro{\FigFiveACenter}{
			(\FigFiveVX*\FigFiveVX+\FigFiveVZ*\FigFiveVZ
			-\FigFiveA*\FigFiveA)/(2*(\FigFiveVX-\FigFiveA))}
		\pgfmathsetmacro{\FigFiveARadius}{abs(\FigFiveA-\FigFiveACenter)}
		\pgfmathsetmacro{\FigFiveAVAngle}{
			atan2(\FigFiveVZ,\FigFiveVX-\FigFiveACenter)}
		\pgfmathsetmacro{\FigFiveBCenter}{
			(\FigFiveVX*\FigFiveVX+\FigFiveVZ*\FigFiveVZ
			-\FigFiveB*\FigFiveB)/(2*(\FigFiveVX-\FigFiveB))}
		\pgfmathsetmacro{\FigFiveBRadius}{abs(\FigFiveB-\FigFiveBCenter)}
		\pgfmathsetmacro{\FigFiveBVAngle}{
			atan2(\FigFiveVZ,\FigFiveVX-\FigFiveBCenter)}
		\pgfmathsetmacro{\FigFiveCCenter}{
			(\FigFiveUX*\FigFiveUX+\FigFiveUZ*\FigFiveUZ
			-\FigFiveC*\FigFiveC)/(2*(\FigFiveUX-\FigFiveC))}
		\pgfmathsetmacro{\FigFiveCRadius}{abs(\FigFiveC-\FigFiveCCenter)}
		\pgfmathsetmacro{\FigFiveCUAngle}{
			atan2(\FigFiveUZ,\FigFiveUX-\FigFiveCCenter)}
		\pgfmathsetmacro{\FigFiveUAngle}{atan2(\FigFiveUZ,\FigFiveUX)}

		\pgfmathsetmacro{\FigFiveMidCenter}{0.5*(\FigFiveP+\FigFiveM)}
		\pgfmathsetmacro{\FigFiveMidRadius}{0.5*(\FigFiveM-\FigFiveP)}
		\pgfmathsetmacro{\FigFiveVOnMidAngle}{
			atan2(\FigFiveVZ,\FigFiveVX-\FigFiveMidCenter)}
		\pgfmathsetmacro{\FigFiveUOnMidAngle}{
			atan2(\FigFiveUZ,\FigFiveUX-\FigFiveMidCenter)}

		\coordinate (O) at (0,0);
		\coordinate (A) at (\FigFiveA,0);
		\coordinate (P) at (\FigFiveP,0);
		\coordinate (B) at (\FigFiveB,0);
		\coordinate (C) at (\FigFiveC,0);
		\coordinate (M) at (\FigFiveM,0);
		\coordinate (Q) at (\FigFiveBraneAngle:\FigFiveQ);
		\coordinate (QEnd) at
			(\FigFiveBraneAngle:\FigFiveYAxisLength);
		\coordinate (QCut) at
			(\FigFiveCutLeftX,\FigFiveIslandZCut);
		\coordinate (U) at (\FigFiveUX,\FigFiveUZ);
		\coordinate (V) at (\FigFiveVX,\FigFiveVZ);

		\ifQuadripartiteIslandView
			\path[QuadripartiteARegion,draw=none] (O) -- (A)
				arc[start angle=180,end angle=\FigFiveAVAngle,
					radius=\FigFiveARadius]
				arc[start angle=\FigFiveVOnMidAngle,
					end angle=\FigFiveUOnMidAngle,
					radius=\FigFiveMidRadius]
				arc[start angle=\FigFiveUAngle,
					end angle=\FigFiveBraneAngle,radius=\FigFiveQ] -- cycle;
			\path[QuadripartiteBRegion,draw=none] (A) -- (B)
				arc[start angle=0,end angle=\FigFiveBVAngle,
					radius=\FigFiveBRadius]
				arc[start angle=\FigFiveAVAngle,end angle=180,
					radius=\FigFiveARadius] -- cycle;
			\path[QuadripartiteCRegion,draw=none] (B) -- (C)
				arc[start angle=0,end angle=\FigFiveCUAngle,
					radius=\FigFiveCRadius]
				arc[start angle=\FigFiveUOnMidAngle,
					end angle=\FigFiveVOnMidAngle,
					radius=\FigFiveMidRadius]
				arc[start angle=\FigFiveBVAngle,end angle=0,
					radius=\FigFiveBRadius] -- cycle;
			\path[QuadripartiteDRegion,draw=none] (C)
				-- (\FigFiveXAxisLength,0)
				-- (\FigFiveXAxisLength,\FigFiveIslandZCut)
				-- (QCut) -- (Q)
				arc[start angle=\FigFiveBraneAngle,
					end angle=\FigFiveUAngle,radius=\FigFiveQ]
				arc[start angle=\FigFiveCUAngle,end angle=0,
					radius=\FigFiveCRadius] -- cycle;

			\draw[QuadripartiteAColor,line width=0.9pt] (O) -- (A)
				node[midway,below=4pt] {$A$};
			\draw[QuadripartiteBColor,line width=0.9pt] (A) -- (B)
				node[midway,below=4pt] {$B$};
			\draw[QuadripartiteCColor,line width=0.9pt] (B) -- (C)
				node[midway,below=4pt] {$C$};
			\draw[QuadripartiteDColor,line width=0.9pt]
				(C) -- (\FigFiveXAxisLength,0)
				node[midway,below=4pt] {$D$};
			\draw[QuadripartiteAColor,line width=0.9pt] (O) -- (Q);
			\draw[QuadripartiteDColor,line width=0.9pt] (Q) -- (QEnd)
				node[above left,text=black] {$Q$};
			\node[QuadripartiteIslandLabel,text=QuadripartiteAColor]
				at (\FigFiveBraneAngle:{0.50*\FigFiveQ}) {$I_A^{(4)}$};
			\node[QuadripartiteIslandLabel,text=QuadripartiteDColor]
				at (\FigFiveBraneAngle:{0.50*(\FigFiveQ+\FigFiveYAxisLength)})
				{$I_D^{(4)}$};
		\else
			\draw[line width=0.6pt] (O) -- (\FigFiveXAxisLength,0);
			\draw[red,line width=0.8pt]
				(O) -- (\FigFiveBraneAngle:\FigFiveYAxisLength)
				node[above left,text=red] {$Q$};
			\node[below=\QuadripartiteRegionLabelOffset]
				at ({0.5*\FigFiveA},0) {$A$};
			\node[below=\QuadripartiteRegionLabelOffset]
				at ({0.5*(\FigFiveA+\FigFiveB)},0) {$B$};
			\node[below=\QuadripartiteRegionLabelOffset]
				at ({0.5*(\FigFiveB+\FigFiveC)},0) {$C$};
			\node[below=\QuadripartiteRegionLabelOffset]
				at ({0.5*(\FigFiveC+\FigFiveXAxisLength)},0) {$D$};
		\fi

		\ifQuadripartiteIslandView\else
			\draw[densely dashed,gray,line width=0.55pt] (P)
				arc[start angle=180,end angle=0,radius=\FigFiveMidRadius];
		\fi

		\draw[blue,line width=1pt] (A)
			arc[start angle=180,end angle=\FigFiveAVAngle,
				radius=\FigFiveARadius];
		\draw[blue,line width=1pt] (B)
			arc[start angle=0,end angle=\FigFiveBVAngle,
				radius=\FigFiveBRadius];
		\draw[blue,line width=1pt] (C)
			arc[start angle=0,end angle=\FigFiveCUAngle,
				radius=\FigFiveCRadius];
		\draw[blue,line width=1pt] (Q)
			arc[start angle=\FigFiveBraneAngle,end angle=\FigFiveUAngle,
				radius=\FigFiveQ];
		\draw[blue,line width=1pt] (V)
			arc[start angle=\FigFiveVOnMidAngle,end angle=\FigFiveUOnMidAngle,
				radius=\FigFiveMidRadius];

		\fill (O) circle[radius=1.3pt] node[below left=2pt] {$O$};
		\fill (A) circle[radius=1.6pt] node[below=2pt] {$b_1$};
		\ifQuadripartiteIslandView\else
			\fill[gray] (P) circle[radius=1.3pt] node[below,text=black,yshift=5pt,left=-2pt] {$p$};
		\fi
		\fill (B) circle[radius=1.6pt] node[below=2pt] {$b_2$};
		\fill (C) circle[radius=1.6pt] node[below=2pt] {$b_3$};
		\ifQuadripartiteIslandView\else
			\fill[gray] (M) circle[radius=1.3pt] node[below,text=black,yshift =6pt,left=-15pt] {$m$};
		\fi
		\fill (Q) circle[radius=1.6pt]
			node[QuadripartiteQLabel] {$q$};
		\fill (U) circle[radius=1.6pt];
		\node[above left=-10pt,yshift=11pt] at (U) {$U$};
		\fill (V) circle[radius=1.6pt];
		\node[above left=-3pt,yshift=-1pt] at (V) {$V$};
	\end{tikzpicture}%
}

\newcommand{\DrawQuadripartitePhaseB}{%
	\def\FigSixTheta{30}
	\ifQuadripartiteIslandView
		\def\FigSixA{2}
		\def\FigSixB{3}
		\def\FigSixC{4}
		\def\FigSixP{3.3944487245}
	\else
		\def\FigSixA{2}
		\def\FigSixB{3}
		\def\FigSixC{4}
		\def\FigSixP{3.3944487245}
	\fi
	\begin{tikzpicture}[x=0.68cm,y=0.68cm,>=stealth]
		\pgfmathsetmacro{\FigSixBraneAngle}{90+\FigSixTheta}
		\ifQuadripartiteIslandView
			\def\FigSixXAxisLength{\QuadripartiteShortXAxisLength}
			\def\FigSixYAxisLength{\QuadripartitePhaseBYAxisLength}
		\else
			\def\FigSixXAxisLength{\QuadripartiteFigureSevenShortXAxisLength}
			\def\FigSixYAxisLength{\QuadripartiteFigureSevenPhaseBYAxisLength}
		\fi
		\pgfmathsetmacro{\FigSixIslandZCut}{
			\FigSixYAxisLength*sin(\FigSixBraneAngle)}
		\pgfmathsetmacro{\FigSixCutLeftX}{
			\FigSixYAxisLength*cos(\FigSixBraneAngle)}
		\pgfmathsetmacro{\FigSixM}{
			(2*\FigSixC*\FigSixB-(\FigSixC+\FigSixB)*\FigSixP)
			/(\FigSixC+\FigSixB-2*\FigSixP)}
		\pgfmathsetmacro{\FigSixQ}{sqrt(\FigSixA*\FigSixP)}

		\pgfmathsetmacro{\FigSixUX}{
			-\FigSixQ*(\FigSixA+\FigSixP-4*\FigSixQ)
			/(2*(\FigSixA+\FigSixP-\FigSixQ))}
		\pgfmathsetmacro{\FigSixUZ}{
			sqrt(\FigSixA*\FigSixP-\FigSixUX*\FigSixUX)}
		\pgfmathsetmacro{\FigSixVDen}{
			2*(\FigSixM*\FigSixM+\FigSixB*\FigSixB
			+\FigSixC*\FigSixC-\FigSixM*\FigSixB
			-\FigSixB*\FigSixC-\FigSixM*\FigSixC)}
		\pgfmathsetmacro{\FigSixVX}{
			((\FigSixM*\FigSixM+\FigSixB*\FigSixC)
			*(\FigSixB+\FigSixC)
			+\FigSixM*(\FigSixB*\FigSixB
			-6*\FigSixB*\FigSixC+\FigSixC*\FigSixC))
			/\FigSixVDen}
		\pgfmathsetmacro{\FigSixVZ}{
			sqrt(3)*(\FigSixB-\FigSixM)*(\FigSixC-\FigSixB)
			*(\FigSixC-\FigSixM)/\FigSixVDen}

		\pgfmathsetmacro{\FigSixACenter}{
			(\FigSixUX*\FigSixUX+\FigSixUZ*\FigSixUZ
			-\FigSixA*\FigSixA)/(2*(\FigSixUX-\FigSixA))}
		\pgfmathsetmacro{\FigSixARadius}{abs(\FigSixA-\FigSixACenter)}
		\pgfmathsetmacro{\FigSixAUAngle}{
			atan2(\FigSixUZ,\FigSixUX-\FigSixACenter)}
		\pgfmathsetmacro{\FigSixBCenter}{
			(\FigSixVX*\FigSixVX+\FigSixVZ*\FigSixVZ
			-\FigSixB*\FigSixB)/(2*(\FigSixVX-\FigSixB))}
		\pgfmathsetmacro{\FigSixBRadius}{abs(\FigSixB-\FigSixBCenter)}
		\pgfmathsetmacro{\FigSixBVAngle}{
			atan2(\FigSixVZ,\FigSixVX-\FigSixBCenter)}
		\pgfmathsetmacro{\FigSixCCenter}{
			(\FigSixVX*\FigSixVX+\FigSixVZ*\FigSixVZ
			-\FigSixC*\FigSixC)/(2*(\FigSixVX-\FigSixC))}
		\pgfmathsetmacro{\FigSixCRadius}{abs(\FigSixC-\FigSixCCenter)}
		\pgfmathsetmacro{\FigSixCVAngle}{
			atan2(\FigSixVZ,\FigSixVX-\FigSixCCenter)}
		\pgfmathsetmacro{\FigSixUAngle}{atan2(\FigSixUZ,\FigSixUX)}

		\pgfmathsetmacro{\FigSixMidCenter}{0.5*(\FigSixM+\FigSixP)}
		\pgfmathsetmacro{\FigSixMidRadius}{0.5*(\FigSixP-\FigSixM)}
		\pgfmathsetmacro{\FigSixUOnMidAngle}{
			atan2(\FigSixUZ,\FigSixUX-\FigSixMidCenter)}
		\pgfmathsetmacro{\FigSixVOnMidAngle}{
			atan2(\FigSixVZ,\FigSixVX-\FigSixMidCenter)}

		\coordinate (O) at (0,0);
		\coordinate (M) at (\FigSixM,0);
		\coordinate (A) at (\FigSixA,0);
		\coordinate (B) at (\FigSixB,0);
		\coordinate (P) at (\FigSixP,0);
		\coordinate (C) at (\FigSixC,0);
		\coordinate (Q) at (\FigSixBraneAngle:\FigSixQ);
		\coordinate (QEnd) at
			(\FigSixBraneAngle:\FigSixYAxisLength);
		\coordinate (QCut) at
			(\FigSixCutLeftX,\FigSixIslandZCut);
		\coordinate (U) at (\FigSixUX,\FigSixUZ);
		\coordinate (V) at (\FigSixVX,\FigSixVZ);

		\ifQuadripartiteIslandView
			\path[QuadripartiteARegion,draw=none] (O) -- (A)
				arc[start angle=180,end angle=\FigSixAUAngle,
					radius=\FigSixARadius]
				arc[start angle=\FigSixUAngle,
					end angle=\FigSixBraneAngle,radius=\FigSixQ] -- cycle;
			\path[QuadripartiteBRegion,draw=none] (A) -- (B)
				arc[start angle=180,end angle=\FigSixBVAngle,
					radius=\FigSixBRadius]
				arc[start angle=\FigSixVOnMidAngle,
					end angle=\FigSixUOnMidAngle,
					radius=\FigSixMidRadius]
				arc[start angle=\FigSixAUAngle,end angle=180,
					radius=\FigSixARadius] -- cycle;
			\path[QuadripartiteCRegion,draw=none] (B) -- (C)
				arc[start angle=0,end angle=\FigSixCVAngle,
					radius=\FigSixCRadius]
				arc[start angle=\FigSixBVAngle,end angle=180,
					radius=\FigSixBRadius] -- cycle;
			\path[QuadripartiteDRegion,draw=none] (C)
				-- (\FigSixXAxisLength,0)
				-- (\FigSixXAxisLength,\FigSixIslandZCut)
				-- (QCut) -- (Q)
				arc[start angle=\FigSixBraneAngle,
					end angle=\FigSixUAngle,radius=\FigSixQ]
				arc[start angle=\FigSixUOnMidAngle,
					end angle=\FigSixVOnMidAngle,
					radius=\FigSixMidRadius]
				arc[start angle=\FigSixCVAngle,end angle=0,
					radius=\FigSixCRadius] -- cycle;

			\draw[QuadripartiteAColor,line width=0.9pt] (O) -- (A)
				node[midway,below=4pt] {$A$};
			\draw[QuadripartiteBColor,line width=0.9pt] (A) -- (B)
				node[midway,below=4pt] {$B$};
			\draw[QuadripartiteCColor,line width=0.9pt] (B) -- (C)
				node[midway,below=4pt] {$C$};
			\draw[QuadripartiteDColor,line width=0.9pt]
				(C) -- (\FigSixXAxisLength,0)
				node[midway,below=4pt] {$D$};
			\draw[QuadripartiteAColor,line width=0.9pt] (O) -- (Q);
			\draw[QuadripartiteDColor,line width=0.9pt] (Q) -- (QEnd)
				node[above left,text=black] {$Q$};
			\node[QuadripartiteIslandLabel,text=QuadripartiteAColor]
				at (\FigSixBraneAngle:{0.50*\FigSixQ}) {$I_A^{(4)}$};
			\node[QuadripartiteIslandLabel,text=QuadripartiteDColor]
				at (\FigSixBraneAngle:{0.50*(\FigSixQ+\FigSixYAxisLength)})
				{$I_D^{(4)}$};
		\else
			\draw[line width=0.6pt] (O) -- (\FigSixXAxisLength,0);
			\draw[red,line width=0.8pt]
				(O) -- (\FigSixBraneAngle:\FigSixYAxisLength)
				node[above left,text=red] {$Q$};
			\node[below=\QuadripartiteRegionLabelOffset]
				at ({0.5*\FigSixA},0) {$A$};
			\node[below=\QuadripartiteRegionLabelOffset]
				at ({0.5*(\FigSixA+\FigSixB)},0) {$B$};
			\node[below=\QuadripartiteRegionLabelOffset]
				at ({0.5*(\FigSixB+\FigSixC)},0) {$C$};
			\node[below=\QuadripartiteRegionLabelOffset]
				at ({0.5*(\FigSixC+\FigSixXAxisLength)},0) {$D$};
		\fi

		\ifQuadripartiteIslandView\else
			\draw[densely dashed,gray,line width=0.55pt] (M)
				arc[start angle=180,end angle=0,radius=\FigSixMidRadius];
		\fi

		\draw[blue,line width=1pt] (Q)
			arc[start angle=\FigSixBraneAngle,end angle=\FigSixUAngle,
				radius=\FigSixQ];
		\draw[blue,line width=1pt] (A)
			arc[start angle=180,end angle=\FigSixAUAngle,
				radius=\FigSixARadius];
		\draw[blue,line width=1pt] (B)
			arc[start angle=180,end angle=\FigSixBVAngle,
				radius=\FigSixBRadius];
		\draw[blue,line width=1pt] (C)
			arc[start angle=0,end angle=\FigSixCVAngle,
				radius=\FigSixCRadius];
		\draw[blue,line width=1pt] (U)
			arc[start angle=\FigSixUOnMidAngle,end angle=\FigSixVOnMidAngle,
				radius=\FigSixMidRadius];

		\fill (O) circle[radius=1.3pt] node[below left=2pt] {$O$};
		\ifQuadripartiteIslandView\else
			\fill[gray] (M) circle[radius=1.3pt]
				node[below,text=black,left=-2pt,yshift=5pt] {$m$};
		\fi
		\fill (A) circle[radius=1.6pt] node[below=2pt] {$b_1$};
		\fill (B) circle[radius=1.6pt] node[below=2pt] {$b_2$};
		\ifQuadripartiteIslandView\else
			\fill[gray] (P) circle[radius=1.3pt]
				node[below,text=black,left=-11pt,yshift=5pt] {$p$};
		\fi
		\fill (C) circle[radius=1.6pt] node[below=2pt] {$b_3$};
		\fill (Q) circle[radius=1.6pt]
			node[QuadripartiteQLabel] {$q$};
		\fill (U) circle[radius=1.6pt];
		\node[above left=-12pt,yshift=10pt ] at (U) {$U$};
		\fill (V) circle[radius=1.6pt];
		\node[above right=-6pt,yshift=6pt] at (V) {$V$};
	\end{tikzpicture}%
}

\def\FigSevenTheta{30}

\def\FigSevenA{2}

\def\FigSevenB{3}

\def\FigSevenC{4}

\newcommand{\DrawQuadripartitePhaseC}{%
		\ifQuadripartiteIslandView\else
			\def\FigSevenA{2}
			\def\FigSevenB{3}
			\def\FigSevenC{4}
		\fi
		\begin{tikzpicture}[x=0.68cm,y=0.68cm,>=stealth]
			\pgfmathsetmacro{\FigSevenCAngle}{90+\FigSevenTheta}
			\ifQuadripartiteIslandView
				\def\FigSevenCXAxisLength{\QuadripartiteShortXAxisLength}
				\def\FigSevenCYAxisLength{\QuadripartiteYAxisLength}
			\else
				\def\FigSevenCXAxisLength{\QuadripartiteFigureSevenShortXAxisLength}
				\def\FigSevenCYAxisLength{\QuadripartiteFigureSevenPhaseCYAxisLength}
			\fi
			\pgfmathsetmacro{\FigSevenCIslandZCut}{
				\FigSevenCYAxisLength*sin(\FigSevenCAngle)}
			\pgfmathsetmacro{\FigSevenCCutLeftX}{
				\FigSevenCYAxisLength*cos(\FigSevenCAngle)}
			\pgfmathsetmacro{\FigSevenCQOne}{sqrt(\FigSevenA*\FigSevenB)}
			\pgfmathsetmacro{\FigSevenCQTwo}{\FigSevenC}
			\pgfmathsetmacro{\FigSevenCVX}{
				-\FigSevenCQOne*(\FigSevenA+\FigSevenB
				-4*\FigSevenCQOne)
				/(2*(\FigSevenA+\FigSevenB-\FigSevenCQOne))}
			\pgfmathsetmacro{\FigSevenCVZ}{
				sqrt(\FigSevenA*\FigSevenB
				-\FigSevenCVX*\FigSevenCVX)}

			\pgfmathsetmacro{\FigSevenCACenter}{
				(\FigSevenCVX*\FigSevenCVX+\FigSevenCVZ*\FigSevenCVZ
				-\FigSevenA*\FigSevenA)
				/(2*(\FigSevenCVX-\FigSevenA))}
			\pgfmathsetmacro{\FigSevenCARadius}{
				abs(\FigSevenA-\FigSevenCACenter)}
			\pgfmathsetmacro{\FigSevenCAVAngle}{
				atan2(\FigSevenCVZ,\FigSevenCVX-\FigSevenCACenter)}
			\pgfmathsetmacro{\FigSevenCBCenter}{
				(\FigSevenCVX*\FigSevenCVX+\FigSevenCVZ*\FigSevenCVZ
				-\FigSevenB*\FigSevenB)
				/(2*(\FigSevenCVX-\FigSevenB))}
			\pgfmathsetmacro{\FigSevenCBRadius}{
				abs(\FigSevenB-\FigSevenCBCenter)}
			\pgfmathsetmacro{\FigSevenCBVAngle}{
				atan2(\FigSevenCVZ,\FigSevenCVX-\FigSevenCBCenter)}
			\pgfmathsetmacro{\FigSevenCVAngle}{
				atan2(\FigSevenCVZ,\FigSevenCVX)}

			\coordinate (CO) at (0,0);
			\coordinate (CA) at (\FigSevenA,0);
			\coordinate (CB) at (\FigSevenB,0);
			\coordinate (CC) at (\FigSevenC,0);
			\coordinate (CQOne) at (\FigSevenCAngle:\FigSevenCQOne);
			\coordinate (CQTwo) at (\FigSevenCAngle:\FigSevenCQTwo);
			\coordinate (CQEnd) at
				(\FigSevenCAngle:\FigSevenCYAxisLength);
			\coordinate (CQCut) at
				(\FigSevenCCutLeftX,\FigSevenCIslandZCut);
			\coordinate (CV) at (\FigSevenCVX,\FigSevenCVZ);

			\ifQuadripartiteIslandView
				\path[QuadripartiteARegion,draw=none] (CO) -- (CA)
					arc[start angle=180,end angle=\FigSevenCAVAngle,
						radius=\FigSevenCARadius]
					arc[start angle=\FigSevenCVAngle,
						end angle=\FigSevenCAngle,
						radius=\FigSevenCQOne] -- cycle;
				\path[QuadripartiteBRegion,draw=none] (CA) -- (CB)
					arc[start angle=0,end angle=\FigSevenCBVAngle,
						radius=\FigSevenCBRadius]
					arc[start angle=\FigSevenCAVAngle,end angle=180,
						radius=\FigSevenCARadius] -- cycle;
				\path[QuadripartiteCRegion,draw=none] (CB) -- (CC)
					arc[start angle=0,end angle=\FigSevenCAngle,
						radius=\FigSevenCQTwo]
					-- (CQOne)
					arc[start angle=\FigSevenCAngle,
						end angle=\FigSevenCVAngle,
						radius=\FigSevenCQOne]
					arc[start angle=\FigSevenCBVAngle,end angle=0,
						radius=\FigSevenCBRadius] -- cycle;
				\path[QuadripartiteDRegion,draw=none] (CC)
					-- (\FigSevenCXAxisLength,0)
					-- (\FigSevenCXAxisLength,\FigSevenCIslandZCut)
					-- (CQCut) -- (CQTwo)
					arc[start angle=\FigSevenCAngle,end angle=0,
						radius=\FigSevenCQTwo] -- cycle;

				\draw[QuadripartiteAColor,line width=0.9pt] (CO) -- (CA)
					node[midway,below=4pt] {$A$};
				\draw[QuadripartiteBColor,line width=0.9pt] (CA) -- (CB)
					node[midway,below=4pt] {$B$};
				\draw[QuadripartiteCColor,line width=0.9pt] (CB) -- (CC)
					node[midway,below=4pt] {$C$};
				\draw[QuadripartiteDColor,line width=0.9pt]
					(CC) -- (\FigSevenCXAxisLength,0)
					node[midway,below=4pt] {$D$};
				\draw[QuadripartiteAColor,line width=0.9pt] (CO) -- (CQOne);
				\draw[QuadripartiteCColor,line width=0.9pt] (CQOne) -- (CQTwo);
				\draw[QuadripartiteDColor,line width=0.9pt] (CQTwo) -- (CQEnd)
					node[above left,text=black] {$Q$};
				\node[QuadripartiteIslandLabel,text=QuadripartiteAColor]
					at (\FigSevenCAngle:{0.50*\FigSevenCQOne}) {$I_A^{(4)}$};
				\node[QuadripartiteIslandLabel,text=QuadripartiteCColor]
					at (\FigSevenCAngle:{0.50*(\FigSevenCQOne+\FigSevenCQTwo)})
					{$I_C^{(4)}$};
				\node[QuadripartiteIslandLabel,text=QuadripartiteDColor]
					at (\FigSevenCAngle:{0.50*(\FigSevenCQTwo+\FigSevenCYAxisLength)})
					{$I_D^{(4)}$};
			\else
				\draw[line width=0.6pt] (CO) -- (\FigSevenCXAxisLength,0);
				\draw[red,line width=0.8pt]
					(CO) -- (\FigSevenCAngle:\FigSevenCYAxisLength)
					node[above left,text=red] {$Q$};
				\node[below=\QuadripartiteRegionLabelOffset]
					at ({0.5*\FigSevenA},0) {$A$};
				\node[below=\QuadripartiteRegionLabelOffset]
					at ({0.5*(\FigSevenA+\FigSevenB)},0) {$B$};
				\node[below=\QuadripartiteRegionLabelOffset]
					at ({0.5*(\FigSevenB+\FigSevenC)},0) {$C$};
				\node[below=\QuadripartiteRegionLabelOffset]
					at ({0.5*(\FigSevenC+\FigSevenCXAxisLength)},0) {$D$};
			\fi

			\draw[blue,line width=1pt] (CA)
				arc[start angle=180,end angle=\FigSevenCAVAngle,
					radius=\FigSevenCARadius];
			\draw[blue,line width=1pt] (CB)
				arc[start angle=0,end angle=\FigSevenCBVAngle,
					radius=\FigSevenCBRadius];
			\draw[blue,line width=1pt] (CQOne)
				arc[start angle=\FigSevenCAngle,end angle=\FigSevenCVAngle,
					radius=\FigSevenCQOne];
			\draw[blue,line width=1pt] (CC)
				arc[start angle=0,end angle=\FigSevenCAngle,
					radius=\FigSevenCQTwo];

			\fill (CO) circle[radius=1.3pt] node[below left=2pt] {$O$};
			\fill (CA) circle[radius=1.6pt] node[below=2pt] {$b_1$};
			\fill (CB) circle[radius=1.6pt] node[below=2pt] {$b_2$};
			\fill (CC) circle[radius=1.6pt] node[below=2pt] {$b_3$};
			\fill (CQOne) circle[radius=1.6pt]
				node[QuadripartiteIndexedQLabel] {$\sqrt{b_1b_2}$};
			\fill (CQTwo) circle[radius=1.6pt]
				node[QuadripartiteIndexedQLabel] {$b_3$};
			\fill (CV) circle[radius=1.6pt];
			\node[above right=-2pt] at (CV) {$V$};
		\end{tikzpicture}%
}

\newcommand{\DrawQuadripartitePhaseD}{%
		\ifQuadripartiteIslandView\else
			\def\FigSevenA{2}
			\def\FigSevenB{3}
			\def\FigSevenC{4}
		\fi
		\begin{tikzpicture}[x=0.68cm,y=0.68cm,>=stealth]
			\pgfmathsetmacro{\FigSevenDAngle}{90+\FigSevenTheta}
			\ifQuadripartiteIslandView
				\def\FigSevenDXAxisLength{\QuadripartiteShortXAxisLength}
				\def\FigSevenDYAxisLength{\QuadripartitePhaseDYAxisLength}
			\else
				\def\FigSevenDXAxisLength{\QuadripartiteFigureSevenShortXAxisLength}
				\def\FigSevenDYAxisLength{\QuadripartiteFigureSevenPhaseDYAxisLength}
			\fi
			\pgfmathsetmacro{\FigSevenDIslandZCut}{
				\FigSevenDYAxisLength*sin(\FigSevenDAngle)}
			\pgfmathsetmacro{\FigSevenDCutLeftX}{
				\FigSevenDYAxisLength*cos(\FigSevenDAngle)}
			\pgfmathsetmacro{\FigSevenDQOne}{\FigSevenA}
			\pgfmathsetmacro{\FigSevenDQTwo}{sqrt(\FigSevenB*\FigSevenC)}
			\pgfmathsetmacro{\FigSevenDVX}{
				-\FigSevenDQTwo*(\FigSevenB+\FigSevenC
				-4*\FigSevenDQTwo)
				/(2*(\FigSevenB+\FigSevenC-\FigSevenDQTwo))}
			\pgfmathsetmacro{\FigSevenDVZ}{
				sqrt(\FigSevenB*\FigSevenC
				-\FigSevenDVX*\FigSevenDVX)}

			\pgfmathsetmacro{\FigSevenDBCenter}{
				(\FigSevenDVX*\FigSevenDVX+\FigSevenDVZ*\FigSevenDVZ
				-\FigSevenB*\FigSevenB)
				/(2*(\FigSevenDVX-\FigSevenB))}
			\pgfmathsetmacro{\FigSevenDBRadius}{
				abs(\FigSevenB-\FigSevenDBCenter)}
			\pgfmathsetmacro{\FigSevenDBVAngle}{
				atan2(\FigSevenDVZ,\FigSevenDVX-\FigSevenDBCenter)}
			\pgfmathsetmacro{\FigSevenDCCenter}{
				(\FigSevenDVX*\FigSevenDVX+\FigSevenDVZ*\FigSevenDVZ
				-\FigSevenC*\FigSevenC)
				/(2*(\FigSevenDVX-\FigSevenC))}
			\pgfmathsetmacro{\FigSevenDCRadius}{
				abs(\FigSevenC-\FigSevenDCCenter)}
			\pgfmathsetmacro{\FigSevenDCVAngle}{
				atan2(\FigSevenDVZ,\FigSevenDVX-\FigSevenDCCenter)}
			\pgfmathsetmacro{\FigSevenDVAngle}{
				atan2(\FigSevenDVZ,\FigSevenDVX)}

			\coordinate (DO) at (0,0);
			\coordinate (DA) at (\FigSevenA,0);
			\coordinate (DB) at (\FigSevenB,0);
			\coordinate (DC) at (\FigSevenC,0);
			\coordinate (DQOne) at (\FigSevenDAngle:\FigSevenDQOne);
			\coordinate (DQTwo) at (\FigSevenDAngle:\FigSevenDQTwo);
			\coordinate (DQEnd) at
				(\FigSevenDAngle:\FigSevenDYAxisLength);
			\coordinate (DQCut) at
				(\FigSevenDCutLeftX,\FigSevenDIslandZCut);
			\coordinate (DV) at (\FigSevenDVX,\FigSevenDVZ);

			\ifQuadripartiteIslandView
				\path[QuadripartiteARegion,draw=none] (DO) -- (DA)
					arc[start angle=0,end angle=\FigSevenDAngle,
						radius=\FigSevenDQOne] -- cycle;
				\path[QuadripartiteBRegion,draw=none] (DA) -- (DB)
					arc[start angle=180,end angle=\FigSevenDBVAngle,
						radius=\FigSevenDBRadius]
					arc[start angle=\FigSevenDVAngle,
						end angle=\FigSevenDAngle,
						radius=\FigSevenDQTwo]
					-- (DQOne)
					arc[start angle=\FigSevenDAngle,end angle=0,
						radius=\FigSevenDQOne] -- cycle;
				\path[QuadripartiteCRegion,draw=none] (DB) -- (DC)
					arc[start angle=0,end angle=\FigSevenDCVAngle,
						radius=\FigSevenDCRadius]
					arc[start angle=\FigSevenDBVAngle,end angle=180,
						radius=\FigSevenDBRadius] -- cycle;
				\path[QuadripartiteDRegion,draw=none] (DC)
					-- (\FigSevenDXAxisLength,0)
					-- (\FigSevenDXAxisLength,\FigSevenDIslandZCut)
					-- (DQCut) -- (DQTwo)
					arc[start angle=\FigSevenDAngle,
						end angle=\FigSevenDVAngle,
						radius=\FigSevenDQTwo]
					arc[start angle=\FigSevenDCVAngle,end angle=0,
						radius=\FigSevenDCRadius] -- cycle;

				\draw[QuadripartiteAColor,line width=0.9pt] (DO) -- (DA)
					node[midway,below=4pt] {$A$};
				\draw[QuadripartiteBColor,line width=0.9pt] (DA) -- (DB)
					node[midway,below=4pt] {$B$};
				\draw[QuadripartiteCColor,line width=0.9pt] (DB) -- (DC)
					node[midway,below=4pt] {$C$};
				\draw[QuadripartiteDColor,line width=0.9pt]
					(DC) -- (\FigSevenDXAxisLength,0)
					node[midway,below=4pt] {$D$};
				\draw[QuadripartiteAColor,line width=0.9pt] (DO) -- (DQOne);
				\draw[QuadripartiteBColor,line width=0.9pt] (DQOne) -- (DQTwo);
				\draw[QuadripartiteDColor,line width=0.9pt] (DQTwo) -- (DQEnd)
					node[above left,text=black] {$Q$};
				\node[QuadripartiteIslandLabel,text=QuadripartiteAColor]
					at (\FigSevenDAngle:{0.50*\FigSevenDQOne}) {$I_A^{(4)}$};
				\node[QuadripartiteIslandLabel,text=QuadripartiteBColor]
					at (\FigSevenDAngle:{0.50*(\FigSevenDQOne+\FigSevenDQTwo)})
					{$I_B^{(4)}$};
				\node[QuadripartiteIslandLabel,text=QuadripartiteDColor]
					at (\FigSevenDAngle:{0.50*(\FigSevenDQTwo+\FigSevenDYAxisLength)})
					{$I_D^{(4)}$};
			\else
				\draw[line width=0.6pt] (DO) -- (\FigSevenDXAxisLength,0);
				\draw[red,line width=0.8pt]
					(DO) -- (\FigSevenDAngle:\FigSevenDYAxisLength)
					node[above left,text=red] {$Q$};
				\node[below=\QuadripartiteRegionLabelOffset]
					at ({0.5*\FigSevenA},0) {$A$};
				\node[below=\QuadripartiteRegionLabelOffset]
					at ({0.5*(\FigSevenA+\FigSevenB)},0) {$B$};
				\node[below=\QuadripartiteRegionLabelOffset]
					at ({0.5*(\FigSevenB+\FigSevenC)},0) {$C$};
				\node[below=\QuadripartiteRegionLabelOffset]
					at ({0.5*(\FigSevenC+\FigSevenDXAxisLength)},0) {$D$};
			\fi

			\draw[blue,line width=1pt] (DA)
				arc[start angle=0,end angle=\FigSevenDAngle,
					radius=\FigSevenDQOne];
			\draw[blue,line width=1pt] (DB)
				arc[start angle=180,end angle=\FigSevenDBVAngle,
					radius=\FigSevenDBRadius];
			\draw[blue,line width=1pt] (DC)
				arc[start angle=0,end angle=\FigSevenDCVAngle,
					radius=\FigSevenDCRadius];
			\draw[blue,line width=1pt] (DQTwo)
				arc[start angle=\FigSevenDAngle,end angle=\FigSevenDVAngle,
					radius=\FigSevenDQTwo];

			\fill (DO) circle[radius=1.3pt] node[below left=2pt] {$O$};
			\fill (DA) circle[radius=1.6pt] node[below=2pt] {$b_1$};
			\fill (DB) circle[radius=1.6pt] node[below=2pt] {$b_2$};
			\fill (DC) circle[radius=1.6pt] node[below=2pt] {$b_3$};
			\fill (DQOne) circle[radius=1.6pt]
				node[QuadripartiteIndexedQLabel] {$b_1$};
			\fill (DQTwo) circle[radius=1.6pt]
				node[QuadripartiteIndexedQLabel] {$\sqrt{b_2b_3}$};
			\fill (DV) circle[radius=1.6pt];
			\node[above right=-2pt] at (DV) {$V$};
		\end{tikzpicture}%
}

\newcommand{\DrawQuadripartitePhaseE}{%
	\def\FigEightTheta{30}
	\ifQuadripartiteIslandView
		\def\FigEightA{1.4}
		\def\FigEightB{3.0}
		\def\FigEightC{4.6}
	\else
		\def\FigEightA{2}
		\def\FigEightB{3}
		\def\FigEightC{4}
	\fi
	\begin{tikzpicture}[x=0.68cm,y=0.68cm,>=stealth]
		\pgfmathsetmacro{\FigEightBraneAngle}{90+\FigEightTheta}
		\ifQuadripartiteIslandView
			\def\FigEightXAxisLength{6.2}
			\def\FigEightYAxisLength{6.4}
		\else
			\def\FigEightXAxisLength{\QuadripartiteFigureSevenShortXAxisLength}
			\def\FigEightYAxisLength{\QuadripartiteFigureSevenPhaseEYAxisLength}
		\fi
		\pgfmathsetmacro{\FigEightIslandZCut}{
			\FigEightYAxisLength*sin(\FigEightBraneAngle)}
		\pgfmathsetmacro{\FigEightCutLeftX}{
			\FigEightYAxisLength*cos(\FigEightBraneAngle)}
		\pgfmathsetmacro{\FigEightQOne}{\FigEightA}
		\pgfmathsetmacro{\FigEightQTwo}{\FigEightB}
		\pgfmathsetmacro{\FigEightQThree}{\FigEightC}

		\coordinate (O) at (0,0);
		\coordinate (A) at (\FigEightA,0);
		\coordinate (B) at (\FigEightB,0);
		\coordinate (C) at (\FigEightC,0);
		\coordinate (QOne) at (\FigEightBraneAngle:\FigEightQOne);
		\coordinate (QTwo) at (\FigEightBraneAngle:\FigEightQTwo);
		\coordinate (QThree) at (\FigEightBraneAngle:\FigEightQThree);
		\coordinate (QEnd) at
			(\FigEightBraneAngle:\FigEightYAxisLength);
		\coordinate (QCut) at
			(\FigEightCutLeftX,\FigEightIslandZCut);

		\ifQuadripartiteIslandView
			\path[QuadripartiteARegion,draw=none] (O) -- (A)
				arc[start angle=0,end angle=\FigEightBraneAngle,
					radius=\FigEightQOne] -- cycle;
			\path[QuadripartiteBRegion,draw=none] (A) -- (B)
				arc[start angle=0,end angle=\FigEightBraneAngle,
					radius=\FigEightQTwo]
				-- (QOne)
				arc[start angle=\FigEightBraneAngle,end angle=0,
					radius=\FigEightQOne] -- cycle;
			\path[QuadripartiteCRegion,draw=none] (B) -- (C)
				arc[start angle=0,end angle=\FigEightBraneAngle,
					radius=\FigEightQThree]
				-- (QTwo)
				arc[start angle=\FigEightBraneAngle,end angle=0,
					radius=\FigEightQTwo] -- cycle;
			\path[QuadripartiteDRegion,draw=none] (C)
				-- (\FigEightXAxisLength,0)
				-- (\FigEightXAxisLength,\FigEightIslandZCut)
				-- (QCut) -- (QThree)
				arc[start angle=\FigEightBraneAngle,end angle=0,
					radius=\FigEightQThree] -- cycle;

			\draw[QuadripartiteAColor,line width=0.9pt] (O) -- (A)
				node[midway,below=4pt] {$A$};
			\draw[QuadripartiteBColor,line width=0.9pt] (A) -- (B)
				node[midway,below=4pt] {$B$};
			\draw[QuadripartiteCColor,line width=0.9pt] (B) -- (C)
				node[midway,below=4pt] {$C$};
			\draw[QuadripartiteDColor,line width=0.9pt]
				(C) -- (\FigEightXAxisLength,0)
				node[midway,below=4pt] {$D$};
			\draw[QuadripartiteAColor,line width=0.9pt] (O) -- (QOne);
			\draw[QuadripartiteBColor,line width=0.9pt] (QOne) -- (QTwo);
			\draw[QuadripartiteCColor,line width=0.9pt] (QTwo) -- (QThree);
			\draw[QuadripartiteDColor,line width=0.9pt] (QThree) -- (QEnd)
				node[above left,text=black] {$Q$};
			\node[QuadripartiteIslandLabel,text=QuadripartiteAColor]
				at (\FigEightBraneAngle:{0.50*\FigEightQOne}) {$I_A^{(4)}$};
			\node[QuadripartiteIslandLabel,text=QuadripartiteBColor]
				at (\FigEightBraneAngle:{0.50*(\FigEightQOne+\FigEightQTwo)})
				{$I_B^{(4)}$};
			\node[QuadripartiteIslandLabel,text=QuadripartiteCColor]
				at (\FigEightBraneAngle:{0.50*(\FigEightQTwo+\FigEightQThree)})
				{$I_C^{(4)}$};
			\node[QuadripartiteIslandLabel,text=QuadripartiteDColor]
				at (\FigEightBraneAngle:{0.50*(\FigEightQThree+\FigEightYAxisLength)})
				{$I_D^{(4)}$};
		\else
			\draw[line width=0.6pt] (O) -- (\FigEightXAxisLength,0);
			\draw[red,line width=0.8pt]
				(O) -- (\FigEightBraneAngle:\FigEightYAxisLength)
				node[above left,text=red] {$Q$};
			\node[below=\QuadripartiteRegionLabelOffset]
				at ({0.5*\FigEightA},0) {$A$};
			\node[below=\QuadripartiteRegionLabelOffset]
				at ({0.5*(\FigEightA+\FigEightB)},0) {$B$};
			\node[below=\QuadripartiteRegionLabelOffset]
				at ({0.5*(\FigEightB+\FigEightC)},0) {$C$};
			\node[below=\QuadripartiteRegionLabelOffset]
				at ({0.5*(\FigEightC+\FigEightXAxisLength)},0) {$D$};
		\fi

		\draw[blue,line width=1pt] (A)
			arc[start angle=0,end angle=\FigEightBraneAngle,
				radius=\FigEightQOne];
		\draw[blue,line width=1pt] (B)
			arc[start angle=0,end angle=\FigEightBraneAngle,
				radius=\FigEightQTwo];
		\draw[blue,line width=1pt] (C)
			arc[start angle=0,end angle=\FigEightBraneAngle,
				radius=\FigEightQThree];

		\fill (O) circle[radius=1.3pt] node[below left=2pt] {$O$};
		\fill (A) circle[radius=1.6pt] node[below=2pt] {$b_1$};
		\fill (B) circle[radius=1.6pt] node[below=2pt] {$b_2$};
		\fill (C) circle[radius=1.6pt] node[below=2pt] {$b_3$};
		\fill (QOne) circle[radius=1.6pt]
			node[QuadripartiteIndexedQLabel] {$b_1$};
		\fill (QTwo) circle[radius=1.6pt]
			node[QuadripartiteIndexedQLabel] {$b_2$};
		\fill (QThree) circle[radius=1.6pt]
			node[QuadripartiteIndexedQLabel] {$b_3$};
	\end{tikzpicture}%
}

\begin{figure}[ht]
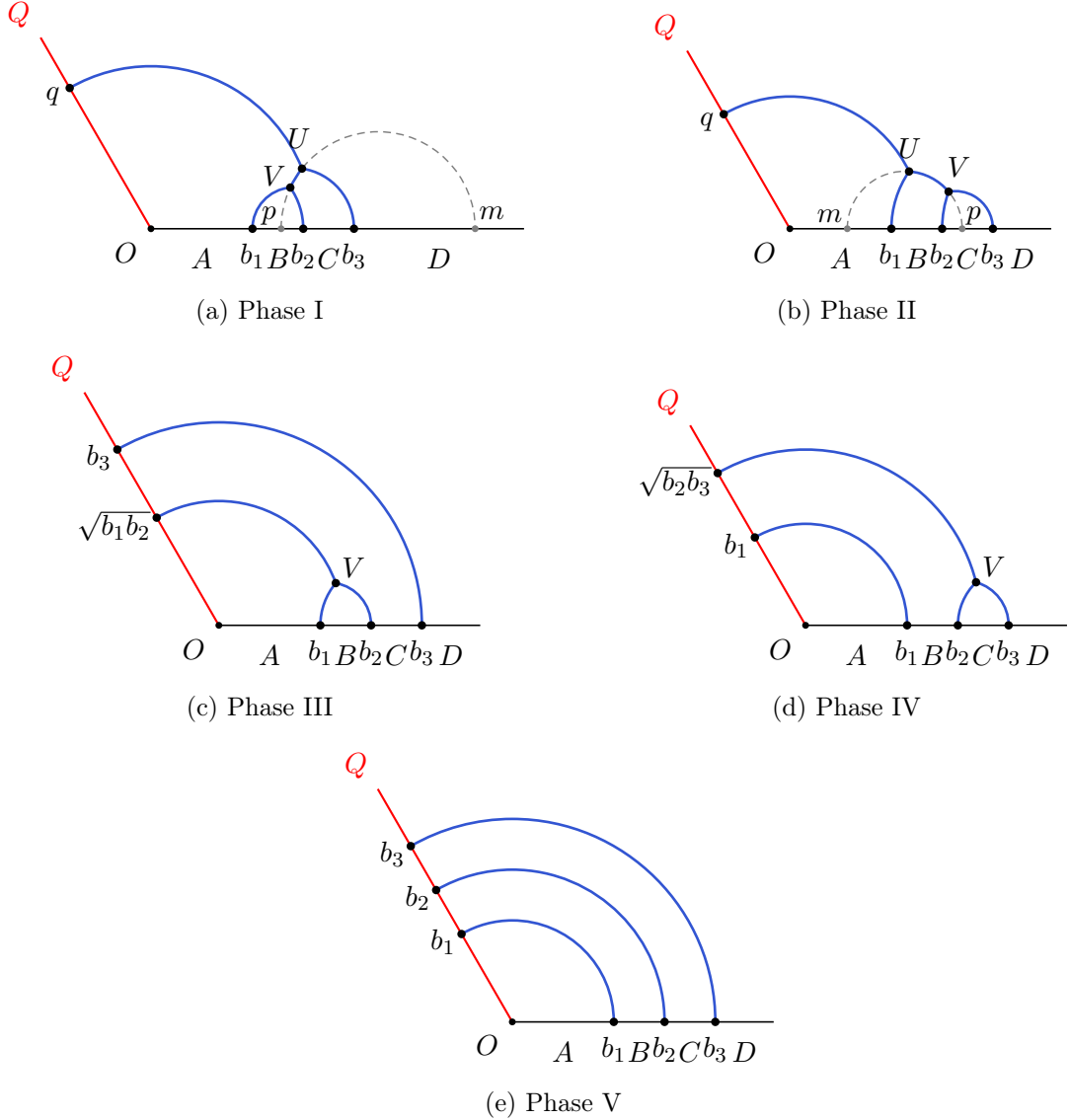

\centering
\begingroup
\colorlet{blue}{UnifiedBraneWebBlue}
\QuadripartiteIslandViewfalse
\begin{subfigure}[t]{0.48\linewidth}
\centering
\DrawQuadripartitePhaseA
\caption{Phase \uppercase\expandafter{\romannumeral 1}}
\label{fig:me4}
\end{subfigure}
\hfill
\begin{subfigure}[t]{0.48\linewidth}
\centering
\DrawQuadripartitePhaseB
\caption{Phase \uppercase\expandafter{\romannumeral 2}}
\label{fig:quadripartite-phase-B}
\end{subfigure}

\medskip
\begin{subfigure}[t]{0.48\linewidth}
\centering
\DrawQuadripartitePhaseC
\caption{Phase \uppercase\expandafter{\romannumeral 3}}
\label{fig:quadripartite-phase-C}
\end{subfigure}
\hfill
\begin{subfigure}[t]{0.48\linewidth}
\centering
\DrawQuadripartitePhaseD
\caption{Phase \uppercase\expandafter{\romannumeral 4}}
\label{fig:quadripartite-phase-D}
\end{subfigure}

\medskip
\begin{subfigure}[t]{0.48\linewidth}
\centering
\DrawQuadripartitePhaseE
\caption{Phase \uppercase\expandafter{\romannumeral 5}}
\label{fig:quadripartite-phase-E}
\end{subfigure}
\endgroup
\caption{The five phases for the quadripartite brane web.  In panels (a) and (b) the dashed arcs are the auxiliary extensions of the internal edge $UV$, which end on the boundary at the auxiliary points $p$ and $m$.}
\label{fig:quadripartite-phases}
\end{figure}

\begin{figure}[ht]
\centering
\begingroup
\colorlet{blue}{UnifiedBraneWebBlue}
\begin{tikzpicture}[x=0.68cm,y=0.68cm,>=stealth]
\def\FigSpecialTheta{30}
\def\FigSpecialBOne{1}
\def\FigSpecialBTwo{2}
\pgfmathsetmacro{\FigSpecialBThree}{
	2*(\FigSpecialBOne+\FigSpecialBTwo)}
\pgfmathsetmacro{\FigSpecialP}{
	0.5*(\FigSpecialBOne+\FigSpecialBTwo)}
\pgfmathsetmacro{\FigSpecialXAxisLength}{\FigSpecialBThree+2}
\def\FigSpecialYAxisLength{4.35}
\pgfmathsetmacro{\FigSpecialBraneAngle}{90+\FigSpecialTheta}
\pgfmathsetmacro{\FigSpecialQ}{
	sqrt(\FigSpecialP*\FigSpecialBThree)}
\pgfmathsetmacro{\FigSpecialVX}{\FigSpecialP}
\pgfmathsetmacro{\FigSpecialVZ}{
	0.5*sqrt(3)*(\FigSpecialBTwo-\FigSpecialBOne)}
\pgfmathsetmacro{\FigSpecialUX}{\FigSpecialP}
\pgfmathsetmacro{\FigSpecialUZ}{
	0.5*sqrt(3)*(\FigSpecialBOne+\FigSpecialBTwo)}
\pgfmathsetmacro{\FigSpecialUAngle}{
	atan2(\FigSpecialUZ,\FigSpecialUX)}

\pgfmathsetmacro{\FigSpecialBOneCenter}{
	(\FigSpecialVX*\FigSpecialVX+\FigSpecialVZ*\FigSpecialVZ
	-\FigSpecialBOne*\FigSpecialBOne)
	/(2*(\FigSpecialVX-\FigSpecialBOne))}
\pgfmathsetmacro{\FigSpecialBOneRadius}{
	abs(\FigSpecialBOne-\FigSpecialBOneCenter)}
\pgfmathsetmacro{\FigSpecialBOneVAngle}{
	atan2(\FigSpecialVZ,\FigSpecialVX-\FigSpecialBOneCenter)}
\pgfmathsetmacro{\FigSpecialBTwoCenter}{
	(\FigSpecialVX*\FigSpecialVX+\FigSpecialVZ*\FigSpecialVZ
	-\FigSpecialBTwo*\FigSpecialBTwo)
	/(2*(\FigSpecialVX-\FigSpecialBTwo))}
\pgfmathsetmacro{\FigSpecialBTwoRadius}{
	abs(\FigSpecialBTwo-\FigSpecialBTwoCenter)}
\pgfmathsetmacro{\FigSpecialBTwoVAngle}{
	atan2(\FigSpecialVZ,\FigSpecialVX-\FigSpecialBTwoCenter)}
\pgfmathsetmacro{\FigSpecialBThreeCenter}{
	(\FigSpecialUX*\FigSpecialUX+\FigSpecialUZ*\FigSpecialUZ
	-\FigSpecialBThree*\FigSpecialBThree)
	/(2*(\FigSpecialUX-\FigSpecialBThree))}
\pgfmathsetmacro{\FigSpecialBThreeRadius}{
	abs(\FigSpecialBThree-\FigSpecialBThreeCenter)}
\pgfmathsetmacro{\FigSpecialBThreeUAngle}{
	atan2(\FigSpecialUZ,\FigSpecialUX-\FigSpecialBThreeCenter)}

\coordinate (O) at (0,0);
\coordinate (BOne) at (\FigSpecialBOne,0);
\coordinate (P) at (\FigSpecialP,0);
\coordinate (BTwo) at (\FigSpecialBTwo,0);
\coordinate (BThree) at (\FigSpecialBThree,0);
\coordinate (Q) at (\FigSpecialBraneAngle:\FigSpecialQ);
\coordinate (U) at (\FigSpecialUX,\FigSpecialUZ);
\coordinate (V) at (\FigSpecialVX,\FigSpecialVZ);

\draw[line width=0.6pt] (O) -- (\FigSpecialXAxisLength,0);
\draw[red,line width=0.8pt] (O)
	-- (\FigSpecialBraneAngle:\FigSpecialYAxisLength)
	node[above left,text=red] {$Q$};

\draw[densely dashed,gray,line width=0.55pt] (P) -- (V);

\draw[blue,line width=1pt] (BOne)
	arc[start angle=180,end angle=\FigSpecialBOneVAngle,
		radius=\FigSpecialBOneRadius];
\draw[blue,line width=1pt] (BTwo)
	arc[start angle=0,end angle=\FigSpecialBTwoVAngle,
		radius=\FigSpecialBTwoRadius];
\draw[blue,line width=1pt] (BThree)
	arc[start angle=0,end angle=\FigSpecialBThreeUAngle,
		radius=\FigSpecialBThreeRadius];
\draw[blue,line width=1pt] (Q)
	arc[start angle=\FigSpecialBraneAngle,
		end angle=\FigSpecialUAngle,radius=\FigSpecialQ];
\draw[blue,line width=1.1pt] (V) -- (U);

\fill (O) circle[radius=1.3pt] node[below left=2pt] {$O$};
\fill (BOne) circle[radius=1.6pt] node[below=2pt] {$b_1$};
\fill[gray] (P) circle[radius=1.3pt] node[below=1pt] {$p$};
\fill (BTwo) circle[radius=1.6pt] node[below=2pt] {$b_2$};
\fill (BThree) circle[radius=1.6pt] node[below=2pt] {$b_3$};
\fill (Q) circle[radius=1.6pt] node[left=2pt] {$q$};
\fill (U) circle[radius=1.6pt];
\node[above left=-10pt,yshift=11pt] at (U) {$U$};
\fill (V) circle[radius=1.6pt];
\node[above left=-3pt,yshift=-1pt] at (V) {$V$};

\node[below=15pt] at ({0.5*\FigSpecialBOne},0) {$A$};
\node[below=15pt]
	at ({0.5*(\FigSpecialBOne+\FigSpecialBTwo)},0) {$B$};
\node[below=15pt]
	at ({0.5*(\FigSpecialBTwo+\FigSpecialBThree)},0) {$C$};
\node[below=15pt]
	at ({0.5*(\FigSpecialBThree+\FigSpecialXAxisLength)},0) {$D$};
\end{tikzpicture}
\caption{Phase~\uppercase\expandafter{\romannumeral 1} in the special case $b_3=2\left(b_1+b_2\right)$. The vertices
	$U$ and
	$V$ lie on a vertical geodesic.}
\label{fig:quadripartite-phase-I-special}
\endgroup
\end{figure}
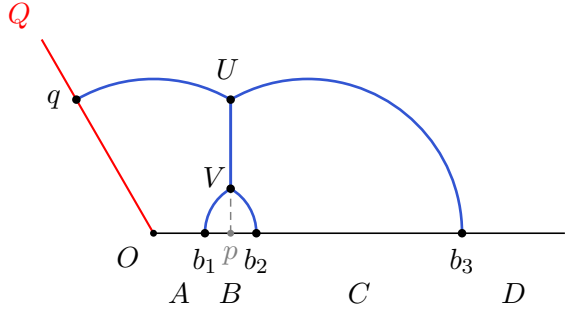

Likewise, the sub-web $(p,b_3)\to U\to q$ is the connected tripartite AdS/BCFT web for the intervals $[0,p]$, $[p,b_3]$, and $[b_3,\infty)$ discussed in the previous subsection.  Equation~\eqref{tri vertex coordinate} therefore gives the coordinates of $U$:
\begin{equation}\label{vertex-U}
	x_U=-\frac{\sqrt{p\: b_3}(p+b_3-4\sqrt{p\: b_3})}{2(p+b_3-\sqrt{p \:b_3})},\qquad z_U=\sqrt{p \:b_3-x_U^2}.
\end{equation}
By construction, $U$ and $V$ lie on the same semicircular geodesic with auxiliary endpoints $(p,m)$.  Applying its circle equation to $V$ gives
\begin{equation}\label{right endpont m}
	\left(x_V-\frac{p+m}{2}\right)^2+z_V^2=\left(\frac{m-p}{2}\right)^2\implies m=\frac{2 b_1 b_2-(b_1+b_2)p}{b_1+b_2-2 p},
\end{equation}
where we used Eq.~\eqref{coordinate vertex-V}.  Applying the same circle equation to $U$ gives
\begin{equation}
	\left(x_U-\frac{p+m}{2}\right)^2+z_U^2=\left(\frac{m-p}{2}\right)^2.
\end{equation}
After substituting Eqs.~\eqref{vertex-U} and \eqref{right endpont m}, we find that the difference between the two sides of the above equation is $-\sqrt p\,(b_3-p)\,\mathcal P/[(b_1+b_2-2p)(p+b_3-\sqrt{p\,b_3})]$, so the condition is equivalent to the vanishing of the \emph{Steiner polynomial}
\begin{equation}\label{Q p solution}
	\mathcal P(\mathsf p)\equiv b_1b_2\left(2\mathsf p-\mathsf b\right)
	-(b_1+b_2)\,\mathsf p\left(\mathsf p^2-\mathsf p\,\mathsf b+\mathsf b^2\right)
	+\mathsf b\,\mathsf p^3\left(2\mathsf b-\mathsf p\right)=0,
\end{equation}
where $\mathsf p\equiv\sqrt p$ and $\mathsf b\equiv\sqrt{b_3}$. This is a quartic equation for $\sqrt p$ and does not admit a useful closed-form solution for generic endpoints. Its relevant root always exists between $b_1$ and $b_2$, since
\begin{equation}\label{eq:steiner-endpoints}
	\mathcal P\big|_{p=b_1}=-\sqrt{b_1}\,(b_2-b_1)(b_3-b_1)<0,\qquad
	\mathcal P\big|_{p=b_2}=\sqrt{b_2}\,(b_2-b_1)(b_3-b_2)>0,
\end{equation}
so there is a root with $b_1<p<b_2$, and it is unique. We denote it by $p=p_{\rm I}(b_1,b_2,b_3)$. The total length is assembled from two pieces. First,
\begin{equation}\label{sub-web 1}
	\begin{aligned}
		\mathcal{L}\left(V,b_1\right)+\mathcal{L}\left(V,b_2\right)=&\log\left[\frac{\left(x_V-b_1\right)^2+z_V^2}{\epsilon \:z_V}\right]+\log\left[\frac{\left(x_V-b_2\right)^2+z_V^2}{\epsilon \:z_V}\right]\\
		=&2\log\left[\frac{2\left(b_2-b_1\right)}{\sqrt{3}\:\epsilon}\right],
	\end{aligned}
\end{equation}
where we used Eqs.~\eqref{coordinate vertex-V} and \eqref{right endpont m}.  The remaining sub-web length, $\mathcal{L}(V,U)+\mathcal{L}(U,b_3)+\mathcal{L}(U,q)$, can be obtained by subtracting the auxiliary segment $\mathcal{L}(V,p)$ from the connected tripartite web associated with $[0,p]$, $[p,b_3]$, and $[b_3,\infty)$ in AdS/BCFT:
\begin{equation}\label{sub-web 2}
	\begin{aligned}
		&\mathcal{L}\left(V,U\right)+\mathcal{L}\left(U,b_3\right)+\mathcal{L}\left(U,q\right)\\
		=&\mathcal{L}\left(U,p\right)+\mathcal{L}\left(U,b_3\right)+\mathcal{L}\left(U,q\right)-\mathcal{L}\left(V,p\right)\\
		=&\log \left[\frac{4(b_3-p)^2}{3\sqrt{3}\epsilon^2}\frac{\sqrt{b_3}+\sqrt{p}}{\sqrt{b_3}-\sqrt{p}}\right]+\operatorname{arccosh}(\sec\theta)-\log\left[\frac{(x_V-p)^2+z_V^2}{\epsilon \:z_V}\right]\\
		=&\log \left[\frac{4(b_3-p)^2}{3\sqrt{3}\epsilon^2}\frac{\sqrt{b_3}+\sqrt{p}}{\sqrt{b_3}-\sqrt{p}}\right]+\operatorname{arccosh}(\sec\theta)-\log\left[\frac{2\sqrt{3}\left(p-b_1\right)\left(b_2-p\right)}{\left(b_2-b_1\right)\epsilon}\right]\\
		=&\log\left[\frac{2}{9\epsilon}\frac{\sqrt{b_3}+\sqrt{p}}{\sqrt{b_3}-\sqrt{p}}\frac{\left(b_2-b_1\right)\left(b_3-p\right)^2}{\left(p-b_1\right)\left(b_2-p\right)}\right]+\operatorname{arccosh}(\sec\theta),
	\end{aligned}
\end{equation}
where we used Eqs.~\eqref{tri-brane-web-length-phase-1}, \eqref{coordinate vertex-V}, and \eqref{right endpont m}.  Combining Eqs.~\eqref{sub-web 1} and \eqref{sub-web 2} gives the total brane web length
\begin{equation}\label{total length quad case 1}
	\mathcal{L}_{\text{web}}=\log\left[\frac{8 (b_2-b_1)^3 \left(\sqrt{b_3}-\sqrt{p}\right) \left(\sqrt{b_3}+\sqrt{p}\right)^3}{27 \epsilon ^3 (p-b_1) (b_2-p)}\right]+\operatorname{arccosh}(\sec\theta),
\end{equation}
and therefore the Phase~I quadripartite multi-entropy is 
\begin{equation}\label{multien quad case 1}
	S_{\text{ph \uppercase\expandafter{\romannumeral 1}}}^{\left(4\right)}\left(A:B:C:D\right)=\frac{c}{6}\log\left[\frac{8 (b_2-b_1)^3 \left(\sqrt{b_3}-\sqrt{p}\right) \left(\sqrt{b_3}+\sqrt{p}\right)^3}{27 \epsilon ^3 (p-b_1) (b_2-p)}\right]+\frac{c}{6}\rho_0.
\end{equation}
In this expression $p=p_{\rm I}(b_1,b_2,b_3)$ is not an independent parameter but the root of Eq.~\eqref{Q p solution}. For the special case $b_3=2(b_1+b_2)$, where $p_{\rm I}=(b_1+b_2)/2$, we have
\begin{equation}
	m\rightarrow \infty,\quad U:\left(\frac{b_1+b_2}{2},\frac{\sqrt{3}}{2}\left(b_1+b_2\right)\right),\quad V:\left(\frac{b_1+b_2}{2},\frac{\sqrt{3}}{2}\left(b_2-b_1\right)\right),
\end{equation}
so the geodesic connecting $U$ and $V$ is vertical, as shown in Fig.~\ref{fig:quadripartite-phase-I-special}.  The corresponding multi-entropy is
\begin{equation}\label{multien quad case 1 special}
	S_{\text{ph \uppercase\expandafter{\romannumeral 1}}}^{\left(4\right)}\left(A:B:C:D\right)\big|_{b_3=2\left(b_1+b_2\right)}=\frac{c}{6}\log\left[\frac{8\left(b_2-b_1\right)\left(b_1+b_2\right)^2}{\epsilon^3}\right]+\frac{c}{6}\rho_0.
\end{equation}

For Phase~\uppercase\expandafter{\romannumeral 2}, shown in Fig.~\ref{fig:quadripartite-phase-B}, the multi-entropy is obtained from Eq.~\eqref{multien quad case 1} by the exchange $b_1\leftrightarrow b_3$,
\begin{equation}\label{multien quad case 2}
 S_{\text{ph \uppercase\expandafter{\romannumeral 2}}}^{\left(4\right)}(A:B:C:D)
 =\frac{c}{6}\log\left[\frac{8 (b_3-b_2)^3 (\sqrt{p}-\sqrt{b_1})(\sqrt{p}+\sqrt{b_1})^3}{27\epsilon^3 (b_3-p)(p-b_2)}\right]+\frac{c}{6}\rho_0.
\end{equation}
Here $p=p_{\rm II}(b_1,b_2,b_3)$ is the root in $(b_2,b_3)$ of the Steiner polynomial obtained from Eq.~\eqref{Q p solution} by the same exchange,
\begin{equation}\label{eq:Qp-phase-II}
\mathcal P_{\rm II}(\mathsf p)\equiv b_2b_3\bigl(2\mathsf p-\mathsf a\bigr)
 -(b_2+b_3)\,\mathsf p\bigl(\mathsf p^2-\mathsf p\,\mathsf a+\mathsf a^2\bigr)
 +\mathsf a\,\mathsf p^3\bigl(2\mathsf a-\mathsf p\bigr)=0,
\end{equation}
with $\mathsf p\equiv\sqrt p$ and $\mathsf a\equiv\sqrt{b_1}$.

\paragraph{Phases \uppercase\expandafter{\romannumeral 3} and \uppercase\expandafter{\romannumeral 4}.} In these phases the quadripartite web factorizes into a connected tripartite sub-web and an independent boundary-to-brane geodesic.  The on-shell multi-entropy is therefore the sum of the corresponding tripartite multi-entropy and an ordinary entanglement entropy.  For Phase~\uppercase\expandafter{\romannumeral 3},
\begin{equation}\label{multien quad case 3}
	\begin{aligned}
		S_{\text{ph \uppercase\expandafter{\romannumeral 3}}}^{\left(4\right)}\left(A:B:C:D\right)=&\:S_{\text{con}}^{\left(3\right)}\left(A:B:C D\right)+S_D\\
		=&\:\frac{c}{6}\log \left[\frac{8b_3(b_2-b_1)^2}{3\sqrt{3}\epsilon^3}\frac{\sqrt{b_2}+\sqrt{b_1}}{\sqrt{b_2}-\sqrt{b_1}}\right]+\frac{c}{3}\rho_0,
	\end{aligned}
\end{equation}
where we used Eqs.~\eqref{HEE AdS/BCFT} and \eqref{tri multi ee phase 1}.  Similarly,
\begin{equation}\label{multien quad case 4}
	\begin{aligned}
		S_{\text{ph \uppercase\expandafter{\romannumeral 4}}}^{\left(4\right)}\left(A:B:C:D\right)=&\:S_{\text{con}}^{\left(3\right)}\left(A B:C: D\right)+S_A\\
		=&\:\frac{c}{6}\log \left[\frac{8b_1(b_3-b_2)^2}{3\sqrt{3}\epsilon^3}\frac{\sqrt{b_3}+\sqrt{b_2}}{\sqrt{b_3}-\sqrt{b_2}}\right]+\frac{c}{3}\rho_0.
	\end{aligned}
\end{equation}

\paragraph{Phase \uppercase\expandafter{\romannumeral 5}.} This phase is completely factorized into three independent boundary-to-brane geodesics, so
\begin{equation}\label{multien quad case 5}
	\begin{aligned}
		S_{\text{ph \uppercase\expandafter{\romannumeral 5}}}^{\left(4\right)}\left(A:B:C:D\right)=&\:S_A+S_{AB}+S_D\\
		=&\:\frac{c}{6}\log \left(\frac{8b_1b_2b_3}{\epsilon^3}\right)+\frac{c}{2}\rho_0.
	\end{aligned}
\end{equation}

The expressions in Eqs.~\eqref{multien quad case 1} and \eqref{multien quad case 2} represent physical saddles only when the corresponding brane webs are geometrically admissible. For Phases~I and II, we impose the following conditions:\footnote{No separate admissibility restriction needs to be imposed on Phases~III and IV when selecting the physical lower envelope of the multi-entropy. According to the discussion in Sec.~\ref{three-party}, before the expression for Phase~III or IV becomes geometrically inadmissible, the always-admissible Phase~V saddle in Eq.~\eqref{multien quad case 5} has already become smaller. Hence the physical quadripartite multi-entropy $S^{(4)}$ never selects a geometrically inadmissible continuation of Phase~III or IV.}
\begin{itemize}
    \item The two trivalent vertices $V$ and $U$ must both lie inside the physical bulk region \eqref{eq:static-physical-wedge}.
    \item The internal edge joining $V$ and $U$ must have positive length, so that the brane web does not cross itself. Equivalently, the two vertices must occur in the correct order along the auxiliary geodesic. For example, in Phase~I, with $p=p_{\rm I}$, one has
    \begin{equation}\label{eq:q4-phase-I-internal-edge}
        \mathcal L(V,U)
        =\mathcal L(U,p)-\mathcal L(V,p)
        =\log\left[
        \frac{(b_2-b_1)(b_3-p)\sqrt{p}}
        {3\sqrt{b_3}(p-b_1)(b_2-p)}
        \right]>0.
    \end{equation}
\end{itemize}
The corresponding conditions for Phase~II follow by the exchange described above. Since $p$ is determined only implicitly by the Steiner polynomial \eqref{Q p solution} and does not admit a useful closed-form expression, these admissibility conditions cannot be rewritten as simple explicit inequalities involving only $b_1$, $b_2$, and $b_3$. They can nevertheless be checked straightforwardly once the relevant root has been determined numerically. We find that whenever the Phase~I saddle violates either condition, at least one of the other, geometrically admissible, phases has a smaller on-shell value. The same conclusion holds for Phase~\uppercase\expandafter{\romannumeral 2}. Moreover, as discussed in Sec.~\ref{three-party}, Phases~III and IV cease to dominate before their expressions become invalid, because the geometrically admissible Phase~V saddle has already become smaller. Consequently, the geometrically inadmissible continuations of these expressions never determine the minimum. The physical quadripartite multi-entropy may therefore be obtained by taking the unrestricted lower envelope
\begin{equation}\label{eq:q4-admissible-minimum}
S^{(4)}(A:B:C:D)
=
\min\left\{S_{\text{ph \uppercase\expandafter{\romannumeral 1}}}^{\left(4\right)},S_{\text{ph \uppercase\expandafter{\romannumeral 2}}}^{\left(4\right)},S_{\text{ph \uppercase\expandafter{\romannumeral 3}}}^{\left(4\right)},S_{\text{ph \uppercase\expandafter{\romannumeral 4}}}^{\left(4\right)},S_{\text{ph \uppercase\expandafter{\romannumeral 5}}}^{\left(4\right)}\right\}.
\end{equation}

\begin{figure}[t]
 \centering
 \includegraphics[width=\linewidth]{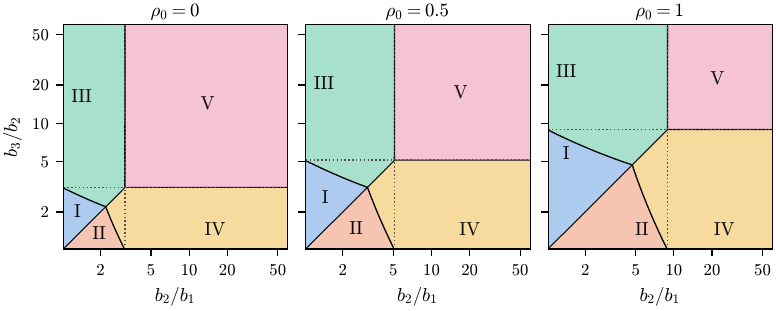}
 \caption{Phase diagram of $S^{\left(4\right)}$ as a function of the endpoint ratios for three values of $\rho_0$. The dotted lines mark the tripartite threshold $\eta_{c,2}(\rho_0)$.}
 \label{fig:q4-phase-diagram}
\end{figure}

Figure~\ref{fig:q4-phase-diagram} shows the phase diagram of the quadripartite multi-entropy as a function of the two ratios $b_2/b_1$ and $b_3/b_2$. Since Phases~\uppercase\expandafter{\romannumeral 1} and \uppercase\expandafter{\romannumeral 2}, as well as Phases~\uppercase\expandafter{\romannumeral 3} and \uppercase\expandafter{\romannumeral 4}, are related by the exchange $b_1\leftrightarrow b_3$, the diagram is symmetric about the diagonal. When only $b_3/b_2$ ($b_2/b_1$) is large, the corresponding outer (inner) geodesic detaches and the remaining $Y$ web gives Phase~III (IV). The fully connected webs dominate when both ratios are moderate, with Phase~I dominating above the diagonal and Phase~II below it, since the trivalent vertex $V$ joins the pair of adjacent entangling points with the smaller ratio.

\section{BCFT analysis}\label{sec:field-theory}

The brane web calculation predicts not only the multi-entropy but also candidate OPE or boundary-OPE channels for the replicated BCFT correlator. In this section, we test this dictionary directly in the large-$c$ limit: a change of bulk topology should correspond to a change of the semiclassical channel controlling the replica correlator. The twist-operator representation follows the multi-entropy replica construction \cite{Gadde:2022cqi,Gadde:2023zzj,Harper:2024ker}; the BCFT correlators are treated with the standard doubling trick and semiclassical monodromy method \cite{Cardy:1991tv,McAvity:1995zd,Fitzpatrick:2014vua,Sully:2020pza}. The monodromy problem reproduces the endpoint dependence dictated by the connected brane web, and the OPE limit fixes the normalization of the block. 

\subsection{Tripartite case}\label{section:Tripartite-field}

For three parties, Eq.~\eqref{definition multi-entropy} gives
\begin{equation}\label{S3 field con}
	S^{\left(3\right)}\left(A:B:C\right)=\lim_{n\rightarrow 1}\frac{1}{1-n}\frac{1}{n}\log\left[Z^{\left(3\right)}_{n^{2}}/\left(Z^{\left(3\right)}_{1}\right)^{n^{2}}\right],
\end{equation}
where the replicated partition function can be computed by the BCFT twist correlator
\begin{equation}\label{Z nn 3}
	Z^{\left(3\right)}_{n^2}=\left\langle\sigma_{g_1}(b_1)\sigma_{g_2}(b_2)\right\rangle_{{\rm BCFT}^{\otimes n^2}}.
\end{equation}
Under the standard BCFT doubling trick \cite{Cardy:1991tv,McAvity:1995zd,Sully:2020pza}, each bulk twist of scaling dimension $\Delta_n^{(3)}$ is represented by a pair of chiral twists carrying mutually inverse replica permutations, each with conformal dimension $h_n^{(3)}=\Delta_n^{(3)}/2$. The BCFT two-point function therefore becomes a reflection-symmetric chiral four-point function on the full plane,
\begin{equation}\label{double-trick}
	Z^{\left(3\right)}_{n^2}=\left\langle\sigma_{g_1}(b_1)\sigma_{g_2}(b_2)\right\rangle_{{\rm BCFT}^{\otimes n^2}}=\left\langle\sigma_{g_2^{-1}}(-b_2)\sigma_{g_1^{-1}}(-b_1)\sigma_{g_1^{}}(b_1)\sigma_{g_2^{}}(b_2)\right\rangle_{{\rm CFT}^{\otimes n^2}}.
\end{equation}
For the connected phase, as shown in the left panel of Fig.~\ref{fig:tripartite-phases-Q-brane}, we expand this four-point function in the bulk-OPE channel selected by the connected web:
\begin{align}\label{four chiral CFT corr}
	\left\langle\sigma_{g_2^{-1}}(-b_2)\sigma_{g_1^{-1}}(-b_1)\sigma_{g_1^{}}(b_1)\sigma_{g_2^{}}(b_2)\right\rangle_{{\rm CFT}^{\otimes n^2}}=\left(\frac{\epsilon}{b_2-b_1}\right)^{2\Delta_n^{(3)}}\sum_i \mathcal{A}_i^{(b)}\,\CC_i\,\CF_i(\eta),
\end{align}
where the cutoff $\epsilon$ makes the correlator dimensionless, $\CC_i$ is the bulk OPE coefficient, and $\mathcal A_i^{(b)}$ is the BCFT one-point coefficient of the exchanged bulk primary (equivalently its boundary-identity BOE coefficient).

The exchanged primary on the connected branch is the twist associated with the relative permutation $g_1g_2$, with conformal dimension $h_n^{(3)}=\Delta_n^{(3)}/2$. The corresponding large-$c$ Virasoro block follows from the semiclassical monodromy method \cite{Fitzpatrick:2014vua,Sully:2020pza}:
\begin{align}\label{semi-classical conformal BCFT}
	\CF_{\Delta}(\eta)=\left(\frac{1+\sqrt{1-\eta}}{2\sqrt{\eta}}\right)^{-2h_n^{\left(3\right)}}=\left(\frac{1+\sqrt{1-\eta}}{2\sqrt{\eta}}\right)^{-\Delta_n^{\left(3\right)}},
\end{align}
where the cross-ratio is
\begin{align}\label{cross-ratio bulk channel}
	\eta=\frac{z_{12}z_{34}}{z_{13}z_{24}}=\left(\frac{b_2-b_1}{b_2+b_1}\right)^2\,.
\end{align}
Substituting this cross-ratio and Eq.~\eqref{semi-classical conformal BCFT} into Eq.~\eqref{four chiral CFT corr} gives
\begin{equation}\label{two-point-function-tri}
	\left\langle\sigma_{g_1}(b_1)\sigma_{g_2}(b_2)\right\rangle_{{\rm BCFT}^{\otimes n^2}}=\left(\frac{\epsilon}{b_2-b_1}\right)^{2\Delta_n^{(3)}}\mathcal{A}_{\Delta_n^{(3)}}^{(b)}\CC_{\Delta_n^{(3)}}\left(\frac{\sqrt{b_2}+\sqrt{b_1}}{2\left(\sqrt{b_2}-\sqrt{b_1}\right)}\right)^{-\Delta_n^{\left(3\right)}},
\end{equation}
where the coefficient $\CC_{\Delta_n^{(3)}}$ (more generally, $\CC_{\Delta_n^{(q)}}$) can be fixed by matching the field-theory result with the bulk calculation in AdS/CFT \cite{Gadde:2022cqi,Gadde:2023zzj,Harper:2024ker},
\begin{equation}\label{C coefficient}
	\lim_{n\rightarrow 1}\frac{1}{1-n}\frac{1}{n^{q-2}}\log\CC_{\Delta_n^{(q)}}=\frac{c}{2}\log\left(\frac{2}{\sqrt{3}}\right).
\end{equation}
Furthermore, for an ordinary $n$-cycle twist in the bipartite replica construction, the boundary-identity BOE coefficient contains $g_b^{1-n}$ \cite{Sully:2020pza}. A $q$-partite elementary twist contains $n^{q-2}$ disjoint $n$-cycles \cite{Gadde:2022cqi,Gadde:2023zzj}; therefore
\begin{equation}\label{A coefficient}
 \mathcal{A}_{\Delta_n^{(q)}}^{(b)}=g_b^{\,n^{q-2}(1-n)}.
\end{equation}
Substituting Eqs.~\eqref{two-point-function-tri}, \eqref{C coefficient} and \eqref{A coefficient} into Eq.~\eqref{S3 field con} gives the connected-channel result,
\begin{equation}
	\begin{aligned}
		S^{\left(3\right)}\left(A:B:C\right)=&\frac{c}{6}\left[2\log\left(\frac{b_2-b_1}{\epsilon}\right)+\log\left(\frac{\sqrt{b_2}+\sqrt{b_1}}{2\left(\sqrt{b_2}-\sqrt{b_1}\right)}\right)+3\log\left(\frac{2}{\sqrt{3}}\right)\right]+\log g_b\\
		=&\frac{c}{6}\log \left[\frac{4(b_2-b_1)^2}{3\sqrt{3}\epsilon^2}\frac{\sqrt{b_2}+\sqrt{b_1}}{\sqrt{b_2}-\sqrt{b_1}}\right]+\log g_b,
	\end{aligned}
\end{equation}
which reproduces the connected bulk result \eqref{tri multi ee phase 1}. 

For the disconnected phase, shown in the right panel of Fig.~\ref{fig:tripartite-phases-Q-brane}, the relevant semiclassical channel is the boundary-OPE (BOE) channel.  Each twist fuses with its reflected image through the identity block, as in the standard holographic BCFT entropy transition \cite{Sully:2020pza}.  Accordingly, the two-point function \eqref{Z nn 3} factorizes, mirroring the decomposition of the bulk brane web into two boundary-to-brane geodesics:
\begin{equation}\label{factor two one-point}
	\begin{aligned}
		\left\langle\sigma_{g_1}(b_1)\sigma_{g_2}(b_2)\right\rangle_{{\rm BCFT}^{\otimes n^2}}=&\left\langle\sigma_{g_1}(b_1)\right\rangle_{{\rm BCFT}^{\otimes n^2}}\times \left\langle\sigma_{g_2}(b_2)\right\rangle_{{\rm BCFT}^{\otimes n^2}}\\
		=&\left(\frac{\epsilon^2}{4b_1 b_2}\right)^{\Delta_n^{(3)}}\left(\mathcal{A}_{\Delta_n^{(3)}}^{(b)}\right)^2.
	\end{aligned}
\end{equation}
Substitution into Eq.~\eqref{S3 field con} gives
\begin{equation}\label{field-phase B-q-3}
	S^{(3)}(A:B:C)=\frac{c}{6}\log\left(\frac{4b_1b_2}{\epsilon^2}\right)+2\log g_b,
\end{equation}
which agrees with the bulk calculation \eqref{tri multi ee phase 2}.

\subsection{Quadripartite case}\label{subsec:q4-monodromy}

For four parties, the replica definition gives
\begin{equation}\label{S4 field theory}
	S^{\left(4\right)}\left(A:B:C:D\right)=\lim_{n\rightarrow 1}\frac{1}{1-n}\frac{1}{n^2}\log\left[Z^{\left(4\right)}_{n^{3}}/\left(Z^{\left(4\right)}_{1}\right)^{n^{3}}\right],
\end{equation}
where the replicated partition function is
\begin{equation}\label{Z 4 n 3}
	\begin{aligned}
		Z^{\left(4\right)}_{n^{3}}=&\left\langle\sigma_{g_1}(b_1)\sigma_{g_2}(b_2)\sigma_{g_3}(b_3)\right\rangle_{{\rm BCFT}^{\otimes n^3}}\\
		=&\left\langle
	\sigma_{g_3^{-1}}(-b_3)\sigma_{g_2^{-1}}(-b_2)\sigma_{g_1^{-1}}(-b_1)
 	\sigma_{g_1}(b_1)\sigma_{g_2}(b_2)\sigma_{g_3}(b_3)
 \right\rangle_{{\rm CFT}^{\otimes n^3}} ,
	\end{aligned}
\end{equation}
where we used the BCFT doubling trick \cite{Cardy:1991tv,McAvity:1995zd,Sully:2020pza}.  For Phase~\uppercase\expandafter{\romannumeral 1}, shown in Fig.~\ref{fig:me4}, we expand the BCFT three-point function in the nested bulk-OPE channel suggested by the brane web topology.  In this channel $\sigma_{g_1}$ and $\sigma_{g_2}$ first fuse to the bulk twist $\sigma_{g_1g_2}$, which subsequently fuses with $\sigma_{g_3}$ to an operator with a nonvanishing BCFT one-point function.  The corresponding monodromy contours, which encode the semiclassical fusion pattern associated with this nested bulk OPE channel in the standard monodromy formalism \cite{Fitzpatrick:2014vua,Sully:2020pza}, are shown in Fig.~\ref{fig:q4-monodromy-contours}. The two nested exchanged twists consequently have the same chiral weight $h_n^{(4)}=\Delta_n^{(4)}/2$ as the elementary twists.  With the explicit OPE prefactors kept separate, the doubled correlator can be written as
\begin{equation}\label{six chiral expansion}
	\begin{aligned}
		&\left\langle
	\sigma_{g_3^{-1}}(-b_3)\sigma_{g_2^{-1}}(-b_2)\sigma_{g_1^{-1}}(-b_1)
 	\sigma_{g_1}(b_1)\sigma_{g_2}(b_2)\sigma_{g_3}(b_3)
 \right\rangle_{{\rm CFT}^{\otimes n^3}}\\
 =&\:\mathcal{A}_{\Delta_n^{(4)}}^{(b)}\left(\CC_{\Delta_n^{(4)}}\right)^2\mathcal{F}\left(b_i,\epsilon_\Delta,\epsilon_\Delta\right),
	\end{aligned}
\end{equation}
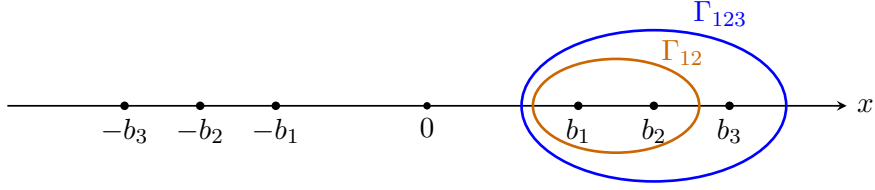
\begin{figure}[ht]
	\centering
    \begingroup
    \def\FigNineA{2}
    \def\FigNineB{3}
    \def\FigNineC{4}
    \begin{tikzpicture}[x=1cm,y=1cm,>=stealth]
        \pgfmathsetmacro{\FigNineInnerCenter}{
            0.5*(\FigNineA+\FigNineB)}
        \pgfmathsetmacro{\FigNineInnerXRadius}{
            0.5*(\FigNineB-\FigNineA)+0.60}
        \pgfmathsetmacro{\FigNineInnerYRadius}{0.62}
        \pgfmathsetmacro{\FigNineOuterCenter}{
            0.5*(\FigNineA+\FigNineC)}
        \pgfmathsetmacro{\FigNineOuterXRadius}{
            0.5*(\FigNineC-\FigNineA)+0.75}
        \pgfmathsetmacro{\FigNineOuterYRadius}{1.00}

        \draw[->,line width=0.65pt]
            ({-\FigNineC-1.55},0) -- ({\FigNineC+1.55},0)
            node[right] {$x$};

        \draw[orange!80!black,line width=1pt]
            (\FigNineInnerCenter,0)
            ellipse[x radius=\FigNineInnerXRadius cm,
                y radius=\FigNineInnerYRadius cm];
        \node[above=1pt,text=orange!80!black,xshift=25pt,yshift=-7pt]
            at (\FigNineInnerCenter,\FigNineInnerYRadius)
            {$\Gamma_{12}$};
        \draw[blue,line width=1pt]
            (\FigNineOuterCenter,0)
            ellipse[x radius=\FigNineOuterXRadius cm,
                y radius=\FigNineOuterYRadius cm];
        \node[above=1pt,text=blue,xshift=25pt,yshift=-6pt]
            at (\FigNineOuterCenter,{\FigNineOuterYRadius+0.10})
            {$\Gamma_{123}$};

        \fill (-\FigNineC,0) circle[radius=1.6pt]
            node[below=1pt] {$-b_3$};
        \fill (-\FigNineB,0) circle[radius=1.6pt]
            node[below=1pt] {$-b_2$};
        \fill (-\FigNineA,0) circle[radius=1.6pt]
            node[below=1pt] {$-b_1$};
        \fill (0,0) circle[radius=1.4pt]
            node[below=1pt] {$0$};
        \fill (\FigNineA,0) circle[radius=1.6pt]
            node[below=1pt] {$b_1$};
        \fill (\FigNineB,0) circle[radius=1.6pt]
            node[below=1pt] {$b_2$};
        \fill (\FigNineC,0) circle[radius=1.6pt]
            node[below=1pt] {$b_3$};
    \end{tikzpicture}
    \endgroup
	\caption{Monodromy contours for the doubled six-point function in the nested bulk-OPE channel.}
	\label{fig:q4-monodromy-contours}
\end{figure}
where the quantity $\mathcal{F}(b_i,\epsilon_\Delta,\epsilon_\Delta)$ denotes the conformal block of this channel. Note that only the leading semiclassical contribution is displayed. We compute the conformal block using the standard monodromy method \cite{Fitzpatrick:2014vua}, with the parameter
\begin{equation}
	\epsilon_\Delta=\frac{6}{c}h_n^{\left(4\right)}=\frac{3\Delta_n^{(4)}}{c}
 =\frac{n}{4}\left(n^2-1\right).
\end{equation}
Near the replica limit relevant for $S^{(4)}$,
\begin{equation}
	\epsilon_\Delta|_{n\rightarrow 1}\ll 1,
\end{equation}
which supplies the perturbative parameter for the linearized Fuchsian problem \cite{Fitzpatrick:2014vua}.  The projective stress tensor is
\begin{align}\label{eq:q4-projective-T}
 T(z)=\sum_{i=1}^{3}\left[
 \frac{\epsilon_\Delta}{(z-b_i)^2}+\frac{\epsilon_\Delta}{(z+b_i)^2}
 +\frac{C_i}{z-b_i}-\frac{C_i}{z+b_i}\right],
\end{align}
where reflection symmetry has been imposed and $C_i$ are the accessory parameters associated with the channel contribution\footnote{The factor $1/2$ comes from the reflection symmetry.},
\begin{equation}\label{C-conformal block}
	C_i=-\frac{1}{2}\frac{\partial f\left(b_i,\epsilon_\Delta,\epsilon_\Delta\right)}{\partial b_i},\quad f\left(b_i,\epsilon_\Delta,\epsilon_\Delta\right)=-\frac{6}{c}\log\mathcal{F}\left(b_i,\epsilon_\Delta,\epsilon_\Delta\right).
\end{equation}
The required asymptotic behavior $T(z)\sim z^{-4}$ as $z\to\infty$ imposes the Ward constraint
\begin{equation}\label{Ward}
 b_1C_1+b_2C_2+b_3C_3+3\epsilon_\Delta=0.
\end{equation}
We solve the associated Fuchsian equation perturbatively,
\begin{equation}
	\frac{d^2\psi\left(z\right)}{dz^2}+T\left(z\right)\psi\left(z\right)=0.
\end{equation}
The accessory parameters $C_i$ can be determined by requiring the solutions of this equation to have prescribed monodromy around the relevant contours. In the light-operator regime $n\to1$ there are no heavy operators, so the stress tensor has no zeroth-order contribution in $\epsilon_\Delta$:
\begin{equation}
	T^{\left(0\right)}=0,\quad T^{\left(1\right)}=T\left(z\right),\quad \Rightarrow \psi^{\left(0\right)}\left(z\right)=1,\: z.
\end{equation} 
The first-order monodromy matrix associated with the contour $\Gamma$ is
\begin{equation}
 \mathbb M^{\left(1\right)}(\Gamma)=2\pi i
 \begin{pmatrix}J_1&J_2\\-J_0&-J_1\end{pmatrix},
\end{equation}
where $J_{0,1,2}$ are three contour moments of $T\left(z\right)$,
\begin{equation}
 J_m(\Gamma)=\frac{1}{2\pi i}\oint_\Gamma dz\,z^mT(z),\qquad m=0,1,2.
\end{equation}
The eigenvalues are fixed by the conformal weight of the exchanged twist and therefore constrain the accessory parameters $C_i$\footnote{See~\cite{Fitzpatrick:2014vua} for the relation between exchanged dimensions and monodromy eigenvalues.}.  For the two nested contours shown in Fig.~\ref{fig:q4-monodromy-contours}, the constraints are
\begin{align}
 \Gamma_{12}:\quad
 &\bigl(2\epsilon_\Delta-(b_2-b_1)C_1\bigr)
  \bigl(2\epsilon_\Delta+(b_2-b_1)C_2\bigr)=\epsilon_\Delta^2,
 \label{eq:Contour-Gamma12}\\
 \Gamma_{123}:\quad
 &\bigl(3\epsilon_\Delta+b_1C_1+b_2C_2+b_3C_3\bigr)^2\notag\\
 &-(C_1+C_2+C_3)
 \bigl[2\epsilon_\Delta(b_1+b_2+b_3)+b_1^2C_1+b_2^2C_2+b_3^2C_3\bigr]
 =\epsilon_\Delta^2.
 \label{eq:Contour-Gamma123}
\end{align}
The three equations \eqref{Ward}, \eqref{eq:Contour-Gamma12}, and \eqref{eq:Contour-Gamma123} determine the three accessory parameters implicitly, but solving these nonlinear equations directly is not illuminating.  The useful observation is instead that the endpoint derivatives of the on-shell web action \eqref{total length quad case 1} satisfy all three constraints:
\begin{equation}\label{C L web}
	C_i=-\epsilon_{\Delta}\frac{d\mathcal{L}_{\text{web}}}{d b_i},
\end{equation}
where the total derivatives $d\mathcal{L}_{\text{web}}/d b_i$ are computed as
\begin{equation}
	\frac{d\mathcal{L}_{\text{web}}}{d b_i}=\frac{\partial \mathcal{L}_{\text{web}}}{\partial b_i}+\frac{\partial \mathcal{L}_{\text{web}}}{\partial \mathsf p}\frac{\partial \mathsf p}{\partial b_i},\quad \frac{\partial \mathsf p}{\partial b_i}=-\frac{\partial_{b_i}\mathcal P(\mathsf p)}{\partial_{\mathsf p}\mathcal P(\mathsf p)},
\end{equation}
where $\mathcal P$ is the Steiner polynomial of Eq.~\eqref{Q p solution}. We now check that the $C_i$ given by the right-hand side of Eq.~\eqref{C L web} indeed satisfy the constraints \eqref{Ward}, \eqref{eq:Contour-Gamma12} and \eqref{eq:Contour-Gamma123}.  From Eq.~\eqref{total length quad case 1}, we obtain
\begin{equation}\label{eq:web-stationarity}
 \frac{\partial\mathcal L_{\rm web}}{\partial\mathsf p}
 =-\frac{2\,\mathcal P(\mathsf p)}{(\mathsf p^2-b_1)(\mathsf p^2-b_2)(\mathsf b-\mathsf p)(\mathsf b+\mathsf p)}=0,
\end{equation}
where we have used the on-shell condition $\mathcal P(\mathsf p)=0$. Consequently, the total endpoint derivative of the on-shell action equals its partial derivative at fixed $p$, and Eq.~\eqref{C L web} gives
\begin{equation}
	\begin{aligned}
		\frac{C_1}{\epsilon_\Delta}&=\frac{3}{b_2-b_1}-\frac{1}{p-b_1},\\
 \frac{C_2}{\epsilon_\Delta}&=-\frac{3}{b_2-b_1}+\frac{1}{b_2-p},\\
 \frac{C_3}{\epsilon_\Delta}&=-\frac{2\sqrt{b_3}-\sqrt p}{\sqrt{b_3}(b_3-p)}.
	\end{aligned}
\end{equation}
Substituting these expressions into the constraints \eqref{Ward}, \eqref{eq:Contour-Gamma12} and \eqref{eq:Contour-Gamma123}, we obtain
\begin{equation}
	\begin{aligned}
		\text{LHS}-\text{RHS of } \eqref{Ward}=&\:-\frac{\mathsf p\,\mathcal P(\mathsf p)}{(\mathsf p^2-b_1)(\mathsf p^2-b_2)(\mathsf b^2-\mathsf p^2)}=0,\\
		\text{LHS}-\text{RHS of } \eqref{eq:Contour-Gamma12}=&\:0,\\
		\text{LHS}-\text{RHS of } \eqref{eq:Contour-Gamma123}=&\:\frac{\mathcal P(\mathsf p)}{\mathsf b\,(\mathsf p^2-b_1)(\mathsf p^2-b_2)}=0,
	\end{aligned}
\end{equation}
where we have used the on-shell condition $\mathcal P(\mathsf p)=0$ again. Comparing Eq.~\eqref{C L web} with Eq.~\eqref{C-conformal block}, we find
\begin{equation}\label{conformal block}
 \begin{aligned}
 f\left(b_i,\epsilon_\Delta,\epsilon_\Delta\right)
 &=2\epsilon_\Delta\left(\mathcal L_{\rm web}+f_0\right),\\
 \mathcal F\left(b_i,\epsilon_\Delta,\epsilon_\Delta\right)
 &=\exp\!\left[-\Delta_n^{(4)}\left(\mathcal L_{\rm web}+f_0\right)\right],
 \end{aligned}
\end{equation}
where $f_0$ is a constant independent of the insertion points $b_i$ and is invisible to the monodromy equations. We fix it by the normalization of the conformal block. In the OPE limit $b_2\to b_1$ the canonically normalized six-point conformal block must reduce to the leading OPE factor of the pair $(b_1,b_2)$ times the four-point conformal block \eqref{semi-classical conformal BCFT} of Sec.~\ref{section:Tripartite-field}, now for the fused twist $\sigma_{g_1g_2}$ at $b_1$ and $\sigma_{g_3}$ at $b_3$, i.e.
\begin{equation}\label{eq:q4-OPE-limit-block}
 \mathcal F\;\xrightarrow[\;b_2\to b_1\;]{}\;
 \left(\frac{\epsilon}{b_2-b_1}\right)^{\Delta_n^{(4)}}
 \left(\frac{\epsilon}{b_3-b_1}\right)^{2\Delta_n^{(4)}}
 \left(\frac{\sqrt{b_3}+\sqrt{b_1}}{2\left(\sqrt{b_3}-\sqrt{b_1}\right)}\right)^{-\Delta_n^{(4)}},
\end{equation}
or, equivalently,
\begin{equation}
	\mathcal{L}_{\text{web}}+f_0\xrightarrow[\;b_2\to b_1\;]{}\;\log\left[\frac{(b_2-b_1)\left(\sqrt{b_3}-\sqrt{b_1}\right)\left(\sqrt{b_3}+\sqrt{b_1}\right)^3}{2\epsilon^3}\right].
\end{equation}
On the other hand, in this limit, $p=(b_1+b_2)/2+\mathcal O\bigl((b_2-b_1)^2\bigr)$, and Eq.~\eqref{total length quad case 1} becomes
\begin{equation}\label{eq:q4-OPE-limit-web}
 \mathcal L_{\rm web}\;\xrightarrow[\;b_2\to b_1\;]{}\;
 \log\left[\frac{32\,(b_2-b_1)\left(\sqrt{b_3}-\sqrt{b_1}\right)\left(\sqrt{b_3}+\sqrt{b_1}\right)^3}{27\,\epsilon^3}\right]+\rho_0 .
\end{equation}
Combining Eqs.~\eqref{eq:q4-OPE-limit-block} and \eqref{eq:q4-OPE-limit-web} with Eq.~\eqref{conformal block}, we can find
\begin{equation}\label{eq:q4-f0}
 f_0=-6\log\left(\frac{2}{\sqrt3}\right)-\rho_0.
\end{equation}
Substituting this result and Eq.~\eqref{conformal block} into Eq.~\eqref{six chiral expansion}, we find that the quadripartite multi-entropy in this channel is
\begin{equation}\label{eq:q4-CFT-generic}
 \begin{aligned}
 S^{(4)}(A:B:C:D)
 =&\lim_{n\to1}\frac{1}{1-n}\frac{1}{n^2}
 \log\left[\mathcal A_{\Delta_n^{(4)}}^{(b)}
 \left(\CC_{\Delta_n^{(4)}}\right)^2
 \exp\!\left(-\Delta_n^{(4)}(\mathcal L_{\rm web}+f_0)\right)\right]\\
 =&\frac{c}{6}\mathcal L_{\rm web},
 \end{aligned}
\end{equation}
which agrees with the bulk calculation \eqref{multien quad case 1}. For the special case $b_3=2(b_1+b_2)$, the constraints \eqref{Ward}, \eqref{eq:Contour-Gamma12} and \eqref{eq:Contour-Gamma123} can be solved explicitly:
\begin{equation}\label{eq:q4-symmetric-accessories}
 C_1=\frac{\epsilon_\Delta}{b_2-b_1},\qquad
 C_2=-\frac{\epsilon_\Delta}{b_2-b_1},\qquad
 C_3=-\frac{2\epsilon_\Delta}{b_3}
 =-\frac{\epsilon_\Delta}{b_1+b_2}.
\end{equation}
Along this locus $db_3=2(db_1+db_2)$, so Eqs.~\eqref{C-conformal block} and \eqref{eq:q4-symmetric-accessories} give
\begin{equation}
 df=-2\left(C_1db_1+C_2db_2+C_3db_3\right)
 =2\epsilon_\Delta\,d\log\!\left[(b_2-b_1)(b_1+b_2)^2\right].
\end{equation}
With the normalization \eqref{eq:q4-OPE-limit-block}, the block on the special locus is
\begin{equation}
 f_{\rm sym}(b_1,b_2;\epsilon_\Delta)
 =2\epsilon_\Delta\log\!\left[
 \frac{27}{8}\,
 \frac{(b_2-b_1)(b_1+b_2)^2}{\epsilon^3}
 \right],
 \label{eq:q4-symmetric-block}
\end{equation}
and the CFT answer is
\begin{equation*}\label{eq:q4-symmetric-field-answer}
 S^{(4)}(A:B:C:D)\big|_{b_3=2(b_1+b_2)}
 =\frac{c}{6}\log\!\left[
 \frac{8(b_2-b_1)(b_1+b_2)^2}{\epsilon^3}
 \right]+\log g_b,
\end{equation*}
in agreement with the bulk result \eqref{multien quad case 1 special}.

For Phase~II, shown in Fig.~\ref{fig:quadripartite-phase-B}, the nested channel is the one in which $\sigma_{g_2}$ and $\sigma_{g_3}$ fuse first, and the relevant contours are $\Gamma_{23}$ and $\Gamma_{123}$. Similarly, the accessory parameters $C_i=-\epsilon_\Delta\,\partial_{b_i}\mathcal L_{\rm web}^{\rm II}$ obtained from Eq.~\eqref{multien quad case 2} satisfy the Ward identity together with the $\Gamma_{23}$ and $\Gamma_{123}$ conditions.  The same normalization argument then reproduces Eq.~\eqref{multien quad case 2}.

For Phases~III and IV the web topology suggests a factorization of Eq.~\eqref{Z 4 n 3} into a two-point function and a one-point function.  For Phase~III,
\begin{equation}\label{factor phase 3}
	\left\langle\sigma_{g_1}(b_1)\sigma_{g_2}(b_2)\sigma_{g_3}(b_3)\right\rangle_{{\rm BCFT}^{\otimes n^3}}
	=\left\langle\sigma_{g_1}(b_1)\sigma_{g_2}(b_2)\right\rangle_{{\rm BCFT}^{\otimes n^3}}\times \left\langle\sigma_{g_3}(b_3)\right\rangle_{{\rm BCFT}^{\otimes n^3}},
\end{equation}
where the two factors are those in Eqs.~\eqref{two-point-function-tri} and \eqref{factor two one-point}:
\begin{equation}\label{two and one}
	\begin{aligned}
		\left\langle\sigma_{g_1}(b_1)\sigma_{g_2}(b_2)\right\rangle_{{\rm BCFT}^{\otimes n^3}}=&\:\left(\frac{\epsilon}{b_2-b_1}\right)^{2\Delta_n^{(4)}}\mathcal{A}_{\Delta_n^{(4)}}^{(b)}\CC_{\Delta_n^{(4)}}\left(\frac{\sqrt{b_2}+\sqrt{b_1}}{2\left(\sqrt{b_2}-\sqrt{b_1}\right)}\right)^{-\Delta_n^{\left(4\right)}},\\
		\left\langle\sigma_{g_3}(b_3)\right\rangle_{{\rm BCFT}^{\otimes n^3}}=&\:\mathcal{A}_{\Delta_n^{(4)}}^{(b)}\left(\frac{\epsilon}{2b_3}\right)^{\Delta_n^{(4)}}.
	\end{aligned}
\end{equation}
Substituting these into Eq.~\eqref{S4 field theory}, we find that the quadripartite multi-entropy in this phase is
\begin{equation}
 S^{(4)}(A:B:C:D)
 =\frac{c}{6}\log \left[
 \frac{8b_3(b_2-b_1)^2}{3\sqrt{3}\epsilon^3}
 \frac{\sqrt{b_2}+\sqrt{b_1}}{\sqrt{b_2}-\sqrt{b_1}}
 \right]+2\log g_b,
\end{equation}
which reproduces the bulk calculation \eqref{multien quad case 3}.  Phase~\uppercase\expandafter{\romannumeral 4} follows by the corresponding exchange $b_1\leftrightarrow b_3$.

For Phase~V the correlator factorizes completely into one-point functions:
\begin{equation}
	\begin{aligned}
		&\left\langle\sigma_{g_1}(b_1)\sigma_{g_2}(b_2)\sigma_{g_3}(b_3)\right\rangle_{{\rm BCFT}^{\otimes n^3}}\\
		=&\left\langle\sigma_{g_1}(b_1)\right\rangle_{{\rm BCFT}^{\otimes n^3}}\times \left\langle\sigma_{g_2}(b_2)\right\rangle_{{\rm BCFT}^{\otimes n^3}}\times \left\langle\sigma_{g_3}(b_3)\right\rangle_{{\rm BCFT}^{\otimes n^3}}\\
		=&\left(\mathcal{A}_{\Delta_n^{(4)}}^{(b)}\right)^3 \left(\frac{\epsilon^3}{8b_1b_2b_3}\right)^{\Delta_n^{(4)}}.
	\end{aligned}
\end{equation}
Using Eqs.~\eqref{factor two one-point} and \eqref{S4 field theory}, we obtain
\begin{equation}
	S^{\left(4\right)}\left(A:B:C:D\right)=\frac{c}{6}\log \left(\frac{8b_1b_2b_3}{\epsilon^3}\right)+3\log g_b,
\end{equation}
which agrees with the bulk calculation \eqref{multien quad case 5}.


\section{Eternal black hole and the multi-entropy curve}\label{sec:eternal-black-hole}
The thermofield-double state provides an equilibrium setting in which brane-ending and wormhole-crossing multiway cuts compete.  The two-sided black-hole/TFD geometry and the Hartman--Maldacena growth of two-sided extremal surfaces are standard \cite{Maldacena:2001kr,Hartman:2013qma}; multipartite extensions and black-hole multi-entropy curves have been explored recently in several complementary settings \cite{Iizuka:2024pzm,Anegawa:2025prn,Balasubramanian:2025jhq, Anegawa:2026yee}.  We work in the left--right reflection-symmetric, planar, principal non-winding sector throughout.  The BTZ metric, the EOW-brane profile and the ordinary two-sided entropy have already been reviewed in Sec.~\ref{sec:review}; these ingredients are the same ones used in holographic BCFT analyses of black-hole interiors and entropy transitions \cite{Rozali:2019day,Sully:2020pza}.  We first determine all five gravitational web lengths, treating the connected cross-boundary saddle by a direct Kruskal extremization.  We then compare these saddles with the semiclassical eight-point fusion trees suggested by the bulk topologies and finally determine their lower envelope and Page-like time dependence.

\subsection{Gravity analysis}\label{subsec:thermal-gravity}

\begin{figure}[t]
\centering
\resizebox{0.90\linewidth}{!}{%
\begin{minipage}{\linewidth}
\centering
	\begingroup
	\colorlet{blue}{UnifiedBraneWebBlue}
	\def\FigTenTheta{30}
	\def\FigTenA{2}
	\def\FigTenB{4}
	\def\FigTenAxisExtension{1.30}
	\def\FigTenCLabelOffset{0.85}
	\def\FigTenHalfGap{4.1}
	\def\FigTenBraneLength{4.6}
	\def\FigTenHalfGapSlight{3.5}
	\def\FigTenBraneLengthSlight{3.8}
	\def\FigTenHalfGapCompact{2.8}
	\def\FigTenBraneLengthCompact{3.0}
	\def\FigTenHalfGapTall{4.5}
	\def\FigTenBraneLengthTall{5.0}
	\def\FigTenPanelWidth{0.40\linewidth}
	\def\FigTenPairGap{0.015\linewidth}
	\def\FigTenUnit{0.70cm}
	\tikzset{
		ThermalQLabel/.style={left=0pt,yshift=-1pt},
		ThermalIndexedQLabel/.style={left=0pt,xshift=1pt,yshift=-1pt}
	}
	\pgfmathsetmacro{\FigTenBraneAngle}{90+\FigTenTheta}
	\pgfmathsetmacro{\FigTenUpperBraneAngle}{270-\FigTenTheta}
	\pgfmathsetmacro{\FigTenMid}{0.5*(\FigTenA+\FigTenB)}
	\pgfmathsetmacro{\FigTenHalfWidth}{0.5*(\FigTenB-\FigTenA)}
	\pgfmathsetmacro{\FigTenSteinerDepth}{sqrt(3)*\FigTenHalfWidth}
	\pgfmathsetmacro{\FigTenSteinerRadius}{2*\FigTenHalfWidth}

	\pgfmathsetmacro{\FigTenW}{sqrt(\FigTenA*\FigTenB)}
	\pgfmathsetmacro{\FigTenPhaseOneVX}{
		-\FigTenW*(\FigTenA+\FigTenB-4*\FigTenW)
		/(2*(\FigTenA+\FigTenB-\FigTenW))}
	\pgfmathsetmacro{\FigTenPhaseOneVZ}{
		sqrt(\FigTenW*\FigTenW
		-\FigTenPhaseOneVX*\FigTenPhaseOneVX)}
	\pgfmathsetmacro{\FigTenPhaseOneLeftCenter}{
		(\FigTenPhaseOneVX*\FigTenPhaseOneVX
		+\FigTenPhaseOneVZ*\FigTenPhaseOneVZ
		-\FigTenA*\FigTenA)
		/(2*(\FigTenPhaseOneVX-\FigTenA))}
	\pgfmathsetmacro{\FigTenPhaseOneLeftRadius}{
		abs(\FigTenA-\FigTenPhaseOneLeftCenter)}
	\pgfmathsetmacro{\FigTenPhaseOneLeftVAngle}{
		atan2(\FigTenPhaseOneVZ,
		\FigTenPhaseOneVX-\FigTenPhaseOneLeftCenter)}
	\pgfmathsetmacro{\FigTenPhaseOneRightCenter}{
		(\FigTenPhaseOneVX*\FigTenPhaseOneVX
		+\FigTenPhaseOneVZ*\FigTenPhaseOneVZ
		-\FigTenB*\FigTenB)
		/(2*(\FigTenPhaseOneVX-\FigTenB))}
	\pgfmathsetmacro{\FigTenPhaseOneRightRadius}{
		abs(\FigTenB-\FigTenPhaseOneRightCenter)}
	\pgfmathsetmacro{\FigTenPhaseOneRightVAngle}{
		atan2(\FigTenPhaseOneVZ,
		\FigTenPhaseOneVX-\FigTenPhaseOneRightCenter)}
	\pgfmathsetmacro{\FigTenPhaseOneVAngle}{
		atan2(\FigTenPhaseOneVZ,\FigTenPhaseOneVX)}

	\newcommand{\DrawThermalPatchFrame}{%
		\coordinate (UpperO) at (0,\FigTenHalfGap);
		\coordinate (UpperA) at (\FigTenA,\FigTenHalfGap);
		\coordinate (UpperB) at (\FigTenB,\FigTenHalfGap);
		\coordinate (LowerO) at (0,-\FigTenHalfGap);
		\coordinate (LowerA) at (\FigTenA,-\FigTenHalfGap);
		\coordinate (LowerB) at (\FigTenB,-\FigTenHalfGap);
		\draw[line width=0.6pt] (UpperO) --
			({\FigTenB+\FigTenAxisExtension},\FigTenHalfGap);
		\draw[line width=0.6pt] (LowerO) --
			({\FigTenB+\FigTenAxisExtension},-\FigTenHalfGap);
		\draw[red,line width=0.8pt] (UpperO) -- ++
			(\FigTenUpperBraneAngle:\FigTenBraneLength)
			node[above left,text=red] {$Q$};
		\draw[red,line width=0.8pt] (LowerO) -- ++
			(\FigTenBraneAngle:\FigTenBraneLength)
			node[below left,text=red] {$Q$};
		\node[above=1pt] at ({0.5*\FigTenA},\FigTenHalfGap) {$A_L$};
		\node[above=1pt]
			at ({0.5*(\FigTenA+\FigTenB)},\FigTenHalfGap) {$B_L$};
		\node[above=1pt]
			at ({\FigTenB+\FigTenCLabelOffset},\FigTenHalfGap) {$C_L$};
		\node[below=1pt] at ({0.5*\FigTenA},-\FigTenHalfGap) {$A_R$};
		\node[below=1pt]
			at ({0.5*(\FigTenA+\FigTenB)},-\FigTenHalfGap) {$B_R$};
		\node[below=1pt]
			at ({\FigTenB+\FigTenCLabelOffset},-\FigTenHalfGap) {$C_R$};
	}
	\newcommand{\MarkThermalPatchPoints}{%
		\fill (UpperO) circle[radius=1.2pt]
			node[above left=1pt] {$O$};
		\fill (LowerO) circle[radius=1.2pt]
			node[below left=1pt] {$O$};
		\fill (UpperA) circle[radius=1.4pt]
			node[above=1pt] {$b_1$};
		\fill (UpperB) circle[radius=1.4pt]
			node[above=1pt] {$b_2$};
		\fill (LowerA) circle[radius=1.4pt]
			node[below=1pt] {$b_1$};
		\fill (LowerB) circle[radius=1.4pt]
			node[below=1pt] {$b_2$};
	}
	\begin{subfigure}[t]{\FigTenPanelWidth}
		\centering
		\begin{tikzpicture}[x=\FigTenUnit,y=\FigTenUnit,>=stealth,
			every node/.style={font=\small}]
			\def\FigTenHalfGap{\FigTenHalfGapSlight}
			\def\FigTenBraneLength{\FigTenBraneLengthSlight}
			\DrawThermalPatchFrame
			\coordinate (PhaseThreeUpperV) at
				(\FigTenMid,{\FigTenHalfGap-\FigTenSteinerDepth});
			\coordinate (PhaseThreeLowerV) at
				(\FigTenMid,{-\FigTenHalfGap+\FigTenSteinerDepth});
			\begin{scope}[shift={(0,\FigTenHalfGap)},yscale=-1]
				\draw[blue,line width=0.9pt] (\FigTenA,0)
					arc[start angle=180,end angle=120,
						radius=\FigTenSteinerRadius];
				\draw[blue,line width=0.9pt] (\FigTenB,0)
					arc[start angle=0,end angle=60,
						radius=\FigTenSteinerRadius];
			\end{scope}
			\begin{scope}[shift={(0,-\FigTenHalfGap)}]
				\draw[blue,line width=0.9pt] (\FigTenA,0)
					arc[start angle=180,end angle=120,
						radius=\FigTenSteinerRadius];
				\draw[blue,line width=0.9pt] (\FigTenB,0)
					arc[start angle=0,end angle=60,
						radius=\FigTenSteinerRadius];
			\end{scope}
			\draw[blue,line width=0.9pt]
				(PhaseThreeUpperV) -- (PhaseThreeLowerV);
			\MarkThermalPatchPoints
			\fill (PhaseThreeUpperV) circle[radius=1.4pt]
				node[right=2pt,xshift=-2pt,yshift=-4pt] {$V_2$};
			\fill (PhaseThreeLowerV) circle[radius=1.4pt]
				node[right=2pt,xshift=-2pt,yshift=4pt] {$V_1$};
		\end{tikzpicture}
		\caption{Phase \uppercase\expandafter{\romannumeral 1}}
		\label{thermal phase 1}
	\end{subfigure}
	\hspace{\FigTenPairGap}
	\begin{subfigure}[t]{\FigTenPanelWidth}
		\centering
		\begin{tikzpicture}[x=\FigTenUnit,y=\FigTenUnit,>=stealth,
			every node/.style={font=\small}]
			\def\FigTenHalfGap{\FigTenHalfGapSlight}
			\def\FigTenBraneLength{\FigTenBraneLengthSlight}
			\DrawThermalPatchFrame
			\begin{scope}[shift={(0,\FigTenHalfGap)},yscale=-1]
				\coordinate (PhaseOneUpperV) at
					(\FigTenPhaseOneVX,\FigTenPhaseOneVZ);
				\coordinate (PhaseOneUpperQ) at
					(\FigTenBraneAngle:\FigTenW);
				\draw[blue,line width=0.9pt] (\FigTenA,0)
					arc[start angle=180,end angle=\FigTenPhaseOneLeftVAngle,
						radius=\FigTenPhaseOneLeftRadius];
				\draw[blue,line width=0.9pt] (\FigTenB,0)
					arc[start angle=0,end angle=\FigTenPhaseOneRightVAngle,
						radius=\FigTenPhaseOneRightRadius];
				\draw[blue,line width=0.9pt] (PhaseOneUpperQ)
					arc[start angle=\FigTenBraneAngle,
						end angle=\FigTenPhaseOneVAngle,radius=\FigTenW];
			\end{scope}
			\begin{scope}[shift={(0,-\FigTenHalfGap)}]
				\coordinate (PhaseOneLowerV) at
					(\FigTenPhaseOneVX,\FigTenPhaseOneVZ);
				\coordinate (PhaseOneLowerQ) at
					(\FigTenBraneAngle:\FigTenW);
				\draw[blue,line width=0.9pt] (\FigTenA,0)
					arc[start angle=180,end angle=\FigTenPhaseOneLeftVAngle,
						radius=\FigTenPhaseOneLeftRadius];
				\draw[blue,line width=0.9pt] (\FigTenB,0)
					arc[start angle=0,end angle=\FigTenPhaseOneRightVAngle,
						radius=\FigTenPhaseOneRightRadius];
				\draw[blue,line width=0.9pt] (PhaseOneLowerQ)
					arc[start angle=\FigTenBraneAngle,
						end angle=\FigTenPhaseOneVAngle,radius=\FigTenW];
			\end{scope}
			\MarkThermalPatchPoints
			\fill (PhaseOneUpperQ) circle[radius=1.2pt]
				node[ThermalQLabel] {$q$};
			\fill (PhaseOneLowerQ) circle[radius=1.2pt]
				node[ThermalQLabel] {$q$};
			\fill (PhaseOneUpperV) circle[radius=1.4pt]
				node[below right=1pt,xshift=-4pt,yshift=1pt] {$V_2$};
			\fill (PhaseOneLowerV) circle[radius=1.4pt]
				node[above right=1pt,xshift=-4pt,yshift=-1pt] {$V_1$};
		\end{tikzpicture}
		\caption{Phase \uppercase\expandafter{\romannumeral 2}}
		\label{thermal phase 2}
	\end{subfigure}
	\par\medskip
	\begin{center}
	\begin{subfigure}[t]{\FigTenPanelWidth}
		\centering
		\begin{tikzpicture}[x=\FigTenUnit,y=\FigTenUnit,>=stealth,
			every node/.style={font=\small}]
			\def\FigTenHalfGap{\FigTenHalfGapTall}
			\def\FigTenBraneLength{\FigTenBraneLengthTall}
			\DrawThermalPatchFrame
			\begin{scope}[shift={(0,\FigTenHalfGap)},yscale=-1]
				\coordinate (PhaseTwoUpperQA) at
					(\FigTenBraneAngle:\FigTenA);
				\coordinate (PhaseTwoUpperQB) at
					(\FigTenBraneAngle:\FigTenB);
				\draw[blue,line width=0.9pt] (\FigTenA,0)
					arc[start angle=0,end angle=\FigTenBraneAngle,
						radius=\FigTenA];
				\draw[blue,line width=0.9pt] (\FigTenB,0)
					arc[start angle=0,end angle=\FigTenBraneAngle,
						radius=\FigTenB];
			\end{scope}
			\begin{scope}[shift={(0,-\FigTenHalfGap)}]
				\coordinate (PhaseTwoLowerQA) at
					(\FigTenBraneAngle:\FigTenA);
				\coordinate (PhaseTwoLowerQB) at
					(\FigTenBraneAngle:\FigTenB);
				\draw[blue,line width=0.9pt] (\FigTenA,0)
					arc[start angle=0,end angle=\FigTenBraneAngle,
						radius=\FigTenA];
				\draw[blue,line width=0.9pt] (\FigTenB,0)
					arc[start angle=0,end angle=\FigTenBraneAngle,
						radius=\FigTenB];
			\end{scope}
			\MarkThermalPatchPoints
			\fill (PhaseTwoUpperQA) circle[radius=1.2pt]
				node[ThermalIndexedQLabel] {$q_1$};
			\fill (PhaseTwoUpperQB) circle[radius=1.2pt]
				node[ThermalIndexedQLabel] {$q_2$};
			\fill (PhaseTwoLowerQA) circle[radius=1.2pt]
				node[ThermalIndexedQLabel] {$q_1$};
			\fill (PhaseTwoLowerQB) circle[radius=1.2pt]
				node[ThermalIndexedQLabel] {$q_2$};
		\end{tikzpicture}
		\caption{Phase \uppercase\expandafter{\romannumeral 3}}
		\label{thermal phase 3}
	\end{subfigure}
	\end{center}
	\par\medskip
	\begin{subfigure}[t]{\FigTenPanelWidth}
		\centering
		\begin{tikzpicture}[x=\FigTenUnit,y=\FigTenUnit,>=stealth,
			every node/.style={font=\small}]
			\def\FigTenHalfGap{\FigTenHalfGapCompact}
			\def\FigTenBraneLength{\FigTenBraneLengthCompact}
			\DrawThermalPatchFrame
			\draw[blue,line width=0.9pt]
				(UpperA) -- (LowerA);
			\draw[blue,line width=0.9pt]
				(UpperB) -- (LowerB);
			\MarkThermalPatchPoints
		\end{tikzpicture}
		\caption{Phase \uppercase\expandafter{\romannumeral 4}}
		\label{thermal phase 4}
	\end{subfigure}
	\hspace{\FigTenPairGap}
	\begin{subfigure}[t]{\FigTenPanelWidth}
		\centering
		\begin{tikzpicture}[x=\FigTenUnit,y=\FigTenUnit,>=stealth,
			every node/.style={font=\small}]
			\def\FigTenHalfGap{\FigTenHalfGapCompact}
			\def\FigTenBraneLength{\FigTenBraneLengthCompact}
			\DrawThermalPatchFrame
			\begin{scope}[shift={(0,\FigTenHalfGap)},yscale=-1]
				\coordinate (PhaseFiveUpperQ) at
					(\FigTenBraneAngle:\FigTenA);
				\draw[blue,line width=0.9pt] (\FigTenA,0)
					arc[start angle=0,end angle=\FigTenBraneAngle,
						radius=\FigTenA];
			\end{scope}
			\begin{scope}[shift={(0,-\FigTenHalfGap)}]
				\coordinate (PhaseFiveLowerQ) at
					(\FigTenBraneAngle:\FigTenA);
				\draw[blue,line width=0.9pt] (\FigTenA,0)
					arc[start angle=0,end angle=\FigTenBraneAngle,
						radius=\FigTenA];
			\end{scope}
			\draw[blue,line width=0.9pt]
				(\FigTenB,\FigTenHalfGap) --
				(\FigTenB,-\FigTenHalfGap);
			\MarkThermalPatchPoints
			\fill (PhaseFiveUpperQ) circle[radius=1.2pt]
				node[ThermalQLabel] {$q$};
			\fill (PhaseFiveLowerQ) circle[radius=1.2pt]
				node[ThermalQLabel] {$q$};
		\end{tikzpicture}
		\caption{Phase \uppercase\expandafter{\romannumeral 5}}
		\label{thermal phase 5}
	\end{subfigure}
\endgroup
	\caption{The five thermal brane web phases for the reflection-symmetric tripartite TFD configuration.}
	\label{fig:thermalq3-poincare-phases}
\end{minipage}%
}
\end{figure}
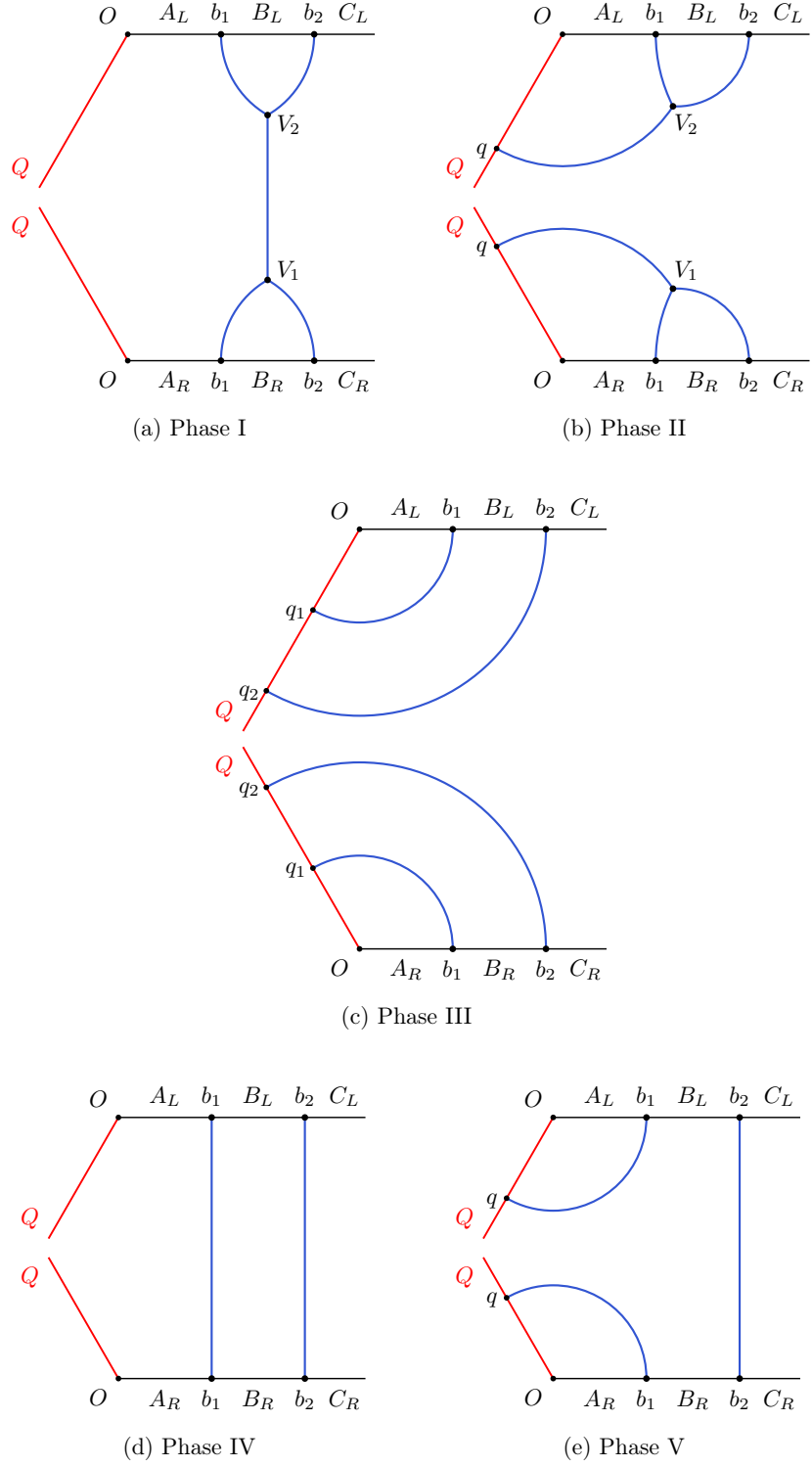

We consider the reflection-symmetric tripartition
\begin{equation}\label{eq:thermal-tripartition}
 A=[0,b_1]_L\cup[0,b_1]_R,\qquad
 B=[b_1,b_2]_L\cup[b_1,b_2]_R,\qquad
 C=[b_2,\infty)_L\cup[b_2,\infty)_R.
\end{equation}
Within the sector specified above there are five reduced brane web topologies, displayed in Fig.~\ref{fig:thermalq3-poincare-phases}. Phase~I is the connected double-Y web through the Einstein--Rosen bridge, Phase~II consists of two local brane-ending Y webs, Phase~III is fully brane-factorized, Phase~IV contains two wormhole-crossing walls, and Phase~V is the mixed brane/wormhole configuration.  We now determine their on-shell lengths directly.

\paragraph{Connected double-Y Phase~I.}

This is the only candidate whose two trivalent vertices move through the Lorentzian black-hole geometry.  It is the multipartite analogue of a Hartman--Maldacena surface: four boundary legs are joined by two equal-tension trivalent vertices, and the internal edge crosses the Einstein--Rosen bridge. The web does not end on $Q$, so its length is a property of the two-sided BTZ geometry alone; in other words, this phase also exists in AdS/CFT. Because Schwarzschild coordinates are singular at the horizon, it is convenient to extremize the web directly in Kruskal coordinates,
\begin{equation}\label{eq:kruskal-metric}
 ds^2=-\frac{4\,du\,dv}{(1+uv)^2}
 +\frac1{z_h^2}\frac{(1-uv)^2}{(1+uv)^2}\,dx^2,
 \qquad -1<uv<1 .
\end{equation}
The Schwarzschild-to-Kruskal maps in the four regions of the maximally extended geometry are
\begin{equation}\label{eq:kruskal-transform}
\begin{aligned}
 {\rm I}:&\quad
 u=-\sqrt{\frac{z_h-z}{z_h+z}}\,e^{-t/z_h},
 &v=+\sqrt{\frac{z_h-z}{z_h+z}}\,e^{t/z_h},\\
 {\rm II}:&\quad
 u=+\sqrt{\frac{z-z_h}{z+z_h}}\,e^{-t/z_h},
 &v=+\sqrt{\frac{z-z_h}{z+z_h}}\,e^{t/z_h},\\
 {\rm III}:&\quad
 u=+\sqrt{\frac{z_h-z}{z_h+z}}\,e^{-t/z_h},
 &v=-\sqrt{\frac{z_h-z}{z_h+z}}\,e^{t/z_h},\\
 {\rm IV}:&\quad
 u=-\sqrt{\frac{z-z_h}{z+z_h}}\,e^{-t/z_h},
 &v=-\sqrt{\frac{z-z_h}{z+z_h}}\,e^{t/z_h}.
\end{aligned}
\end{equation}

\begin{figure}[htbp]
	\centering
	\begin{tikzpicture}[x=1cm,y=1cm,>=stealth]
		\draw[black,thick] (0,0) -- (0,6);
		\draw[black,thick] (6,0) -- (6,6);
		
		\draw[red,thick]
		plot[domain=0:6,samples=61,sharp plot]
		(\x,{6+0.055*sin(900*\x)});
		\draw[red,thick]
		plot[domain=0:6,samples=61,sharp plot]
		(\x,{0.055*sin(900*\x)});
		
		\draw[black,dashed] (0,0) -- (6,6);
		\draw[black,dashed] (0,6) -- (6,0);
		
		\draw[gray,very thick,->] (3,3) -- (0,6);
		\draw[gray,very thick,->] (3,3) -- (6,6);
		\node[above right] at (0.55,5.25) {$u$};
		\node[above left]  at (5.45,5.25) {$v$};
		
		\node[font=\bfseries] at (5.0,3.0) {I};
		\node[font=\bfseries] at (3.0,5.0) {II};
		\node[font=\bfseries] at (1.0,3.0) {III};
		\node[font=\bfseries] at (3.0,1.0) {IV};
		
		\draw[gray,thick,->]
		plot[domain=-1.90:1.90,samples=64,smooth]
		({3+sqrt(1.15^2+(\x)^2)},{3+\x});
		\draw[gray,thick,->]
		plot[domain=1.90:-1.90,samples=64,smooth]
		({3-sqrt(1.15^2+(\x)^2)},{3+\x});
		\node[] at (5.25,4.55) {$t$};
		\node[] at (1.15,1.85) {$t$};
	\end{tikzpicture}
	\caption{Penrose diagram of the eternal black hole.}
	\label{fig:ads3-penrose-null}
\end{figure}
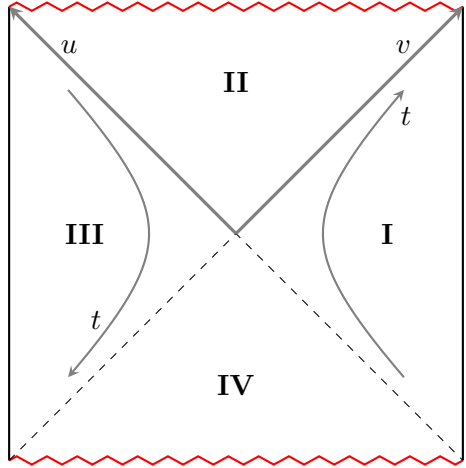

Regions I and III are the right and left exteriors, while II and IV are the black- and white-hole interiors; see Fig.~\ref{fig:ads3-penrose-null} for the Penrose diagram. For two arbitrary bulk points $P=(u,v,x)$ and $P'=(u',v',x')$ the geodesic distance is
\begin{equation}\label{eq:kruskal-distance}
 \cosh\mathcal L(P,P')=
 \frac{2(uv'+vu')+(1-uv)(1-u'v')
 \cosh\!\left(\frac{x-x'}{z_h}\right)}{(1+uv)(1+u'v')} .
\end{equation}
At the cutoff surface $z=\epsilon$ it is useful to keep the exact factor
\begin{equation}
 a_\epsilon\equiv\sqrt{\frac{z_h-\epsilon}{z_h+\epsilon}}
 =1-\frac{\epsilon}{z_h}+O(\epsilon^2).
\end{equation}
With the left exterior parametrized by $t_L=-t$ in the Kruskal chart (the Euclidean half-period is already encoded by the continuation to region III), the endpoint coordinates are
\begin{equation}
 (u_R,v_R)=\bigl(-a_\epsilon e^{-t/z_h},\,a_\epsilon e^{t/z_h}\bigr),
 \qquad
 (u_L,v_L)=\bigl(a_\epsilon e^{t/z_h},\,-a_\epsilon e^{-t/z_h}\bigr).
\end{equation}
The left--right reflection relates the two bulk vertices by $(u_2,v_2)=(v_1,u_1)$. Since the web does not end on the brane, its length functional is also invariant under $x\to b_1+b_2-x$, which implies that both vertices lie at the midpoint:
\begin{equation}\label{eq:thermal-vertex-symmetry}
 x_{V_1}=x_{V_2}=\frac{b_1+b_2}{2}.
\end{equation}
Using Eq.~\eqref{eq:kruskal-distance} and retaining the leading cutoff dependence, the sum of the four boundary-to-vertex legs and the internal vertex-to-vertex geodesic is
\begin{align}\label{eq:thermal-connected-functional}
\mathcal L_{\rm I}(u_1,v_1)={}&4\log\!\left[
\frac{2z_h}{\epsilon(1+u_1v_1)}
\left(e^{t/z_h}u_1-e^{-t/z_h}v_1
 +(1-u_1v_1)\cosh\!\frac{b_2-b_1}{2z_h}\right)\right]
\notag\\
&+2\operatorname{arsinh}\left(\frac{v_1-u_1}{1+u_1 v_1}\right),
\end{align}
where the second term is the internal geodesic length and we have used the non-crossing condition $u_1<v_1$. The extremality conditions
\begin{equation}\label{eq:thermal-uv-extremality}
 \partial_{u_1}\mathcal L_{\rm I}=\partial_{v_1}\mathcal L_{\rm I}=0
\end{equation}
simplify to
\begin{equation}\label{simplified extremal conditions}
	\frac{u_1+v_1}{1-u_1v_1}=\frac{\sinh\!\left(\frac{t}{z_h}\right)}{\cosh\!\left(\frac{b_2-b_1}{2z_h}\right)},\quad \quad \frac{v_1-u_1}{1+u_1v_1}=\sinh\chi,
\end{equation}
where we have introduced the shorthand
\begin{equation}\label{eq:thermal-chi}
 \chi\equiv
 \log\!\left[
 \frac{
 \sqrt{\cosh^2(t/z_h)+\sinh^2\!\left(\frac{b_2-b_1}{2z_h}\right)}
 +\cosh(t/z_h)}
 {\sqrt3\,\sinh\!\left(\frac{b_2-b_1}{2z_h}\right)}\right].
\end{equation}
From Eq.~\eqref{eq:thermal-connected-functional} we find $\mathcal{L}\left(V_1,V_2\right)=2\chi$, so geometric admissibility requires $\chi>0$. Interestingly, $\chi$ is directly related to the TFD cross-ratio
\eqref{eq:thermal-connected-block},
\begin{equation}
	\eta_{\rm TFD}=
 \frac{\sinh^2\!\left(\frac{b_2-b_1}{2z_h}\right)}
 {\cosh^2(\frac{t}{z_h})+\sinh^2\!\left(\frac{b_2-b_1}{2z_h}\right)}\Rightarrow e^\chi=\frac{1+\sqrt{1-\eta_{\rm TFD}}}{\sqrt{3\eta_{\rm TFD}}}.
\end{equation}
Thus a nondegenerate connected non-crossing web exists precisely when
\begin{equation}\label{eta constraint}
 \eta_{\rm TFD}<\frac34
 \qquad\Longleftrightarrow\qquad
 \sinh\!\left(\frac{b_2-b_1}{2z_h}\right)
 <\sqrt3\cosh\!\left(\frac{t}{z_h}\right).
\end{equation}
At equality the internal edge shrinks to zero.  Mapping the corresponding symmetric ideal quadrilateral to the Poincar\'e disk gives, for the angle $\vartheta$ between the two external legs inherited from either trivalent junction,
\begin{equation}
 \cos\vartheta=1-2\eta_{\rm TFD}.
\end{equation}
Hence $\eta_{\rm TFD}=3/4$ is exactly the limiting $2\pi/3$ configuration. The coincident point is therefore the endpoint of the two-trivalent-vertex branch, not an additional four-valent phase. This geometric interpretation also explains why the formal smooth solution is absent when the inequality \eqref{eta constraint} is violated.

Note that when the separation $b_2-b_1$ is small enough, this condition holds at all times $t$. Conversely, when the separation is large, it imposes a lower bound on $t$:
\begin{equation}\label{geo admiss condition}
	\begin{aligned}
		\sinh\left(\frac{b_2-b_1}{2z_h}\right)\leq &\sqrt{3},\quad \chi>0 \Rightarrow t\in\left(0,\infty\right),\\
		\sinh\left(\frac{b_2-b_1}{2z_h}\right)> &\sqrt{3},\quad \chi>0 \Rightarrow t>\widetilde t_\star=z_h\operatorname{arccosh}\!\left[
 \frac1{\sqrt3}\sinh\!\left(\frac{b_2-b_1}{2z_h}\right)\right].
	\end{aligned}
\end{equation}
The solutions of the extremal conditions \eqref{simplified extremal conditions} can be expressed in terms of Schwarzschild coordinates. If $V_1$ lies in the right exterior, region I, then
\begin{equation}\label{eq:thermal-region-I-uv}
 u_1=-\sqrt{\frac{z_h-z_V}{z_h+z_V}}e^{-t_V/z_h},
 \qquad
 v_1=\sqrt{\frac{z_h-z_V}{z_h+z_V}}e^{t_V/z_h},
\end{equation}
and the two algebraic conditions imply
\begin{align}
 z_V={}&z_h\operatorname{sech}\chi
 \sqrt{1+\sinh^2\!\left(\frac{t}{z_h}\right)
 \operatorname{sech}^2\!\left(\frac{b_2-b_1}{2z_h}\right)},
 \label{eq:thermal-zV}\\
 t_V={}&z_h\operatorname{arccosh}\!\left(
 \frac{z_V\sinh\chi}{\sqrt{z_h^2-z_V^2}}\right).
 \label{eq:thermal-tV-I}
\end{align}
If $V_1$ lies in the black-hole interior, region II,
\begin{equation}\label{eq:thermal-region-II-uv}
 u_1=\sqrt{\frac{z_V-z_h}{z_V+z_h}}e^{-t_V/z_h},
 \qquad
 v_1=\sqrt{\frac{z_V-z_h}{z_V+z_h}}e^{t_V/z_h},
\end{equation}
with the same radial coordinate \eqref{eq:thermal-zV} and
\begin{equation}\label{eq:thermal-tV-II}
 t_V=z_h\operatorname{arsinh}\!\left(
 \frac{z_V\sinh\chi}{\sqrt{z_V^2-z_h^2}}\right).
\end{equation}
When the separation satisfies
\begin{equation}
 \sinh\!\left(\frac{b_2-b_1}{2z_h}\right)\leq \frac1{\sqrt2},
\end{equation}
the vertex $V_1$ lies in the right exterior, region I, at all times. However, when
\begin{equation}
 \frac1{\sqrt2}<\sinh\!\left(\frac{b_2-b_1}{2z_h}\right)\leq \sqrt3,
\end{equation}
the vertex $V_1$ moves from the right exterior (region I) into the black-hole interior (region II) as $t$ increases. The horizon crossing $z_V=z_h$ occurs at
\begin{equation}\label{eq:thermal-tstar}
 t_\star=\frac{z_h}{2}\log\!\left[
 \frac{\sqrt3}{
 \sinh\!\left(\frac{b_2-b_1}{z_h}-\operatorname{arsinh}\sqrt3\right)}\right].
\end{equation}
For larger separations the geometrically admissible branch instead begins at $\widetilde t_\star$ given in Eq.~\eqref{geo admiss condition}, and the vertex $V_1$ always lies inside the black hole.
For compactness, define the three limiting radial values
\begin{equation}\label{eq:thermal-Xi}
\Xi_1=\frac{\sqrt3\sinh\!\left(\frac{b_2-b_1}{2z_h}\right)}
{2\cosh\!\left(\frac{b_2-b_1}{2z_h}\right)-1},\qquad
\Xi_2=\sqrt3\tanh\!\left(\frac{b_2-b_1}{2z_h}\right),\qquad
\Xi_3=\frac2{\sqrt3}\tanh\!\left(\frac{b_2-b_1}{2z_h}\right).
\end{equation}
The location of the geometrically admissible saddle is summarized in Table~\ref{tab:thermal-connected-ranges}.
\begin{table}[t]
\centering
\renewcommand{\arraystretch}{1.15}
\begin{tabular}{c c c c}
\hline
$\sinh\!\left(\frac{b_2-b_1}{2z_h}\right)$ & $t$ & $z_V/z_h$ & region of $V_1$\\
\hline
$(0,1/\sqrt2]$ & $(0,\infty)$ & $(\Xi_1,\Xi_2)$ & I\\
$(1/\sqrt2,\sqrt3]$ & $(0,t_\star)$ & $(\Xi_1,1)$ & I\\
 & $(t_\star,\infty)$ & $(1,\Xi_2)$ & II\\
$(\sqrt3,\infty)$ & $(\widetilde t_\star,\infty)$ & $(\Xi_3,\Xi_2)$ & II\\
\hline
\end{tabular}
\caption{Location of the vertex $V_1$ in the connected Phase~I saddle. When the separation exceeds the last threshold, no separated two-vertex solution exists for $0<t<\widetilde t_\star$.}
\label{tab:thermal-connected-ranges}
\end{table}

Substituting Eqs.~\eqref{eq:thermal-region-I-uv} and \eqref{eq:thermal-region-II-uv} into Eq.~\eqref{eq:thermal-connected-functional}, we obtain a unified expression for the length of the Phase~I brane web,
\begin{align}\label{eq:thermal-LI}
\mathcal L_{\rm I}={}&2\log\!\left[
\frac{4\beta^2}{3\sqrt3\,\pi^2\epsilon^2}
\sinh\!\left(\frac{\pi(b_2-b_1)}{\beta}\right)\right]
\notag\\
&+2\log\!\left[
\cosh\!\left(\frac{2\pi t}{\beta}\right)
+\sqrt{\cosh^2\!\left(\frac{2\pi t}{\beta}\right)
+\sinh^2\!\left(\frac{\pi(b_2-b_1)}{\beta}\right)}\right].
\end{align}
Note that once the admissibility condition \eqref{eta constraint} is violated, the expression \eqref{eq:thermal-LI} can only be understood as a formal analytic continuation rather than as a physical saddle.

\paragraph{Local brane-ending Phases~II and III.}
Both configurations factorize into identical time-independent webs on the two exteriors.  The induced metric on a constant-time BTZ slice is
\begin{equation}\label{eq:thermal-slice-metric}
 ds^2=\frac1{z^2}\left(\frac{dz^2}{1-z^2/z_h^2}+dx^2\right).
\end{equation}
This hyperbolic slice is isometric to the Poincar\'e half-plane. A convenient map is
\begin{equation}\label{eq:BTZ-Poincare-map}
 x=z_h\operatorname{arctanh}\!\left(
 \frac{2\widetilde x}{1+\widetilde x^2+\widetilde z^2}\right),\qquad
 z=\frac{2z_h\widetilde z}
 {\sqrt{(1-\widetilde x^2-\widetilde z^2)^2+4\widetilde z^2}},
\end{equation}
under which the EOW brane \eqref{BTZ EOW} is mapped to the straight Poincar\'e brane
\begin{equation}
 \widetilde z=-\cot\theta\,\widetilde x,
 \qquad \cot\theta=\kappa=\frac{1}{\sinh\rho_0}.
\end{equation}
This is the vacuum brane of Sec.~\ref{sec:review} with the same angle $\theta$.
Near the asymptotic boundary,
\begin{equation}\label{eq:thermal-boundary-map}
 \widetilde x=\tanh\!\left(\frac{x}{2z_h}\right),\qquad
 \widetilde z=\frac{z}{2z_h\cosh^2(x/2z_h)}+O(z^3),
\end{equation}
so a fixed BTZ regulator $z=\epsilon$ becomes the position-dependent Poincar\'e regulator
\begin{equation}\label{eq:thermal-cutoff-map}
 \widetilde\epsilon(x)=
 \frac{\epsilon}{2z_h\cosh^2(x/2z_h)}.
\end{equation}
Applying \eqref{eq:thermal-boundary-map}--\eqref{eq:thermal-cutoff-map} to the two vacuum tripartite branches in Sec.~\ref{three-party} gives
\begin{align}
\mathcal L_{\rm II}={}&4\log\!\left[
\frac{\beta}{\pi\epsilon}
\sinh\!\left(\frac{\pi(b_2-b_1)}{\beta}\right)\right]
+2\log\!\left[
\frac{\sqrt{\tanh(\frac{\pi b_2}{\beta})}+\sqrt{\tanh(\frac{\pi b_1}{\beta})}}
{2\left(\sqrt{\tanh(\frac{\pi b_2}{\beta})}-\sqrt{\tanh(\frac{\pi b_1}{\beta})}\right)}\right]
\notag\\
&+6\log\frac{2}{\sqrt3}+2\rho_0,
\label{eq:thermal-LII}\\[1mm]
\mathcal L_{\rm III}={}&2\log\!\left[
\frac{\beta^2}{\pi^2\epsilon^2}
\sinh\!\left(\frac{2\pi b_1}{\beta}\right)
\sinh\!\left(\frac{2\pi b_2}{\beta}\right)\right]
+4\rho_0.
\label{eq:thermal-LIII}
\end{align}
The first is the thermal image of the connected vacuum $Y$ web on each exterior; the second is the thermal image of the completely brane-factorized vacuum configuration, as shown in Figs.~\ref{thermal phase 2} and \ref{thermal phase 3}, respectively.


\paragraph{Wormhole-supported Phases~IV and V.}
The remaining two saddles are sums of ordinary two-sided geodesics.  Phase~IV contains two Hartman--Maldacena walls, while Phase~V combines one brane-ending pair with one Hartman--Maldacena wall.  Using the ordinary two-sided lengths reviewed in Sec.~\ref{sec:review},
\begin{align}
\mathcal L_{\rm IV}={}&4\log\!\left[
\frac{\beta}{\pi\epsilon}
\cosh\!\left(\frac{2\pi t}{\beta}\right)\right],
\label{eq:thermal-LIV}\\
\mathcal L_{\rm V}={}&2\log\!\left[
\frac{\beta}{\pi\epsilon}
\sinh\!\left(\frac{2\pi b_1}{\beta}\right)\right]
+2\log\!\left[
\frac{\beta}{\pi\epsilon}
\cosh\!\left(\frac{2\pi t}{\beta}\right)\right]
+2\rho_0.
\label{eq:thermal-LV}
\end{align}
Equations~\eqref{eq:thermal-LI}, \eqref{eq:thermal-LII}, \eqref{eq:thermal-LIII}, \eqref{eq:thermal-LIV} and \eqref{eq:thermal-LV} are the five on-shell geometric candidates, Phases~I--V, whose competition is analyzed in Sec.~\ref{subsec:black-hole-curve}. As in Sec.~\ref{four-party}, one can check that whenever the Phase~I brane web is geometrically inadmissible, the geometrically admissible Phase~IV saddle has a smaller on-shell length; see the right panel of Fig.~\ref{fig:thermal-pagecurve-roman} for an example. Hence the formal continuation of Phase~I cannot affect the physical lower envelope, and the time-dependent tripartite holographic multi-entropy is
\begin{equation}\label{eq:thermal-lower-envelope}
 S^{(3)}(t)=\frac{c}{6}
 \min_{X\in\{\mathrm{I},\mathrm{II},\mathrm{III},\mathrm{IV},\mathrm{V}\}}\mathcal L_X(t).
\end{equation}

\subsection{Field theory calculation}\label{subsec:thermal-cft}

We now compare the five gravitational saddles with candidate semiclassical channels in the replicated thermal BCFT. The thermal cylinder is parametrized by the complex coordinate $\xi=x+i\tau_E$, where $\tau_E=it$ is the Euclidean time and is periodically identified as $\tau_E\sim\tau_E+\beta$. The conformal boundary is located at $x=0$. The replicated partition function $Z_{n^2}^{\left(3\right)}$ in Eq.~\eqref{S3 field con} can be computed from the thermal correlator
\begin{equation}
	Z_{n^2}^{\left(3\right)}=\left\langle\prod_{i=1}^{4}\sigma_{g_i}(\xi_i,\bar\xi_i)\right\rangle_{{\rm T-BCFT}^{\otimes n^2}},
\end{equation}
where the subscript ``T-BCFT'' denotes the thermal background and $\xi_i$ are the complex coordinates of the insertions,
\begin{equation}\label{eq:thermal-insertions}
\begin{aligned}
 (x_1,t_1)&=(b_1,t), & (x_2,t_2)&=(b_2,t),\\
 (x_3,t_3)&=(b_2,-t+i\beta/2), &
 (x_4,t_4)&=(b_1,-t+i\beta/2).
\end{aligned}
\end{equation}
Since correlators are much easier to compute on the right half-plane in the vacuum state, we map the thermal half-cylinder $(x\geq 0)$ to the right half-plane by the conformal transformation
\begin{equation}\label{eq:thermal-map}
 w=\tanh\!\left(\frac{\pi\xi}{\beta}\right),
\end{equation}
where $w$ is the complex coordinate on the right half-plane. Under this transformation the boundary metric becomes
\begin{equation}\label{conformal factor}
	ds^2=d\xi d\bar{\xi}=\frac{\beta^2}{\pi^2}\cosh^2\left(\frac{\pi \xi}{\beta}\right)\cosh^2\left(\frac{\pi \bar{\xi}}{\beta}\right)dwd\bar{w}\Rightarrow \Omega\left(\xi,\bar{\xi}\right)=\frac{\beta}{\pi}\cosh\left(\frac{\pi \xi}{\beta}\right)\cosh\left(\frac{\pi \bar{\xi}}{\beta}\right),
\end{equation}
where $\Omega\left(\xi,\bar{\xi}\right)$ denotes the Weyl factor. Under the transformation \eqref{eq:thermal-map}, the twist correlator transforms as
\begin{align}\label{eq:thermal-correlator-map}
&\left\langle\prod_{i=1}^{4}\sigma_{g_i}(\xi_i,\bar\xi_i)\right\rangle_{{\rm T-BCFT}^{\otimes n^2}}
=
\prod_{i=1}^{4}\Omega(\xi_i,\bar\xi_i)^{-\Delta_n^{(3)}}
\left\langle\prod_{i=1}^{4}\sigma_{g_i}(w_i,\bar w_i)\right\rangle_{{\rm P-BCFT}^{\otimes n^2}},
\end{align}
where the subscript ``P-BCFT'' on the right-hand side indicates that the correlator is evaluated on the right half-plane. For the connected Phase~I, the bulk web selects a connected bulk-OPE channel in which the operators on each exterior are fused first. More explicitly, $\sigma_{g_1}$ fuses with $\sigma_{g_2}$, while $\sigma_{g_3}$ fuses with $\sigma_{g_4}$, and the two resulting intermediate twists subsequently fuse with each other into the identity:
\begin{equation}
 \left(\sigma_{g_1}\times\sigma_{g_2}\rightarrow\sigma_{g_1g_2}\right)
 \times
 \left(\sigma_{g_3}\times\sigma_{g_4}\rightarrow\sigma_{g_3g_4}\right)\rightarrow \text{id},
\end{equation}
where we have used
\begin{equation}
	g_4=g_1^{-1},\quad g_3=g_2^{-1}\quad \Rightarrow \quad  g_3g_4=\left(g_1g_2\right)^{-1}.
\end{equation}

The same structure becomes manifest after applying the doubling trick \cite{Cardy:1991tv,McAvity:1995zd,Sully:2020pza},
\begin{equation}
	\begin{aligned}
		\left\langle\prod_{i=1}^{4}
 \sigma_{g_i}(w_i,\bar w_i)\right\rangle_{{\rm P\text{-}BCFT}^{\otimes n^2}}
 =&
 \left\langle\prod_{i=1}^{4}
 \sigma_{g_i}(w_i)\,
 \sigma_{g_i^{-1}}(-\bar w_i)\right\rangle_{{\rm CFT}^{\otimes n^2}}\\
  =&
 \left\langle\prod_{i=1}^{4}
 \sigma_{g_i}(w_i)\right\rangle_{{\rm CFT}^{\otimes n^2}} \times \left\langle\prod_{i=1}^{4}
 \sigma_{g_i^{-1}}(-\bar w_i)\right\rangle_{{\rm CFT}^{\otimes n^2}}\\
 =&\left\langle\prod_{i=1}^{4}
 \sigma_{g_i}(w_i,\bar w_i)\right\rangle_{{\rm CFT}^{\otimes n^2}},
	\end{aligned}
\end{equation}
which implies that in this channel the BCFT correlator reduces to the CFT correlator in the $s$-channel,
\begin{equation}\label{eq:thermal-connected-CFT}
	\begin{aligned}
		\left\langle\prod_{i=1}^{4}
 \sigma_{g_i}(w_i,\bar w_i)\right\rangle_{{\rm P\text{-}BCFT}^{\otimes n^2}}=&\left\langle\prod_{i=1}^{4}
 \sigma_{g_i}(w_i,\bar w_i)\right\rangle_{{\rm CFT}^{\otimes n^2}}
\\
=&
\left(\frac{\epsilon^2}{w_{21}\bar w_{21}}\right)^{\Delta_n^{(3)}}
\left(\frac{\epsilon^2}{w_{34}\bar w_{34}}\right)^{\Delta_n^{(3)}}
\CC_{\Delta_n^{(3)}}^2
\CF_{\Delta_n^{(3)}}(\eta)\,
\overline{\CF}_{\Delta_n^{(3)}}(\bar\eta),
	\end{aligned}
\end{equation}
where $w_{ij}=w_{i}-w_j$, the bulk OPE coefficient $\mathcal{C}_{\Delta_n^{\left(3\right)}}$ is given by \eqref{C coefficient}, and the conformal block $\mathcal{F}$ is given by \eqref{semi-classical conformal BCFT},
\begin{equation}\label{eq:thermal-connected-block}
 \CF_{\Delta_n^{(3)}}(\eta)=
 \left(\frac{1+\sqrt{1-\eta}}{2\sqrt\eta}\right)^{-\Delta_n^{(3)}},\quad \eta=\frac{w_{12}w_{34}}{w_{13}w_{24}}= \frac{\sinh^2\left[\frac{\pi(b_2 - b_1)}{\beta}\right]}{\cosh^2\left(\frac{2\pi t}{\beta}\right) + \sinh^2\left[\frac{\pi(b_2 - b_1)}{\beta}\right]}.
\end{equation}
Substituting Eq.~\eqref{eq:thermal-connected-CFT} into Eq.~\eqref{S3 field con}, we obtain the tripartite multi-entropy in this phase,
\begin{align}\label{eq:thermal-CFT-connected-result}
S^{(3)}\left(A:B:C\right)={}&\frac{c}{3}
 \log\!\left[\frac{4\beta^2}{3\sqrt3\pi^2\epsilon^2}
 \sinh\!\left(\frac{\pi(b_2-b_1)}{\beta}\right)\right]
\notag\\
&+\frac{c}{3}\log\!\left[
 \cosh\!\left(\frac{2\pi t}{\beta}\right)
 +\sqrt{\cosh^2\!\left(\frac{2\pi t}{\beta}\right)
 +\sinh^2\!\left(\frac{\pi(b_2-b_1)}{\beta}\right)}\right].
\end{align}

For Phases~II and III, the four-point correlator \eqref{eq:thermal-correlator-map} factorizes into two two-point correlators, one on each asymptotic boundary:
\begin{equation}\label{eq:thermal-G12G34}
	\begin{aligned}
		\left\langle\prod_{i=1}^{4}\sigma_{g_i}(w_i,\bar w_i)\right\rangle_{{\rm P\text{-}BCFT}^{\otimes n^2}}=\left\langle\sigma_{g_1}(w_1,\bar w_1)\sigma_{g_2}(w_2,\bar w_2)\right\rangle_{{\rm P\text{-}BCFT}^{\otimes n^2}}\\
		\times \left\langle\sigma_{g_3}(w_3,\bar w_3)\sigma_{g_4}(w_4,\bar w_4)\right\rangle_{{\rm P\text{-}BCFT}^{\otimes n^2}}.
	\end{aligned}
\end{equation}
For Phase~II, each two-point correlator is in the bulk OPE channel discussed in Sec.~\ref{section:Tripartite-field}; for example,
\begin{equation}\label{two-point function bulk channel TFD}
	\begin{aligned}
		&\left\langle\sigma_{g_1}(w_1,\bar w_1)\sigma_{g_2}(w_2,\bar w_2)\right\rangle_{{\rm P\text{-}BCFT}^{\otimes n^2}}\\
 =&\left[\frac{\epsilon^2}{(w_1-w_2)(\bar w_1-\bar w_2)}\right]^{\Delta_n^{(3)}}
 \mathcal A_{\Delta_n^{(3)}}^{(b)}\CC_{\Delta_n^{(3)}}
 \CF_{\Delta_n^{(3)}}(\eta_{\text{bulk}}),
	\end{aligned}
\end{equation}
where the semiclassical block and the cross-ratio in the bulk OPE channel are
given by Eqs.~\eqref{semi-classical conformal BCFT} and \eqref{cross-ratio bulk channel}, respectively,
\begin{equation}\label{eq:thermal-two-point-crossratio}
 \CF_{\Delta_n^{(3)}}(\eta_{\text{bulk}})=
 \left(\frac{1+\sqrt{1-\eta_{\text{bulk}}}}{2\sqrt{\eta_{\text{bulk}}}}\right)^{-\Delta_n^{(3)}},\quad \eta_{\text{bulk}}=\frac{(w_1-w_2)(\bar w_1-\bar w_2)}{(w_1+\bar w_2)(w_2+\bar w_1)}.
\end{equation}
Substituting Eqs.~\eqref{two-point function bulk channel TFD} and \eqref{eq:thermal-G12G34} into Eq.~\eqref{S3 field con}, we obtain the tripartite multi-entropy in this phase,
\begin{equation}
	\begin{aligned}
		S^{(3)}\left(A:B:C\right)=&\frac{2c}{3}\log\!\left[
\frac{\beta}{\pi\epsilon}
\sinh\!\left(\frac{\pi(b_2-b_1)}{\beta}\right)\right]
+c\log\frac{2}{\sqrt3}+2\log g_b\\
&+\frac{c}{3}\log\!\left[
\frac{\sqrt{\tanh(\frac{\pi b_2}{\beta})}+\sqrt{\tanh(\frac{\pi b_1}{\beta})}}
{2\left(\sqrt{\tanh(\frac{\pi b_2}{\beta})}-\sqrt{\tanh(\frac{\pi b_1}{\beta})}\right)}\right].
	\end{aligned}
\end{equation}

For Phase~III, the two two-point correlators in Eq.~\eqref{eq:thermal-G12G34} are in the boundary OPE channel discussed in Sec.~\ref{section:Tripartite-field}:
\begin{equation}\label{two-point function boundary channel TFD}
	\left\langle\sigma_{g_1}(w_1,\bar w_1)\sigma_{g_2}(w_2,\bar w_2)\right\rangle_{{\rm P\text{-}BCFT}^{\otimes n^2}}=\left(\frac{\epsilon}{w_1+\bar w_1}\right)^{\Delta_n^{(3)}}
	\left(\frac{\epsilon}{w_2+\bar w_2}\right)^{\Delta_n^{(3)}}
	g_b^{\,2n(1-n)}.
\end{equation}
Substituting this and Eq.~\eqref{eq:thermal-G12G34} into Eq.~\eqref{S3 field con}, we obtain the tripartite multi-entropy in this phase,
\begin{equation}\label{eq:thermal-CFT-brane-factorized}
	S^{(3)}\left(A:B:C\right)=
	\frac{c}{3}\log\!\left[
	\frac{\beta^2}{\pi^2\epsilon^2}
	\sinh\!\left(\frac{2\pi b_1}{\beta}\right)
	\sinh\!\left(\frac{2\pi b_2}{\beta}\right)\right]
	+4\log g_b.
\end{equation}

For Phases~IV and V, the four-point correlator \eqref{eq:thermal-correlator-map} instead factorizes as
\begin{equation}\label{eq:thermal-G14G23}
	\begin{aligned}
		\left\langle\prod_{i=1}^{4}\sigma_{g_i}(w_i,\bar w_i)\right\rangle_{{\rm P\text{-}BCFT}^{\otimes n^2}}=\left\langle\sigma_{g_1}(w_1,\bar w_1)\sigma_{g_4}(w_4,\bar w_4)\right\rangle_{{\rm P\text{-}BCFT}^{\otimes n^2}}\\
		\times \left\langle\sigma_{g_2}(w_2,\bar w_2)\sigma_{g_3}(w_3,\bar w_3)\right\rangle_{{\rm P\text{-}BCFT}^{\otimes n^2}}.
	\end{aligned}
\end{equation}
The two-point correlator between $w_2$ and $w_3$ is in the bulk OPE channel,
\begin{equation}\label{twopointfunction}
	\left\langle\sigma_{g_2}(w_2,\bar w_2)\sigma_{g_3}(w_3,\bar w_3)\right\rangle_{{\rm P\text{-}BCFT}^{\otimes n^2}}= \left[\frac{\epsilon^2}{(w_2-w_3)(\bar w_2-\bar w_3)}\right]^{\Delta_n^{(3)}}.
\end{equation}
Note that this is not the same as Eq.~\eqref{two-point function bulk channel TFD}, since $\sigma_{g_2}$ and $\sigma_{g_3}$ fuse into the identity rather than into another twist operator of dimension $\Delta_n^{(3)}$. For Phase~IV, the two-point correlator between $w_1$ and $w_4$ is also in the bulk OPE channel and is given by Eq.~\eqref{twopointfunction} with the replacements $w_2\rightarrow w_1$, $w_3\rightarrow w_4$. Substituting Eqs.~\eqref{twopointfunction} and \eqref{eq:thermal-G14G23} into Eq.~\eqref{S3 field con}, we obtain the tripartite multi-entropy in this phase,
\begin{equation}
	S^{(3)}\left(A:B:C\right)=\frac{2c}{3}\log\!\left[
 \frac{\beta}{\pi\epsilon}\cosh\!\left(\frac{2\pi t}{\beta}\right)\right].
\end{equation}

Similarly, for Phase~V the two-point correlator between $w_1$ and $w_4$ is in the boundary OPE channel and is given by Eq.~\eqref{two-point function boundary channel TFD} with the replacement $w_2\rightarrow w_4$:
\begin{equation}
	\left\langle\sigma_{g_1}(w_1,\bar w_1)\sigma_{g_4}(w_4,\bar w_4)\right\rangle_{{\rm P\text{-}BCFT}^{\otimes n^2}}=\left(\frac{\epsilon}{w_1+\bar w_1}\right)^{\Delta_n^{(3)}}
	\left(\frac{\epsilon}{w_4+\bar w_4}\right)^{\Delta_n^{(3)}}
	g_b^{\,2n(1-n)}.
\end{equation}
Substituting this, Eq.~\eqref{twopointfunction} and Eq.~\eqref{eq:thermal-G14G23} into Eq.~\eqref{S3 field con}, we obtain the tripartite multi-entropy in this phase,
\begin{equation}
	S^{(3)}\left(A:B:C\right)=\frac{c}{3}\log\!\left[
 \frac{\beta^2}{\pi^2\epsilon^2}
 \sinh\!\left(\frac{2\pi b_1}{\beta}\right)
 \cosh\!\left(\frac{2\pi t}{\beta}\right)\right]+2\log g_b.
\end{equation}
All of these field-theory results agree with the bulk calculations of the previous subsection. Consequently, all five principal bulk saddles (Phases~I--V) have natural semiclassical fusion-channel counterparts that reproduce the corresponding tripartite multi-entropy.

\subsection{Five-saddle phase diagram and equilibrium multi-entropy curves}\label{subsec:black-hole-curve}

It is convenient to introduce the dimensionless variables
\begin{equation}\label{eq:thermal-dimensionless}
 \alpha=\frac{\pi b_1}{\beta},\qquad
 \delta=\frac{\pi(b_2-b_1)}{\beta},\qquad
 \tau=\frac{2\pi t}{\beta}.
\end{equation}
In these variables the cross-ratio introduced in Eq.~\eqref{eq:thermal-connected-block} takes the compact form
\begin{equation}\label{eq:thermal-cross-ratio-dimensionless}
 \eta_{\rm TFD}=\frac{\sinh^2\delta}{\cosh^2\tau+\sinh^2\delta}\:.
\end{equation}

The transitions among Phases~III, IV and V are inherited directly from those of the entanglement entropy of the interval $B$,
\begin{equation}\label{eq:ordinary-thermal-crossings}
 \cosh\tau_{\mathrm{IV}/\mathrm{V}}=e^{\rho_0}\sinh(2\alpha),\qquad
 \cosh\tau_{\mathrm{V}/\mathrm{III}}=e^{\rho_0}\sinh\left[2(\alpha+\delta)\right],
\end{equation}
where $\tau_{\mathrm{IV}/\mathrm{V}}$ and $\tau_{\mathrm{V}/\mathrm{III}}$ denote the corresponding transition times.

Transitions involving Phase~I, by contrast, change the connectivity of a trivalent multiway web.  Their independence from the ordinary HRT scales is already visible in the growth rates:
\begin{align}\label{eq:thermal-growth-rates}
 \partial_\tau\mathcal L_{\rm IV}&=4\tanh\tau,
 &\partial_\tau\mathcal L_{\rm V}&=2\tanh\tau,\notag\\
 \partial_\tau\mathcal L_{\rm I}&=\frac{2\sinh\tau}{\sqrt{\cosh^2\tau+\sinh^2\delta}},
 &\partial_\tau\mathcal L_{\rm II}&=\partial_\tau\mathcal L_{\rm III}=0.
\end{align}
For $\tau>0$ (i.e.\ $t>0$), we find
\begin{equation}
 \partial_\tau\mathcal L_{\rm IV}>
 \partial_\tau\mathcal L_{\rm V}>
 \partial_\tau\mathcal L_{\rm I}>0.
\end{equation}
Generally, the dominant saddle changes from faster-growing branch to slower-growing branches, and the multi-entropy eventually saturates on one of the two static phases, Phase~II or Phase~III.

\begin{figure}[t]
 \centering
 \includegraphics[width=0.86\linewidth]{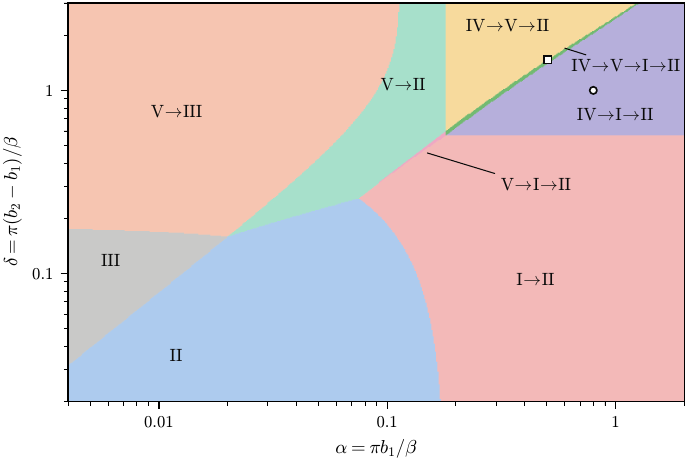}
 \caption{Ordered sequences of dominant saddles at $\rho_0=1$ as $\tau$ increases from zero. Both axes are logarithmic. The circle and the square mark the parameter sets of Fig.~\ref{fig:thermal-pagecurve-roman}(a) and (b), respectively.}
 \label{fig:thermal-sequence-roman}
\end{figure}
At $\rho_0=1$, the following sequences of dominant phases occur as time increases:
\begin{equation}\label{eq:thermal-transition-words}
\begin{gathered}
 \mathrm{V}\to\mathrm{III},\qquad \mathrm{V}\to\mathrm{II},\qquad \mathrm{I}\to\mathrm{II},\qquad
 \mathrm{IV}\to\mathrm{I}\to\mathrm{II},\\
 \mathrm{IV}\to\mathrm{V}\to\mathrm{II},\qquad \mathrm{V}\to\mathrm{I}\to\mathrm{II},\qquad
 \mathrm{IV}\to\mathrm{V}\to\mathrm{I}\to\mathrm{II}.
\end{gathered}
\end{equation}
Which sequence is realized depends on the endpoints $b_1$ and $b_2$,
as shown in Fig.~\ref{fig:thermal-sequence-roman}.
For endpoints close to the BCFT boundary on the thermal scale, $\alpha,\delta\ll1$, the web is always static and there is no phase transition. The web is the pair of local $Y$ webs (II) or the completely brane-factorized web (III), according to whether $b_2/b_1=1+\delta/\alpha$ lies below or above the vacuum threshold $\eta_{c,2}$, with $\eta_{c,2}\simeq8.93$ at $\rho_0=1$. The largest time-dependent domains are $\mathrm{V}\to\mathrm{III}$, $\mathrm I\to\mathrm{II}$ and $\mathrm{IV}\to\mathrm I\to\mathrm{II}$. The two phase-transition sequences that pass through both Phase~V and Phase~I occur only in two narrow regions of parameter space: one between the regions corresponding to $\mathrm{IV}\to\mathrm V\to\mathrm{II}$ and $\mathrm{IV}\to\mathrm I\to\mathrm{II}$, and the other between those corresponding to $\mathrm V\to\mathrm{II}$ and $\mathrm I\to\mathrm{II}$.

\begin{figure}[ht]
 \centering
 \includegraphics[width=\linewidth]{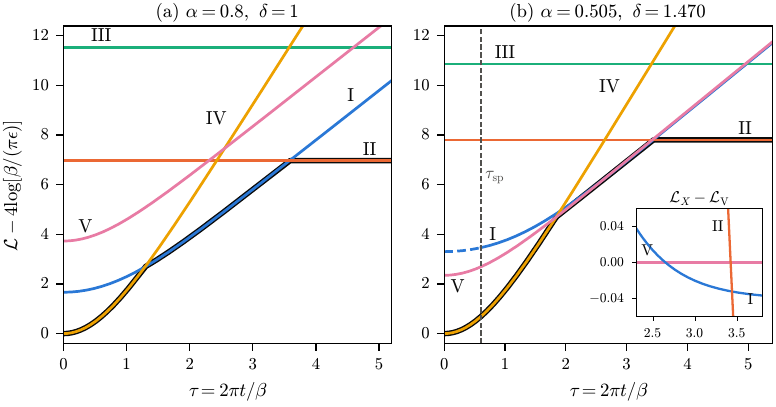}
 \caption{Equilibrium multi-entropy curves at $\rho_0=1$ for the parameter sets \eqref{eq:thermal-generic-point} (a) and \eqref{eq:thermal-representative-point} (b), with the common ultraviolet term subtracted.  The thin curves are the five saddles and the thick black curve is their lower envelope, which follows $\mathrm{IV}\to\mathrm{I}\to\mathrm{II}$ in (a) and $\mathrm{IV}\to\mathrm{V}\to\mathrm{I}\to\mathrm{II}$ in (b).  The vertical dashed line in (b) marks $\tau_{\rm sp}$, where Phase~I first becomes geometrically admissible; for $\tau<\tau_{\rm sp}$ the dashed blue curve is the formal continuation of $\mathcal L_{\rm I}$, which lies above $\mathcal L_{\rm IV}$. The inset shows $\mathcal L_{\rm I}$ and $\mathcal L_{\rm II}$ relative to $\mathcal L_{\rm V}$ in the window where the three saddles compete.}
 \label{fig:thermal-pagecurve-roman}
\end{figure}

The Page-like structure\footnote{The terminology ``Page-like'' refers only to the growth and saturation of this equilibrium lower envelope.  Genuine multi-entropy and related multipartite diagnostics exhibit Page-like structures in genuinely evaporating or dynamical settings as well \cite{Iizuka:2024pzm,Anegawa:2025prn,Balasubramanian:2025jhq,Fujiki:2026qdt,Ju:2026lst,Anegawa:2026yee}, but the eternal black hole considered here is a TFD state, not an evaporating system.  The important physics is that the connected Phase~I introduces a reorganization scale that cannot be reconstructed by nesting the two ordinary HRT transitions.} is most transparent in a one-dimensional time scan; two examples are shown in Fig.~\ref{fig:thermal-pagecurve-roman}.  The first example is
\begin{equation}\label{eq:thermal-generic-point}
 \alpha=0.8,\qquad \delta=1,\qquad \rho_0=1,
\end{equation}
where $\sinh\delta<\sqrt3$, so the connected web exists at all times. The phase transitions in this example are
\begin{equation}\label{eq:thermal-generic-word}
 \mathrm{IV}\longrightarrow\mathrm{I}\longrightarrow\mathrm{II},
 \qquad
 \tau_{\mathrm{IV}/\mathrm{I}}\simeq1.293,\qquad
 \tau_{\mathrm{I}/\mathrm{II}}\simeq3.585.
\end{equation}

Neither of the ordinary HRT transitions \eqref{eq:ordinary-thermal-crossings}, which would occur at $\tau_{\mathrm{IV}/\mathrm V}\simeq2.552$ and $\tau_{\mathrm V/\mathrm{III}}\simeq4.599$, is realized: Phase~V is never dominant, and the saturation time is set by the reconnection of the connected web into two local $Y$ webs.
For the second example, we choose
\begin{equation}\label{eq:thermal-representative-point}
 \alpha=0.505,\qquad \delta=1.470,\qquad \rho_0=1.
\end{equation}
Phase~I becomes admissible only at $\tau_{\rm sp}\simeq0.606$; for $\tau<\tau_{\rm sp}$ its formal continuation lies above Phase~IV, as discussed in Sec.~\ref{subsec:thermal-gravity}, and therefore plays no role. The phase transitions in this example are
\begin{equation}\label{eq:thermal-representative-word}
 \mathrm{IV}\longrightarrow\mathrm{V}\longrightarrow\mathrm{I}\longrightarrow\mathrm{II},
\end{equation}
with transition times
\begin{equation}\label{eq:thermal-representative-times}
 \tau_{\mathrm{IV}/\mathrm{V}}\simeq1.843,\qquad
 \tau_{\mathrm{V}/\mathrm{I}}\simeq2.647,\qquad
 \tau_{\mathrm{I}/\mathrm{II}}\simeq3.436.
\end{equation}
At late times, Phases~I and V are nearly degenerate, and their difference approaches the constant
\begin{equation}\label{eq:thermal-I-V-offset}
 \mathcal L_{\rm I}-\mathcal L_{\rm V}\;\longrightarrow\;
 2\log\left[\frac{8\sinh\delta}{3\sqrt3\,e^{\rho_0}\sinh(2\alpha)}\right]=-0.041.
\end{equation}

\section{Summary and outlook}\label{sec:discussion}
In this work we have developed the semiclassical multi-entropy geometry of AdS$_3$/BCFT$_2$ in a sector where the bulk problem can be formulated as an explicit planar network optimization.  The central effect of the EOW brane is topological rather than merely local.  In ordinary AdS/BCFT the brane enlarges the class of admissible RT/HRT surfaces by allowing them to terminate on $Q$ \cite{Takayanagi:2011zk,Fujita:2011fp,Sully:2020pza}.  For multi-entropy the same relative-homology freedom acts on an entire multiway cut: connected Steiner networks can split into several components, and each component may terminate on the brane.  The resulting phase structure therefore contains information that is not captured by following the RT surfaces of the individual boundary intervals separately.

For static adjacent partitions we classified the reduced planar webs explicitly.  The tripartite problem contains a connected brane-ending $Y$ network and a factorized phase, while the quadripartite problem contains five reduced topologies.  Direct extremization determines the candidate lengths and makes the $2\pi/3$ force-balance condition at each equal-tension trivalent vertex manifest.  We have also separated algebraic stationarity from geometric admissibility: the fully connected quadripartite webs are governed by a single Steiner polynomial whose root always exists, but they enter the lower envelope only when the internal edge has positive length and the outer vertex lies inside the physical AdS/BCFT region.  With these explicit conditions the static quadripartite phase diagram follows; all five topologies dominate somewhere, and the connected domain grows with the brane tension.  The auxiliary hyperbolic construction in Appendix~\ref{app:hyperbolic} gives an independent geometric interpretation of the same saddles, and the counting argument in Appendix~\ref{app:web-counting} explains why the number of planar topologies is governed by a Catalan structure.  One immediate consequence is that the factorization scale of a multipartite web need not coincide with the ordinary RT transition scale.  The EOW brane can therefore reorganize genuinely multipartite correlations before any analogous change is visible in a pairwise entanglement diagnostic.

The BCFT analysis provides a complementary test of this picture.  After doubling, the static replica observables become reflection-symmetric chiral correlators whose semiclassical fusion channels can be compared directly with the brane web topologies; in the thermal problem the same dictionary supplies candidate eight-point channels, with the connected Phase~I reduction remaining a bulk-motivated ansatz.  For the connected quadripartite configuration, the derivatives of the on-shell web action furnish explicit accessory parameters at generic endpoint positions.  Their Ward-identity and linearized-monodromy residuals vanish on the same polynomial locus that determines the physical Steiner modulus.  This is a stronger check than simply reproducing the final logarithmic dependence of the entropy: it ties the internal modulus of the bulk network directly to the semiclassical conformal-block problem.  The calculation also separates kinematics from dynamics. The moduli-dependent block is fixed by the semiclassical channel, while its normalization is fixed by the OPE limit. By contrast, the twist OPE coefficient remains genuine CFT data. Agreement with the classical bulk result requires its leading holographic value, which is the same for each trivalent vertex, and does not follow from conformal symmetry alone.

The thermal TFD problem shows that the same distinction persists in Lorentzian geometry.  Within the reflection-symmetric principal non-winding sector, five reduced saddles compete.  The connected Phase~I saddle is qualitatively special because its two trivalent vertices move through the Kruskal geometry and are joined across the Einstein--Rosen bridge.  Extremizing directly in $(u_1,v_1)$ gives a single smooth branch that passes between the exterior and the black-hole interior and exists only in the Steiner-admissible domain
\begin{equation}
 \eta_{\rm TFD}<\frac34.
\end{equation}
At $\eta_{\rm TFD}=3/4$ the internal edge collapses and the limiting external angle is $2\pi/3$; this is the endpoint of the two-vertex branch, not an additional four-valent phase.  Equally importantly, this existence line is not itself a dominance transition.  In the largest time-dependent domain of the phase diagram with a wormhole-supported early phase, the lower envelope follows $\mathrm{IV}\to\mathrm{I}\to\mathrm{II}$: neither of the two ordinary HRT switches is realized, and both the onset of the slower growth and the saturation time are set by multipartite reconnections. In a narrow region of parameters the envelope instead follows $\mathrm{IV}\to\mathrm{V}\to\mathrm{I}\to\mathrm{II}$. This separation between admissibility, multipartite reconnection and ordinary HRT transitions is the main physical lesson of the thermal analysis. The resulting curve is Page-like only as an equilibrium growth-and-saturation pattern; no evaporation is taking place in the eternal TFD geometry.

There are several limitations to the present analysis.  The web prescription is used on the replica-selected branch near $n=1$ and does not by itself determine the dominant finite-integer-$n$ bulk filling.  Replica-symmetry analyses and random-tensor-network calculations indicate that additional saddles, replica-symmetry breaking or other obstructions can become relevant away from this limit \cite{Penington:2022dhr,Gadde:2024taa,Hu:2026bhg,Akella:2026bci,Iizuka:2026mlw}.  The static completeness statement is likewise restricted to reduced planar disk topologies, and the thermal classification assumes left--right reflection symmetry and the principal non-winding BTZ homotopy class.  Winding sectors, nonsymmetric networks and a genuinely covariant/maximin formulation should enlarge the problem.  A complete eight-point monodromy analysis of the thermal BCFT correlator would also be desirable; here the connected Phase~I result is a bulk-motivated channel ansatz checked against the gravitational saddle rather than a proof of global block dominance.

A particularly important next step is to make gravity itself part of the multipartite variational problem.  We defer to forthcoming work the problem of \emph{multipartite entropy islands} \cite{BasuXu:2026MultipartiteIslands}.  The natural expectation is that the fixed-background brane webs of the present paper are replaced, in a gravitating region, by a generalized multi-entropy functional whose competing saddles include multiway island boundaries and replica-wormhole geometries.  From this viewpoint the relative-homology structures found here are useful prototypes for the more general island problem, but they should not be conflated with it: an endpoint on an AdS/BCFT EOW brane is a boundary condition in the present calculation, whereas an island boundary must arise from extremization of a gravitational generalized-entropy functional. A direct treatment of this distinction, together with the corresponding gravitational replica derivation, is deferred to that work.

A second related direction concerns \emph{reflected multi-entropy}.  Reflected multi-entropy combines the multi-entropy contraction with canonical purification \cite{Yuan:2024yfg}, so its bulk dual is expected to involve multiway cuts of an entanglement wedge rather than the absolute bulk webs studied here.  AdS/BCFT adds a further layer because such cuts can meet the EOW brane, leading to new boundary-ending channels and additional factorization transitions.  We likewise defer to forthcoming work \cite{BasuXu:2026ReflectedMultiEntropy} the reflected-multi-entropy construction in AdS/BCFT and its extension to island phases, where canonical purification, defect/extremal-web conditions and gravitational extremization must be treated simultaneously.  Together, these companion directions should clarify which aspects of the multiway geometry are universal and which depend essentially on the replica preparation of the mixed-state observable.

These extensions also sharpen the relation to the broader landscape of multipartite measures.  Entanglement-wedge cross sections and multipartite entanglement of purification \cite{Takayanagi:2017knl,Nguyen:2017yqw,Bao:2017nhh,Espindola:2018ozt,Bao:2018fso,Bao:2018gck}, reflected entropy and its multipartite generalizations \cite{Dutta:2019gen,Bao:2019zqc}, reflected multi-entropy \cite{Yuan:2024yfg}, multipartite Markov gaps \cite{Iizuka:2025elr}, and genuine multi-entropy or multipartite-signal constructions \cite{Iizuka:2025ioc,Iizuka:2025bcc,Iizuka:2025caq,Balasubramanian:2024ysu,Gadde:2026msg,Ju:2026xfh,Balasubramanian:2026chr} probe different replica sewings, partition structures, or subtractions of lower-order correlations.  Closely related alternative diagnostics of genuine multipartite entanglement, such as latent entropy (L-entropy), provide a useful comparison because they are built from reflected-entropy bounds rather than the multi-entropy replica invariant \cite{Basak:2024uwc,Basak:2026seh}.  The same permutation-replica language connects multi-entropy to multi-invariants, monotones and topological diagnostics \cite{Gadde:2023zni,Gadde:2025csh,Gadde:2024taa,Gadde:2025ybn,Nezami:2016zni,Akella:2026xza,Liu:2024ulq,Sheffer:2025jtc,Sheffer:2025zyr,Yuan:2025dgx,Chang:2026uds,Akella:2026rbe,DelZotto:2026fpw}, while critical, symmetry-resolved, Lifshitz and Gaussian examples provide complementary many-body tests \cite{Harper:2025uui,Iizuka:2025pqq,Berthiere:2025toi,Camargo:2026xrd}.  Black-hole and dynamical applications now range from evaporation-inspired multi-entropy curves to BTZ, quenches, Vaidya collapse and continued Einstein--Rosen-bridge growth after ordinary subsystem entropies have saturated \cite{Iizuka:2024pzm,Anegawa:2025prn,Balasubramanian:2025jhq,Fujiki:2026qdt,Ju:2026lst,Anegawa:2026yee}.  The AdS/BCFT setting provides a common geometric arena in which these distinctions can be made concrete because the EOW brane allows the topology of a candidate cut to change in a controlled way.  Recent studies of four-boundary AdS$_3$ wormholes provide a complementary illustration: at finite Fenchel--Nielsen twist, the $q=4$ multi-entropy can become sensitive to bulk moduli while the complete bipartite RT entropy vector remains unchanged \cite{Iizuka:2026kbo,Anegawa:2026npk}.

The broader lesson is that a multipartite holographic observable is sensitive not only to the lengths of extremal curves but also to how those curves are allowed to join, split and terminate.  In the present fixed-background problem this information is encoded by the topology and Steiner geometry of brane webs; in gravitational island problems it should be supplemented by generalized-entropy extremization and replica wormholes.  Establishing this hierarchy of structures is, in our view, an important step toward a systematic geometric description of multipartite entanglement in holography.

\acknowledgments

We would like to thank Yiwei Zhong for useful discussions and comments on the draft. The authors are supported by the NSFC Grant No. 12447108 and the Shing-Tung Yau Center of Southeast University. MX is supported by the SEU Innovation Capability Enhancement Plan for Doctoral Students
(Grant No.~CXJH\_SEU 26163).

\appendix
\section{Hyperbolic geometry behind the static brane webs}\label{app:hyperbolic}
\tikzset{
 AppendixGeo/.style={draw=UnifiedBraneWebBlue,line width=1.05pt,line cap=round,line join=round},
 AppendixAux/.style={draw=orange!85!black,dashed,line width=0.9pt,line cap=round},
 AppendixBoundary/.style={draw=black,line width=0.7pt},
 AppendixBrane/.style={draw=red,line width=0.9pt},
 AppendixGuide/.style={draw=gray!65,line width=0.55pt}
}


\newcommand{\AppendixQFourSetup}{%
  \coordinate (O) at (0,0);
  \coordinate (bOne) at (1.000000,0);
  \coordinate (bTwo) at (2.200000,0);
  \coordinate (bThree) at (4.200000,0);
  \coordinate (Vapp) at (1.709229,1.030566);
  \coordinate (Uapp) at (1.883888,1.812838);
  \coordinate (qapp) at (0,2.614462);
  \draw[AppendixBoundary] (O)--(5.0,0);
  \draw[AppendixBrane] (O)--(0,3.25);
  \node[red,above left=-1pt] at (0,3.25) {$Q$};
  \foreach \p/\lab in {bOne/b_1,bTwo/b_2,bThree/b_3}{%
    \fill (\p) circle[radius=1.25pt];
    \node[below=2pt] at (\p) {$\lab$};
  }
  \fill (qapp) circle[radius=1.25pt];
  \fill (Vapp) circle[radius=1.25pt];
  \fill (Uapp) circle[radius=1.25pt];
  \node[left=2pt] at (qapp) {$q$};
  \node[above left=1pt] at (1.909229,0.630566) {$V$};
  \node[above right=1pt,left=-5pt] at (1.983888,2.012838) {$U$};
  \node[below=10pt] at (0.50,0) {$A$};
  \node[below=10pt] at (1.60,0) {$B$};
  \node[below=10pt] at (3.20,0) {$C$};
  \node[below=10pt] at (4.60,0) {$D$};
}

\newcommand{\DrawAppendixQFourAuxiliary}{%
\begin{tikzpicture}[x=1.45cm,y=1.45cm,font=\small]
  \AppendixQFourSetup

  \draw[AppendixGeo] (Vapp) arc[start angle=110.928977,end angle=180.000000,radius=1.103361];
  \draw[AppendixGeo] (Vapp) arc[start angle=50.928978,end angle=0.000000,radius=1.327424];
  \draw[AppendixGeo] (Uapp) arc[start angle=163.898926,end angle=170.928977,radius=6.536690];
  \draw[AppendixGeo] (Uapp) arc[start angle=103.898925,end angle=0.000000,radius=1.867517];
  \draw[AppendixGeo] (Uapp) arc[start angle=43.898927,end angle=90.000000,radius=2.614462];
  \draw[AppendixAux] (Uapp) arc[start angle=128.014930,end angle=180.000000,radius=2.300992];
  \draw[AppendixAux] (Uapp) arc[start angle=19.782923,end angle=0.000000,radius=5.356175];
  \node[text=UnifiedBraneWebBlue,left] at (1.5,0.42) {$X$};
  \node[text=UnifiedBraneWebBlue,right] at (1.7,0.41) {$Y$};
  \node[text=UnifiedBraneWebBlue,right] at (1.70,1.4) {$L$};
  \node[text=UnifiedBraneWebBlue,right] at (3.75,1.30) {$Z$};
  \node[text=UnifiedBraneWebBlue,above] at (0.78,2.47) {$W$};
  \node[text=orange!85!black,left,fill=white,inner sep=0.8pt]
     at (1.35,1.30) {$K_{b_1}$};
  \node[text=orange!85!black,right,fill=white,inner sep=0.8pt]
     at (2.14,1.06) {$K_{b_2}$};
  \draw[orange!85!black,line width=0.55pt]
    ($(Uapp)+(218.014930:0.22)$) arc[start angle=218.014930,end angle=253.898926,radius=0.22];
  \node[text=orange!85!black,font=\scriptsize,fill=white,inner sep=0.6pt]
     at (1.64,1.46) {$\lambda$};
\end{tikzpicture}%
}

The static calculations in Sec.~\ref{sec:brane-web} were organized in the coordinates most convenient for direct extremization.  This appendix gives a complementary geometric treatment of the same connected saddles.  Its purpose is twofold.  First, it isolates the hyperbolic-triangle identities that encode the $2\pi/3$ force-balance condition at a trivalent holographic multiway-cut vertex \cite{Gadde:2022cqi,Harper:2024ker,Iizuka:2026qqg}.  Second, it shows explicitly how those identities reproduce the connected tripartite length \eqref{tri-brane-web-length-phase-1} and the nontrivial modulus dependence of the quadripartite Phase~I solution of Sec.~\ref{four-party}.  In the four-party case the alternative parametrization exposes the relation between the main-text modulus $p$ and a normalized brane-endpoint modulus $r$, and shows that the two closure conditions are the same quartic equation.  The detailed hyperbolic derivation is carried out first for the tensionless brane; the finite-tension expression then uses the boundary-entropy shift already established in the main-text extremization.  This appendix should therefore be read as an independent check of the moduli-dependent geometry, not as a second independent derivation of every finite-tension term.  Its use is limited to time-reflection-symmetric components; the Lorentzian thermal section instead treats the connected TFD web directly in Kruskal coordinates.

\subsection{Boundary-anchored hyperbolic triangles}\label{app:hyp-identities}

\begin{figure}[ht]
 \centering
 \begin{tikzpicture}[
   x=1.28cm,y=1.28cm,
   every node/.style={font=\small},
   hyp boundary/.style={draw=black!75,line width=0.7pt},
   hyp edge/.style={draw=UnifiedBraneWebBlue,line width=1.1pt,
     line cap=round,line join=round},
   hyp angle/.style={draw=black!60,line width=0.5pt},
   hyp label/.style={text=UnifiedBraneWebBlue!75!black,
     font=\small\itshape}
 ]
   \begin{scope}
     \fill[black!1.5] (0,0) circle[radius=2.0];
     \draw[hyp boundary] (0,0) circle[radius=2.0];

     \coordinate (P) at (-1.15,0.15);
     \coordinate (Q) at (0,-1.25);
     \coordinate (R) at (1.10,0.65);

     \draw[hyp edge] (P)
       .. controls (-0.55,0.95) and (0.55,1.02) .. (R);
     \draw[hyp edge] (Q)
       .. controls (0.25,-0.45) and (0.65,0.35) .. (R);
     \draw[hyp edge] (P)
       .. controls (-0.70,-0.28) and (-0.22,-0.92) .. (Q);

     \draw[hyp angle] ($(P)+(-43:0.22)$)
       arc[start angle=-43,end angle=53,radius=0.22];
     \draw[hyp angle] ($(Q)+(73:0.22)$)
       arc[start angle=73,end angle=124,radius=0.22];
     \draw[hyp angle] ($(R)+(146:0.22)$)
       arc[start angle=146,end angle=214,radius=0.22];

     \node[hyp label] at (-0.03,1.10) {$A$};
     \node[hyp label] at (0.98,-0.02) {$B$};
     \node[hyp label] at (-0.88,-0.55) {$C$};
     \node[font=\scriptsize] at (-0.83,0.18) {$\beta$};
     \node[font=\scriptsize] at (-0.06,-0.94) {$\alpha$};
     \node[font=\scriptsize] at (0.78,0.65) {$\gamma$};

     \foreach \X in {P,Q,R}
       \fill[black] (\X) circle[radius=1.45pt];
     \node[font=\small] at (0,-2.28) {(a)};
   \end{scope}

   \begin{scope}[xshift=5.80cm]
     \fill[black!1.5] (0,0) circle[radius=2.0];
     \draw[hyp boundary] (0,0) circle[radius=2.0];

     \coordinate (PI) at (-1.64,-1.15);
     \coordinate (QI) at (0.25,-1.15);
     \coordinate (RI) at (1.20,0.65);

     \draw[hyp edge] (PI)
       .. controls (-0.82,0.12) and (0.42,0.95) .. (RI);
     \draw[hyp edge] (QI)
       .. controls (0.50,-0.45) and (0.80,0.25) .. (RI);
     \draw[hyp edge] (PI)
       .. controls (-1.05,-0.55) and (-0.35,-0.55) .. (QI);

     \draw[hyp angle] ($(PI)+(33.66:0.70)$)
       arc[start angle=33.66,end angle=53.67,radius=0.70];
     \draw[hyp angle] ($(QI)+(70:0.22)$)
       arc[start angle=70,end angle=135,radius=0.22];
     \draw[hyp angle] ($(RI)+(159:0.22)$)
       arc[start angle=159,end angle=225,radius=0.22];

     \node[hyp label] at (-0.18,0.85) {$A$};
     \node[hyp label] at (1.12,-0.03) {$B$};
     \node[hyp label] at (-0.45,-0.92) {$C$};
     \node[font=\scriptsize] at (-0.80,-0.58) {$\beta=0$};
     \node[font=\scriptsize] at (0.18,-0.84) {$\alpha$};
     \node[font=\scriptsize] at (0.89,0.53) {$\gamma$};

     \foreach \X in {PI,QI,RI}
       \fill[black] (\X) circle[radius=1.45pt];
     \node[font=\small] at (0,-2.28) {(b)};
   \end{scope}
 \end{tikzpicture}
 \caption{Hyperbolic-triangle notation used throughout Appendix~\ref{app:hyperbolic}.  Panel (a) shows a generic triangle with side lengths $A,B,C$ opposite the angles $\alpha,\beta,\gamma$.  Panel (b) shows the ideal-vertex limit relevant when one or two geodesic legs are anchored on the regulated asymptotic AdS boundary.}
 \label{fig:hyp-trig}
\end{figure}
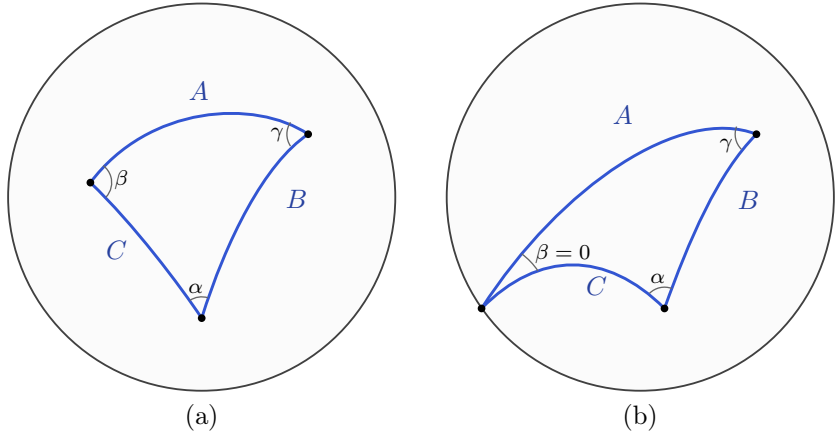

The elementary geometry is summarized in Fig.~\ref{fig:hyp-trig}.  For a hyperbolic triangle with side lengths $A,B,C$ and opposite angles $\alpha,\beta,\gamma$, the two cosine laws may be written as
\begin{align}
 \cosh A&=\cosh B\cosh C-\sinh B\sinh C\cos\alpha,\notag\\
 \cosh C\sin\alpha\sin\beta&=\cos\gamma+\cos\alpha\cos\beta.
\end{align}
The complementary sine law is
\begin{align}
 \frac{\sinh A}{\sin\alpha}
 =\frac{\sinh B}{\sin\beta}
 =\frac{\sinh C}{\sin\gamma}.
\end{align}

The identities become especially efficient when a vertex approaches the asymptotic boundary.  If the vertex opposite $B$ is ideal, $\beta\to0$, and $A,C$ are understood as regulated lengths, the first relation reduces to
\begin{align}
 e^A=e^C\bigl(\cosh B-\sinh B\cos\alpha\bigr)+\mathcal O(\epsilon).
 \label{one-bdy-vertex}
\end{align}
If two vertices are ideal, $\beta=\gamma=0$, the finite part of the regulated length obeys
\begin{align}
 A=B+C+2\log\sin\!\left(\frac{\alpha}{2}\right)+\mathcal O(\epsilon).
 \label{two-bdy-vertices}
\end{align}
For an equal-tension multiway cut, the three incident edges at a trivalent vertex meet at $\alpha=2\pi/3$.  The two boundary-anchored identities then become the simple gluing rules
\begin{align}
 A&=C+\log\!\left(\cosh B+\frac12\sinh B\right),\notag\\
 A&=B+C+2\log\frac{\sqrt3}{2},
 \label{eq:app-steiner-gluing}
\end{align}
up to the common UV-regulator terms already displayed in \eqref{one-bdy-vertex}--\eqref{two-bdy-vertices}.  These are the only hyperbolic identities needed in the two checks below.

\subsection{Connected tripartite web}\label{app:hyp-q3}

Section~\ref{three-party} obtained the connected tripartite saddle by extremizing the three geodesic lengths directly in Poincar\'e coordinates.  The same result follows almost algebraically from \eqref{eq:app-steiner-gluing}.  Referring to Fig.~\ref{fig:phase-A}, let
\begin{equation}
 X=\mathcal L(b_1,V),\qquad
 Y=\mathcal L(b_2,V),\qquad
 W=\mathcal L(V,q),
\end{equation}
where $V$ is the trivalent vertex and $q$ the endpoint on the EOW brane.  We temporarily suppress the common $-\log\epsilon$ contribution of each asymptotic leg and restore it in the final expression.

Applying the one-boundary-vertex relation to the triangles $(b_1,V,q)$ and $(b_2,V,q)$, together with the two-boundary-vertex relation to the triangle $(b_1,V,b_2)$, gives
\begin{align}
 \frac{b_1^2+q^2+2b_1q\sin\theta}{q\cos\theta}
 &=e^X\left(\cosh W+\frac12\sinh W\right),\notag\\
 \frac{b_2^2+q^2+2b_2q\sin\theta}{q\cos\theta}
 &=e^Y\left(\cosh W+\frac12\sinh W\right),\label{eq:app-q3-triangles}\\
 2\log(b_2-b_1)
 &=X+Y+2\log\frac{\sqrt3}{2}.\notag
\end{align}
These three equations encode the complete $2\pi/3$ Steiner condition.  Solving them at fixed brane endpoint $q$ gives
\begin{align}
 W&=\sinh^{-1}\!\left[
 \frac{b_1b_2+q^2+q(b_1+b_2)\sin\theta}
 {q(b_2-b_1)\cos\theta}\right]-\frac12\log3,\notag\\
 X&=\frac12\log\!\left[\frac43(b_2-b_1)^2\right]
 -\frac12\log\!\left(
 \frac{b_2^2+q^2+2b_2q\sin\theta}
 {b_1^2+q^2+2b_1q\sin\theta}\right),\notag\\
 Y&=\frac12\log\!\left[\frac43(b_2-b_1)^2\right]
 +\frac12\log\!\left(
 \frac{b_2^2+q^2+2b_2q\sin\theta}
 {b_1^2+q^2+2b_1q\sin\theta}\right).
 \label{eq:app-q3-XYZ}
\end{align}
The only remaining variational parameter is therefore the position of the brane endpoint.  Extremizing $X+Y+W$ with respect to $q$ fixes
\begin{equation}
 q_\star=\sqrt{b_1b_2},
\end{equation}
precisely as in the coordinate derivation of Sec.~\ref{three-party}.  Restoring the two UV regulators gives
\begin{align}
 \mathcal L_{\rm ext}
 &=\log\!\left[
 \frac{4(b_2-b_1)(\sqrt{b_2}+\sqrt{b_1})^2}
 {3\sqrt3\,\epsilon^2}\right]
 +\log(\sec\theta+\tan\theta).\label{eq:app-q3-final}
\end{align}
Using
$\log(\sec\theta+\tan\theta)=\operatorname{arccosh}(\sec\theta)=\rho_0$, Eq.~\eqref{eq:app-q3-final} is exactly Eq.~\eqref{tri-brane-web-length-phase-1}.  The two derivations therefore have complementary advantages: the main-text extremization gives the bulk vertex coordinates explicitly, whereas the hyperbolic derivation shows that the characteristic square-root dependence and the constant $3\log(2/\sqrt3)$ are direct consequences of the $2\pi/3$ junction geometry.

\subsection{Quadripartite Phase I: one-modulus reduction}\label{app:hyp-q4}

\begin{figure}[ht]
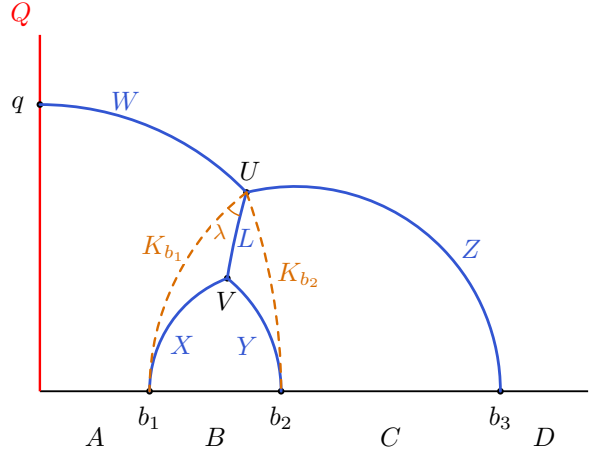

 \centering
 \DrawAppendixQFourAuxiliary
 \caption{Exact Poincar\'e-half-plane realization of the connected quadripartite Phase~I web studied in Sec.~\ref{four-party}, shown for a representative tensionless brane $Q:x=0$.  Each blue segment is a circular geodesic arc centered on $z=0$.  The three legs at each of the two trivalent vertices meet at $2\pi/3$, while the $Uq$ leg intersects $Q$ orthogonally.  The endpoint $q$ is free to move along $Q$, and $L$ denotes the internal geodesic edge. The dashed lines cut out auxiliary triangles used to eliminate the individual external legs in favor of the internal-edge variable $s=e^L$.  The angle $\lambda$ between $Ub_1$ and the internal edge is fixed by the same $2\pi/3$ Steiner condition.}
 \label{fig:auxiliary-triangles}
\end{figure}

The fully connected four-party web of Sec.~\ref{four-party} contains two trivalent vertices and one internal edge.  The main text solves it by extending the internal geodesic to auxiliary asymptotic endpoints $(p,m)$.  Here we derive the same physical saddle in a complementary parametrization that keeps the EOW endpoint explicit.  To expose the geometry cleanly we first set the brane tension to zero, so $Q$ is the vertical geodesic $x=0$, and restore the tilt at the end.  The boundary partition is the same as in \eqref{eq:q4-partition-b},
\begin{equation}
 A=[0,b_1],\qquad B=[b_1,b_2],\qquad
 C=[b_2,b_3],\qquad D=[b_3,\infty).
\end{equation}
The Phase~I topology is (cf.\ Fig.~\ref{fig:auxiliary-triangles})
\begin{equation}
 b_1,b_2\longrightarrow V\longrightarrow U\longrightarrow b_3,q,
\end{equation}
with two Fermat--Torricelli vertices $V=(x_V,z_V)$ and $U=(x_U,z_U)$.  We denote the five physical edge lengths by
\begin{equation}
 X=\mathcal L(b_1,V),\quad
 Y=\mathcal L(b_2,V),\quad
 L=\mathcal L(V,U),\quad
 Z=\mathcal L(U,b_3),\quad
 W=\mathcal L(U,q).
\end{equation}
Their sum is the regulated web action,
\begin{equation}
 \mathcal L_{\rm web}=X+Y+L+Z+W.
 \label{eq:app-q4-total}
\end{equation}

\paragraph{Brane endpoint and the outer Steiner angle.}
For fixed $U$, the length of the final leg is
\begin{equation}
 \mathcal L(U,q)=\operatorname{arccosh}\!\left(
 \frac{x_U^2+z_U^2+q^2}{2z_Uq}\right).
\end{equation}
Extremizing with respect to $q$ gives $q=\sqrt{x_U^2+z_U^2}$, so the geodesic hits $Q$ orthogonally and
\begin{equation}
 W=\operatorname{arccosh}\frac{\sqrt{x_U^2+z_U^2}}{z_U}.
\end{equation}
It is convenient to write
\begin{equation}
 U=(q\tanh W,q\operatorname{sech}W),\qquad
 r\equiv\frac{q}{b_3},\qquad t\equiv e^W.
\end{equation}
The $2\pi/3$ angle between $Uq$ and $Ub_3$ then fixes
\begin{equation}
 t=e^W=\frac{1+r}{\sqrt3(1-r)},
 \qquad
 2-\sqrt3\le r<1,
 \label{W(r)}
\end{equation}
where the lower bound is the condition $W\ge0$ such that $U$ lies on the physical side of the tensionless brane. Thus the brane endpoint already supplies a convenient dimensionless modulus $r$ for the entire outer part of the web.  For a brane of positive tension the web has the same shape, but $U$ may lie in the additional region $-z\tan\theta\le x<0$.  The formulas below then hold with $W$ the signed distance from $U$ to the line $x=0$, the physical brane-ending leg having length $W+\rho_0\ge0$, and the range of the modulus is enlarged to
\begin{equation}
 \frac{\sqrt3\,e^{-\rho_0}-1}{\sqrt3\,e^{-\rho_0}+1}\le r<1,
 \label{eq:app-r-range-tension}
\end{equation}
which is equivalent to the tripartite admissibility condition \eqref{eq:q3-static-admissibility} with $s_U=1/r$ and is no restriction for $\rho_0\ge\log\sqrt3$.

\paragraph{Reducing the internal edge to the same modulus.}
The triangle $(b_1,V,b_2)$ immediately gives
\begin{equation}
 X+Y=2\log\frac{b_2-b_1}{\epsilon}+\log\frac43.
 \label{X+Y}
\end{equation}
Applying the same boundary-anchored identity to the $U$--$b_3$ leg and then using \eqref{W(r)} gives
\begin{equation}
 Z=\log\frac{2b_3(1-r^2)}{\sqrt3\,r\epsilon}.
 \label{Z}
\end{equation}
To determine the internal edge, introduce the two auxiliary geodesics from $b_1$ and $b_2$ to $U$ shown in Fig.~\ref{fig:auxiliary-triangles}.  Their lengths are denoted $K_{b_1}=\mathcal L(U,b_1)$ and $K_{b_2}=\mathcal L(U,b_2)$.

Defining $s=e^L$, the one-boundary-vertex identity gives
\begin{align}
 K_{b_1}&=X+\log\frac{3s+s^{-1}}4,
 \label{Ka}\\
 K_{b_2}&=Y+\log\frac{3s+s^{-1}}4.
 \label{Kb}
\end{align}
The sine law in the triangle $(b_1,V,U)$ fixes the auxiliary angle as
\begin{equation}
 \sin\lambda=
 \frac{4\sin(2\pi/3)}{3s+s^{-1}}.
 \label{sin-lambda}
\end{equation}
The triangles $(b_1,U,b_3)$ and $(b_2,U,b_3)$ then imply
\begin{align}
 2\log\frac{b_3-b_1}{\epsilon}
 &=X+Z+\log\frac{(3s+1)^2}{16s},
 \label{X-Z}\\
 2\log\frac{b_3-b_2}{\epsilon}
 &=Y+Z+\log\frac{(3s-1)^2}{16s}.
 \label{Y-Z}
\end{align}
Adding these relations and using \eqref{X+Y} and \eqref{Z} reduces the internal-edge problem to a quadratic equation,
\begin{equation}
 9(1-r^2)s^2
 -\frac{12(b_3-b_1)(b_3-b_2)}{b_3(b_2-b_1)}rs
 -(1-r^2)=0.
 \label{quadratic-in-s}
\end{equation}
The physical root has $s=e^L>0$ and is therefore
\begin{equation}
 s(r)=
 \frac{2Kr+\sqrt{(1-r^2)^2+4K^2r^2}}
 {3(1-r^2)},
 \qquad
 K\equiv\frac{(b_3-b_1)(b_3-b_2)}{b_3(b_2-b_1)}.
 \label{s-r}
\end{equation}
At this stage every edge length is determined by the single modulus $r$.

\paragraph{Closing the web on the EOW brane.}
The remaining condition comes from the two triangles that contain the brane endpoint $q$.  Using \eqref{Ka}--\eqref{sin-lambda} and $t=e^W$, their finite parts can be written as
\begin{align}
 \log\frac{b_1^2+b_3^2r^2}{b_3r\epsilon}
 &=X+\log\frac{t^2(3s-1)^2+3(s+1)^2}{16st},\notag\\
 \log\frac{b_2^2+b_3^2r^2}{b_3r\epsilon}
 &=Y+\log\frac{t^2(3s+1)^2+3(s-1)^2}{16st}.
 \label{brane-eqns}
\end{align}
Here the angle at $U$ between the auxiliary geodesic and the brane-ending leg is $2\pi/3-\lambda$ for $b_1$ and $2\pi/3+\lambda$ for $b_2$, the opposite assignment to that of the triangles ending on $b_3$ in Eqs.~\eqref{X-Z} and \eqref{Y-Z}.
Eliminating $X,Y,Z,s$ with the preceding relations leaves a single quartic equation for the physical brane modulus,
\begin{equation}
 \mathcal Q_r(r)
 =r^4+\left(\frac{b_1+b_2}{b_3}-2\right)r^3
 -\frac{b_1+b_2}{b_3}r^2
 +\left(\frac{b_1+b_2}{b_3}-\frac{2b_1b_2}{b_3^2}\right)r
 +\frac{b_1b_2}{b_3^2}=0.
 \label{eq:app-r-quartic}
\end{equation}
The physical root $r=r_\star$ lies in the interval $\sqrt{b_1/b_3}<r<\sqrt{b_2/b_3}$, where it is unique in all configurations we have examined.  The web exists when $s(r_\star)>1$ and when $r_\star$ lies in the range \eqref{W(r)} or \eqref{eq:app-r-range-tension}.

\paragraph{Relation to the main-text modulus.}
The connection with Sec.~\ref{four-party} is direct.  In the main-text decomposition, the vertex $U$ belongs to a tripartite sub-web with asymptotic endpoints $(p,b_3)$.  The tripartite result \eqref{tri vertex coordinate} therefore places its EOW endpoint at $q=\sqrt{pb_3}$. Hence
\begin{equation}
 p=\frac{q^2}{b_3}=b_3r^2.
 \label{eq:p-r-map}
\end{equation}
Substituting \eqref{eq:p-r-map}, i.e.\ $\mathsf p=\mathsf b\,r$, into the Steiner polynomial of Eq.~\eqref{Q p solution} gives the identity
\begin{equation}
 \mathcal P(\mathsf b\,r)
 =-b_3^{5/2}\,\mathcal Q_r(r).
 \label{eq:Qp-factorization}
\end{equation}
The hyperbolic closure conditions and the circle condition of Sec.~\ref{four-party} are therefore the same quartic equation, written for the two moduli $r$ and $p$.  Likewise, at the root the internal-edge variable \eqref{s-r} coincides with Eq.~\eqref{eq:q4-phase-I-internal-edge}.  On the symmetric locus $b_3=2(b_1+b_2)$ the root is $r_\star=1/2$.

\paragraph{On-shell length and comparison with Sec.~\ref{four-party}.}
Combining \eqref{X+Y}, \eqref{Z}, \eqref{s-r} and \eqref{W(r)} gives the tensionless on-shell length
\begin{equation}
 \mathcal L_{\rm web}
 =\log\!\left[
 \frac{8b_3(b_2-b_1)^2}{9\epsilon^3}
 \frac{s(r_\star)(1+r_\star)^2}{r_\star}\right].
 \label{eq:app-q4-length-zeroT}
\end{equation}
For a tilted brane the endpoint contribution simply restores the same additive boundary-entropy term that appears in the main-text result,
\begin{equation}
 \mathcal L_{\rm web}
 =\log\!\left[
 \frac{8b_3(b_2-b_1)^2}{9\epsilon^3}
 \frac{s(r_\star)(1+r_\star)^2}{r_\star}\right]
 +\operatorname{arccosh}(\sec\theta).
 \label{bulk-answer}
\end{equation}
Using $p=b_3r_\star^2$ together with $\mathcal Q_r(r_\star)=0$ transforms \eqref{bulk-answer} into the main-text expression \eqref{total length quad case 1}.  The equivalence is particularly transparent on the symmetric locus $b_3=2(b_1+b_2)$, where
\begin{equation}
 r_\star=\frac12,
 \qquad
 p_\star=\frac{b_1+b_2}{2},
 \qquad
 s_\star=\frac{b_1+b_2}{b_2-b_1}.
\end{equation}
Substitution into \eqref{bulk-answer} gives
\begin{equation}
 \mathcal L_{\rm web}
 =\log\!\left[
 \frac{8(b_2-b_1)(b_1+b_2)^2}{\epsilon^3}\right]
 +\operatorname{arccosh}(\sec\theta),
\end{equation}
which is exactly Eq.~\eqref{multien quad case 1 special}.  The hyperbolic derivation thus identifies the geometric origin of the auxiliary modulus and shows that the two parametrizations lead to the same closure condition.


\section{Counting admissible brane web topologies: Catalan structure}\label{app:web-counting}
This appendix provides the topological counting used in Sec.~\ref{par:extended-homology-review}.  The statement is restricted to the \emph{reduced static planar disk sector}: closed web components and degree-two vertices are removed, every bulk chamber must meet a nonempty asymptotic boundary party, and winding, intrinsically Lorentzian, and replica-symmetry-breaking configurations are outside the count.  Within this sector the object being counted is a planar brane web $\mathcal W$ whose fixed asymptotic endpoints are the entangling points between adjacent parties and whose additional endpoints may lie on the EOW brane $Q$ \cite{Gadde:2022cqi,Gadde:2023zzj}.  The result proved below is a connectivity statement, not a dynamical one: the number of reduced connectivity classes is $C_{q-1}$.  A class contributes to the multi-entropy only when its geodesic realization exists, obeys the equal-tension junction conditions, passes the geometric admissibility tests, and is competitive in the on-shell length.

\subsection{Relative homology and decomposition of the brane web}
Consider first a general planar configuration on a disk-like static slice.  We introduce four quantities that will be used only in this appendix:
\begin{itemize}
 \item $N_{\partial}$: the number of fixed web endpoints on the asymptotic boundary;
 \item $N_Q$: the number of web endpoints on the EOW brane $Q$;
 \item $N_{\mathcal W}$: the number of connected components of the brane web $\mathcal W$;
 \item $N_{\mathcal R}$: the number of connected bulk regions obtained after cutting the slice along $\mathcal W$.
\end{itemize}
The total number of endpoints of the web is therefore
\begin{equation}\label{eq:web-endpoint-count}
 N_{\rm end}=N_{\partial}+N_Q.
\end{equation}
After removing closed components and degree-two points, every connected component of a minimal web is a tree.  A tree with $L$ endpoints divides a disk into $L$ regions, or equivalently adds $L-1$ new regions.  Summing this statement over all $N_{\mathcal W}$ connected components gives the planar Euler relation
\begin{equation}\label{eq:web-region-count}
 N_{\mathcal R}=1+N_{\rm end}-N_{\mathcal W}.
\end{equation}
Every nontrivial connected component has at least two endpoints, hence
\begin{equation}\label{eq:web-component-bound}
 N_{\rm end}\ge 2N_{\mathcal W}.
\end{equation}
Finally, in a reduced multi-entropy configuration every connected bulk region must meet the asymptotic boundary.  If the asymptotic boundary is divided into $N_{\rm seg}$ nonempty segments by the fixed web endpoints, then
\begin{equation}\label{eq:web-region-bound}
 N_{\mathcal R}\le N_{\rm seg}.
\end{equation}
These relations are simply bookkeeping identities for the planar brane web; no assumption about the detailed geodesic shape has entered.

For the adjacent $q$-party problem of the main text,
\begin{equation}
 A_1=[0,b_1],\quad A_2=[b_1,b_2],\quad\ldots,\quad
 A_q=[b_{q-1},\infty),
\end{equation}
there are
\begin{equation}
 N_{\partial}=q-1,\qquad N_{\rm seg}=q.
\end{equation}
Because the $q$ parties are distinct, the relative-homology prescription requires one bulk region adjacent to each $A_a$; therefore $N_{\mathcal R}=q$.  Substitution in Eq.~\eqref{eq:web-region-count} gives
\begin{equation}\label{eq:web-components-equal-NQ}
 N_{\mathcal W}=N_Q.
\end{equation}
Combining this with Eq.~\eqref{eq:web-component-bound} yields $N_Q\le q-1$.  For $q>1$ the web has at least one fixed asymptotic endpoint, hence at least one nontrivial connected component.  Equation~\eqref{eq:web-components-equal-NQ} then implies $N_Q\ge1$.  We recover
\begin{equation}
 1\le N_Q\le q-1.
\end{equation}

Equation~\eqref{eq:web-components-equal-NQ} has a useful geometric consequence.  Each connected component of the reduced web carries exactly one endpoint on $Q$.  To see this, suppose a component had two or more brane endpoints.  The path in that component joining two consecutive brane endpoints, together with the intervening segment of $Q$, would enclose a bulk region with no asymptotic boundary party.  This is excluded by the reduced relative-homology condition.  Since the total numbers of connected components and brane endpoints are equal, no component can then have zero brane endpoints either.

Planarity also fixes how the asymptotic endpoints are distributed among the connected components.  Along $Q$, orienting from the end adjacent to $A_1$ toward the end adjacent to $A_q$, the nonempty relative-homology portions must occur in the same order as the parties on the asymptotic boundary.  We may therefore write
\begin{equation}\label{eq:ordered-Q-assignment}
 \mathcal Q_{A_{i_0}}\,|\,\mathcal Q_{A_{i_1}}\,|\cdots|\,
 \mathcal Q_{A_{i_{N_Q}}},
 \qquad
 1=i_0<i_1<\cdots<i_{N_Q}=q,
\end{equation}
where each vertical bar denotes one endpoint of $\mathcal W$ on $Q$.  The connected component ending at the $s$th brane endpoint is anchored on the consecutive set
\begin{equation}\label{eq:component-anchor-block}
 b_{i_{s-1}},\,b_{i_{s-1}+1},\ldots,b_{i_s-1}.
\end{equation}
If instead two endpoints of one component enclosed an endpoint belonging to another component in the asymptotic ordering, their paths to distinct points of $Q$ would interlace and the planar webs would have to cross.  Thus Eq.~\eqref{eq:component-anchor-block} is forced by planarity.

It is convenient to denote the number of asymptotic endpoints in the $s$th connected component by
\begin{equation}\label{eq:ks-definition}
 k_s\equiv i_s-i_{s-1}\ge1.
\end{equation}
The ordered brane assignment is then equivalent to an ordered decomposition
\begin{equation}\label{eq:ordered-composition}
 k_1+k_2+\cdots+k_{N_Q}=q-1.
\end{equation}
This is the only combinatorial information needed below.  Notice that $k_s$ has a direct geometric meaning: the $s$th connected brane web component has $k_s$ fixed endpoints on the asymptotic boundary and one endpoint on $Q$.

For completeness, the numbers of internal junctions and geodesic segments in such a component follow immediately.  Let $V_s$ and $E_s$ denote the numbers of trivalent bulk junctions and web segments, respectively.  The component has $k_s+1$ external endpoints.  Since it is a tree,
\begin{equation}
 E_s=V_s+k_s,
\end{equation}
and summing the valencies gives
\begin{equation}
 3V_s+(k_s+1)=2E_s.
\end{equation}
Hence
\begin{equation}\label{eq:component-VE}
 V_s=k_s-1,\qquad E_s=2k_s-1.
\end{equation}
In particular, $k_s=1$ is a single boundary-to-brane geodesic, $k_s=2$ is a $Y$ web with one trivalent junction, and $k_s=3$ contains two trivalent junctions.

\subsection{Ordinary AdS/CFT and the connected Catalan problem}\label{app:ads-cft-catalan}
Before counting the additional EOW-brane sectors, it is useful to isolate the corresponding topology problem in ordinary AdS$_3$/CFT$_2$.  Consider a pure state on one connected asymptotic boundary, partitioned into $q$ nonempty connected regions
\begin{equation}
 R_1|R_2|\cdots|R_q
\end{equation}
with the ordering understood cyclically.  There are now $q$ asymptotic interfaces between neighboring regions.  In the static planar sector the holographic multi-entropy is represented by a minimal $q$-way cut built from equal-tension geodesic segments.  The multi-entropy literature emphasizes that, when the $R_a$ are connected and lie on a single boundary, the minimal cut is a connected network with equiangular trivalent junctions and contains exactly $q-2$ such junctions \cite{Gadde:2022cqi,Gadde:2023zzj,Iizuka:2025caq,Iizuka:2025bcc,Iizuka:2026qqg}.

The number of possible planar connectivities follows directly from the same local junction structure.  Let $M_q$ denote the number of connected planar trivalent webs whose $q$ external legs end at the $q$ asymptotic interfaces in their fixed cyclic order.  Choose one external leg as a reference leg.  The trivalent junction adjacent to it separates the remaining $q-1$ external legs into two nonempty consecutive groups containing $j$ and $q-1-j$ legs.  Each group is itself joined to that junction by a planar rooted trivalent subtree.  A rooted trivalent subtree with $m$ ordered external legs is counted by $C_{m-1}$.  Therefore
\begin{equation}\label{eq:ads-cft-catalan-recursion}
 M_q=\sum_{j=1}^{q-2}C_{j-1}C_{q-j-2}=C_{q-2},
\end{equation}
which is the standard Catalan recursion.  Equivalently, one may represent the same connectivity classes by triangulations of a convex $q$-gon: the dual graph of a triangulation is precisely a planar trivalent web with the prescribed cyclic ordering.  Thus
\begin{equation}\label{eq:ads-cft-catalan-count}
 \mathcal N_q^{\rm AdS/CFT}=C_{q-2}
 \qquad (q\ge2).
\end{equation}
For $q=3,4,5,6$ this gives $1,2,5,14$ candidate planar topologies, respectively.  In particular, the two ordinary four-party channels are the $C_2=2$ alternatives familiar from the four-party multiway-cut problem, while five connected boundary regions admit $C_3=5$ planar junction patterns.

Equation~\eqref{eq:ads-cft-catalan-count} is not an additional dynamical assumption.  It is a combinatorial consequence of the planar connected multiway-cut geometry already used in the holographic multi-entropy prescription.  The existing literature usually focuses on the minimal cut itself, its $q-2$ trivalent vertices, and the comparison of the competing channels rather than on packaging the channel count as a Catalan sequence; we make the counting explicit here because it provides the cleanest reference point for the AdS/BCFT extension.

The relation to the BCFT problem is immediate.  A fully connected AdS/BCFT web has $q-1$ fixed asymptotic endpoints and one endpoint on $Q$, hence again $q$ external legs and $q-2$ trivalent junctions.  Its possible planar connectivities are therefore also counted by $C_{q-2}$.  What is new in AdS/BCFT is not the local junction combinatorics but the possibility, allowed by relative homology, of having several distinct endpoints on $Q$.  These extra sectors split the web into several connected brane-ending pieces and are responsible for the enlargement from $C_{q-2}$ to $C_{q-1}$ derived below.

\subsection{Catalan counting of connected brane webs in AdS/BCFT}\label{app:catalan-organization}
We now return to AdS/BCFT.  Let $T_k$ be the number of planar connected brane web topologies with $k$ consecutive asymptotic endpoints and one endpoint on $Q$.  Such a component has $k+1$ external legs, so Eq.~\eqref{eq:ads-cft-catalan-count} already suggests $T_k=C_{k-1}$.  The following direct recursion is useful because the same $T_k$ will enter the sectors with several brane endpoints.  The case $k=1$ is unique,
\begin{equation}
 T_1=1.
\end{equation}
For $k\ge2$, consider the trivalent junction adjacent to the brane-ending branch.  Removing that junction separates the ordered asymptotic endpoints into two nonempty consecutive sets containing, say, $j$ and $k-j$ endpoints.  Conversely, joining any admissible topology on the first set to any admissible topology on the second set produces one admissible topology on all $k$ endpoints.  Therefore
\begin{equation}\label{eq:Tk-recursion}
 T_k=\sum_{j=1}^{k-1}T_jT_{k-j},\qquad T_1=1.
\end{equation}
This is exactly the Catalan recursion.  Consequently
\begin{equation}\label{eq:catalan-number}
 T_k=C_{k-1}
 =\frac{1}{k}\binom{2k-2}{k-1}.
\end{equation}
The first few cases have an immediate brane web interpretation:
\begin{equation}
 T_1=1,\qquad T_2=1,\qquad T_3=2,\qquad T_4=5.
\end{equation}
Thus one asymptotic endpoint gives a single geodesic, two endpoints give a unique $Y$ web, three endpoints give the two connected quadripartite phases I and II, and four endpoints give five connected five-party webs.

We can now count configurations with a fixed number $N_Q=r$ of EOW-brane endpoints.  By Eq.~\eqref{eq:ordered-composition}, choose an ordered decomposition
\begin{equation}
 k_1+\cdots+k_r=q-1,
 \qquad k_s\ge1.
\end{equation}
For this fixed decomposition the connected components can be chosen independently, giving
\begin{equation}\label{eq:fixed-composition-count}
 \prod_{s=1}^{r}C_{k_s-1}
\end{equation}
brane web topologies.  Summing over all ordered decompositions gives
\begin{equation}\label{eq:fixed-NQ-count-sum}
 \mathcal N_{q,r}
 =\sum_{\substack{k_1+\cdots+k_r=q-1\\ k_s\ge1}}
   \prod_{s=1}^{r}C_{k_s-1}.
\end{equation}
To evaluate this sum, introduce the Catalan generating function
\begin{equation}
 C(x)=\sum_{n\ge0}C_nx^n,
 \qquad C(x)=1+xC(x)^2,
\end{equation}
and the generating function for one connected brane web component,
\begin{equation}\label{eq:B-generating}
 B(x)=\sum_{k\ge1}T_kx^k=xC(x).
\end{equation}
The coefficient of $x^{q-1}$ in $B(x)^r$ implements precisely the ordered sum in Eq.~\eqref{eq:fixed-NQ-count-sum}, so with $n=q-1$,
\begin{align}
 \mathcal N_{q,r}
 &=[x^n]B(x)^r
  =[x^{n-r}]C(x)^r\notag\\
 &=\frac{r}{2n-r}\binom{2n-r}{n-r},
 \qquad n=q-1.
 \label{eq:catalan-triangle}
\end{align}
In the last step we used the standard coefficient obtained from Lagrange inversion~\cite{Gessel:2016Lagrange,Surya:2023Lagrange},
\begin{equation}
 [x^m]C(x)^r
 =\frac{r}{2m+r}\binom{2m+r}{m},
 \qquad m\ge0.
\end{equation}
The numbers $\mathcal N_{q,r}$ are a Catalan-triangle refinement: they resolve the total number of planar brane web topologies according to the number of EOW-brane endpoints.

Finally, summing over all allowed values of $r$ gives the total number of phases.  Since a contribution to the coefficient of $x^n$ can involve at most $n$ factors of $B(x)$,
\begin{equation}
 \sum_{r\ge1}B(x)^r=\frac{B(x)}{1-B(x)}.
\end{equation}
Using $B(x)=xC(x)$ together with $C(x)=1+xC(x)^2$ gives
\begin{equation}
 \frac{B(x)}{1-B(x)}=C(x)-1.
\end{equation}
Therefore, within the reduced sector just stated,
\begin{equation}\label{eq:total-catalan-phases}
 \mathcal N_q^{\rm AdS/BCFT}=C_{q-1}.
\end{equation}
Combining Eqs.~\eqref{eq:ads-cft-catalan-count} and~\eqref{eq:total-catalan-phases} with $\mathcal N_{q,1}=C_{q-2}$ reproduces Eq.~\eqref{eq:main-catalan-comparison} and isolates the effect of the EOW brane: the fully connected BCFT sector has exactly the ordinary AdS/CFT topology count, and the shift by one Catalan index appears only after summing over the additional relative-homology sectors with $N_Q>1$.

For reference,
\begin{center}
\begin{tabular}{c|c|c|c}
$q$ & ordinary AdS/CFT & $\{\mathcal N_{q,r}\}_{r=1}^{q-1}$ in AdS/BCFT & total AdS/BCFT\\ \hline
2 & $C_0=1$ & $\{1\}$ & $C_1=1$\\
3 & $C_1=1$ & $\{1,1\}$ & $C_2=2$\\
4 & $C_2=2$ & $\{2,2,1\}$ & $C_3=5$\\
5 & $C_3=5$ & $\{5,5,3,1\}$ & $C_4=14$\\
6 & $C_4=14$ & $\{14,14,9,4,1\}$ & $C_5=42$
\end{tabular}
\end{center}

\subsection{A $q=5$ example}\label{app:five-party-example}
Consider the five-party partition
\begin{equation}\label{eq:q5-partition}
 A=[0,b_1],\qquad B=[b_1,b_2],\qquad C=[b_2,b_3],\qquad
 D=[b_3,b_4],\qquad E=[b_4,\infty).
\end{equation}
In ordinary AdS/CFT, five adjacent connected regions on a single boundary would admit only the five connected planar topologies $C_3=5$.  In the present AdS/BCFT problem there are four fixed asymptotic endpoints $b_1,\ldots,b_4$, and the additional sectors with several endpoints on $Q$ enlarge the total to
\begin{equation}
 \mathcal N_5^{\rm AdS/BCFT}=C_4=14.
\end{equation}
The first five of these are precisely the fully connected $N_Q=1$ configurations and therefore have the same connectivity patterns as the five ordinary AdS/CFT five-party webs.  The remaining nine are genuinely new relative-homology sectors made possible by the EOW brane.  The assignment on $Q$ makes this decomposition particularly transparent.  The complete list is
\begin{center}
\small
\begin{tabular}{c|c|c|c}
phases & $N_Q$ & assignment along $Q$ & asymptotic endpoints in each connected component\\ \hline
I--V & 1 & $\mathcal Q_A|\mathcal Q_E$
 & $\{b_1,b_2,b_3,b_4\}$\\ \hline
VI--VII & 2 & $\mathcal Q_A|\mathcal Q_D|\mathcal Q_E$
 & $\{b_1,b_2,b_3\}\;|\;\{b_4\}$\\
VIII & 2 & $\mathcal Q_A|\mathcal Q_C|\mathcal Q_E$
 & $\{b_1,b_2\}\;|\;\{b_3,b_4\}$\\
IX--X & 2 & $\mathcal Q_A|\mathcal Q_B|\mathcal Q_E$
 & $\{b_1\}\;|\;\{b_2,b_3,b_4\}$\\ \hline
XI & 3 & $\mathcal Q_A|\mathcal Q_C|\mathcal Q_D|\mathcal Q_E$
 & $\{b_1,b_2\}\;|\;\{b_3\}\;|\;\{b_4\}$\\
XII & 3 & $\mathcal Q_A|\mathcal Q_B|\mathcal Q_D|\mathcal Q_E$
 & $\{b_1\}\;|\;\{b_2,b_3\}\;|\;\{b_4\}$\\
XIII & 3 & $\mathcal Q_A|\mathcal Q_B|\mathcal Q_C|\mathcal Q_E$
 & $\{b_1\}\;|\;\{b_2\}\;|\;\{b_3,b_4\}$\\ \hline
XIV & 4 & $\mathcal Q_A|\mathcal Q_B|\mathcal Q_C|\mathcal Q_D|\mathcal Q_E$
 & $\{b_1\}\;|\;\{b_2\}\;|\;\{b_3\}\;|\;\{b_4\}$
\end{tabular}
\end{center}
The first row contains five phases because a connected component with four asymptotic endpoints has $T_4=C_3=5$ planar trivalent junction patterns.  Written only as a mnemonic for the order in which consecutive branches meet, these are
\begin{equation}\label{eq:q5-five-connected-associations}
 \begin{split}
 &(((b_1b_2)b_3)b_4),\qquad ((b_1(b_2b_3))b_4),\qquad
 ((b_1b_2)(b_3b_4)),\\
 &(b_1((b_2b_3)b_4)),\qquad (b_1(b_2(b_3b_4))).
 \end{split}
\end{equation}
The parentheses in Eq.~\eqref{eq:q5-five-connected-associations} do not introduce new geometric variables; they only record the planar connectivity of the trivalent junctions.  For $N_Q=2$, the ordered decompositions of the four asymptotic endpoints are $3+1$, $2+2$ and $1+3$.  Their multiplicities are
\begin{equation}
 C_2C_0+C_1C_1+C_0C_2=2+1+2=5.
\end{equation}
For $N_Q=3$, the decompositions $2+1+1$, $1+2+1$ and $1+1+2$ each contribute one topology, giving three phases.  For $N_Q=4$ every component is a single boundary-to-brane geodesic, giving the unique fully factorized phase XIV.  Thus
\begin{equation}
 5+5+3+1=14
\end{equation}
reproduces the Catalan result while keeping the relative-homology meaning of each class manifest.

The corresponding vacuum AdS/BCFT connectivity classes are shown in Figs.~\ref{fig:five-party-r1}--\ref{fig:five-party-r34}.  Every panel is drawn as an exact representative saddle for the tilted brane $Q:x=-z/\sqrt{3}$, corresponding to $\rho_0=\log\sqrt{3}$, with ordered boundary anchors $b_1<b_2<b_3<b_4$.  Every blue segment is a circular Poincar\'e geodesic whose Euclidean center lies on the asymptotic boundary $z=0$; the coordinates of the bulk junctions and brane endpoints are obtained by extremizing the corresponding equal-tension network. Consequently, the three segments incident on each trivalent vertex meet pairwise at $2\pi/3$, and every segment ending on $Q$ intersects the brane orthogonally. These representatives are used only to display the topology: changing the endpoint data or turning on the brane tension deforms the arcs continuously without changing the connectivity class.

\tikzset{
  FivePartyWeb/.style={draw=UnifiedBraneWebBlue,line width=0.95pt,line cap=round,line join=round},
  FivePartyBrane/.style={draw=red,line width=0.85pt},
  FivePartyBoundary/.style={draw=black,line width=0.65pt}
}
\newcommand{\FivePartySetup}{%
  \coordinate (O) at (0,0);
  \coordinate (b1) at (1.000000,0);
  \coordinate (b2) at (2.000000,0);
  \coordinate (b3) at (3.200000,0);
  \coordinate (b4) at (4.700000,0);
  \draw[FivePartyBoundary] (O)--(5.60,0);
  \draw[FivePartyBrane] (O)--(120:5.05);
  \node[red,above left=-1pt,font=\scriptsize] at (120:5.05) {$Q$};
  \fill (O) circle[radius=1.15pt];
  \node[below left=1pt,font=\scriptsize] at (O) {$O$};
  \foreach \p/\lab in {b1/b_1,b2/b_2,b3/b_3,b4/b_4}{%
    \fill (\p) circle[radius=1.15pt];
    \node[below=1.4pt,font=\scriptsize] at (\p) {$\lab$};
  }
  \node[below=8pt,font=\scriptsize] at (0.50,0) {$A$};
  \node[below=8pt,font=\scriptsize] at (1.50,0) {$B$};
  \node[below=8pt,font=\scriptsize] at (2.60,0) {$C$};
  \node[below=8pt,font=\scriptsize] at (3.95,0) {$D$};
  \node[below=8pt,font=\scriptsize] at (5.25,0) {$E$};
}
\newcommand{\FiveMarkVertex}[1]{\fill (#1) circle[radius=1.25pt];}
\newcommand{\FiveMarkQ}[2]{\fill (#1) circle[radius=1.25pt];\node[left=1pt,font=\scriptsize] at (#1) {$#2$};}

\newcommand{\DrawFivePartyI}{%
\begin{tikzpicture}[x=0.52cm,y=0.52cm]
\FivePartySetup
\coordinate (c1q) at (120:3.417784);
\coordinate (c1v0) at (2.941796,1.739853);
\coordinate (c1v1) at (2.742971,1.326522);
\coordinate (c1v2) at (1.828912,0.763175);
\draw[FivePartyWeb] (c1v0) arc[start angle=150.601153,end angle=158.020923,radius=3.544310];
\draw[FivePartyWeb] (c1v1) arc[start angle=98.020924,end angle=145.271237,radius=1.339627];
\draw[FivePartyWeb] (c1v2) arc[start angle=85.271234,end angle=180.000000,radius=0.765782];
\draw[FivePartyWeb] (c1v2) arc[start angle=25.271236,end angle=0.000000,radius=1.787698];
\draw[FivePartyWeb] (c1v1) arc[start angle=38.020923,end angle=0.000000,radius=2.153622];
\draw[FivePartyWeb] (c1v0) arc[start angle=90.601154,end angle=0.000000,radius=1.739949];
\draw[FivePartyWeb] (c1v0) arc[start angle=30.601153,end angle=120.000000,radius=3.417784];
\FiveMarkVertex{c1v0}
\FiveMarkVertex{c1v1}
\FiveMarkVertex{c1v2}
\FiveMarkQ{c1q}{q}
\end{tikzpicture}}

\newcommand{\DrawFivePartyII}{%
\begin{tikzpicture}[x=0.52cm,y=0.52cm]
\FivePartySetup
\coordinate (c1q) at (120:2.895289);
\coordinate (c1v0) at (2.056571,2.037944);
\coordinate (c1v1) at (1.890570,1.282862);
\coordinate (c1v2) at (2.215639,0.922836);
\draw[FivePartyWeb] (c1v0) arc[start angle=164.739356,end angle=170.462803,radius=7.742642];
\draw[FivePartyWeb] (c1v1) arc[start angle=110.462799,end angle=180.000000,radius=1.369263];
\draw[FivePartyWeb] (c1v1) arc[start angle=50.462801,end angle=33.695331,radius=1.663437];
\draw[FivePartyWeb] (c1v2) arc[start angle=153.695333,end angle=180.000000,radius=2.082474];
\draw[FivePartyWeb] (c1v2) arc[start angle=93.695333,end angle=0.000000,radius=0.924759];
\draw[FivePartyWeb] (c1v0) arc[start angle=104.739352,end angle=0.000000,radius=2.107288];
\draw[FivePartyWeb] (c1v0) arc[start angle=44.739356,end angle=120.000000,radius=2.895289];
\FiveMarkVertex{c1v0}
\FiveMarkVertex{c1v1}
\FiveMarkVertex{c1v2}
\FiveMarkQ{c1q}{q}
\end{tikzpicture}}

\newcommand{\DrawFivePartyIII}{%
\begin{tikzpicture}[x=0.52cm,y=0.52cm]
\FivePartySetup
\coordinate (c1q) at (120:2.402221);
\coordinate (c1v0) at (1.774120,1.619618);
\coordinate (c1v1) at (1.592575,0.858555);
\coordinate (c1v2) at (3.382306,1.087018);
\draw[FivePartyWeb] (c1v0) arc[start angle=162.393369,end angle=170.773139,radius=5.354459];
\draw[FivePartyWeb] (c1v1) arc[start angle=110.773138,end angle=180.000000,radius=0.918249];
\draw[FivePartyWeb] (c1v1) arc[start angle=50.773139,end angle=0.000000,radius=1.108316];
\draw[FivePartyWeb] (c1v0) arc[start angle=102.393368,end angle=40.958874,radius=1.658260];
\draw[FivePartyWeb] (c1v2) arc[start angle=160.958873,end angle=180.000000,radius=3.331889];
\draw[FivePartyWeb] (c1v2) arc[start angle=100.958874,end angle=0.000000,radius=1.107209];
\draw[FivePartyWeb] (c1v0) arc[start angle=42.393369,end angle=120.000000,radius=2.402221];
\FiveMarkVertex{c1v0}
\FiveMarkVertex{c1v1}
\FiveMarkVertex{c1v2}
\FiveMarkQ{c1q}{q}
\end{tikzpicture}}

\newcommand{\DrawFivePartyIV}{%
\begin{tikzpicture}[x=0.52cm,y=0.52cm]
\FivePartySetup
\coordinate (c1q) at (120:1.830274);
\coordinate (c1v0) at (1.079147,1.478291);
\coordinate (c1v1) at (2.682094,1.308873);
\coordinate (c1v2) at (2.656773,1.036900);
\draw[FivePartyWeb] (c1v0) arc[start angle=173.870694,end angle=180.000000,radius=13.845218];
\draw[FivePartyWeb] (c1v0) arc[start angle=113.870693,end angle=54.062765,radius=1.616570];
\draw[FivePartyWeb] (c1v1) arc[start angle=174.062762,end angle=175.299612,radius=12.653575];
\draw[FivePartyWeb] (c1v2) arc[start angle=115.299615,end angle=180.000000,radius=1.146905];
\draw[FivePartyWeb] (c1v2) arc[start angle=55.299612,end angle=0.000000,radius=1.261221];
\draw[FivePartyWeb] (c1v1) arc[start angle=114.062762,end angle=0.000000,radius=1.433440];
\draw[FivePartyWeb] (c1v0) arc[start angle=53.870692,end angle=120.000000,radius=1.830274];
\FiveMarkVertex{c1v0}
\FiveMarkVertex{c1v1}
\FiveMarkVertex{c1v2}
\FiveMarkQ{c1q}{q}
\end{tikzpicture}}

\newcommand{\DrawFivePartyV}{%
\begin{tikzpicture}[x=0.52cm,y=0.52cm]
\FivePartySetup
\coordinate (c1q) at (120:1.604389);
\coordinate (c1v0) at (1.158076,1.110371);
\coordinate (c1v1) at (1.997606,0.992917);
\coordinate (c1v2) at (3.244992,0.935968);
\draw[FivePartyWeb] (c1v0) arc[start angle=163.795241,end angle=180.000000,radius=3.978812];
\draw[FivePartyWeb] (c1v0) arc[start angle=103.795241,end angle=60.276283,radius=1.143352];
\draw[FivePartyWeb] (c1v1) arc[start angle=0.276284,end angle=0.000000,radius=205.911946];
\draw[FivePartyWeb] (c1v1) arc[start angle=120.276283,end angle=54.495769,radius=1.149736];
\draw[FivePartyWeb] (c1v2) arc[start angle=174.495768,end angle=180.000000,radius=9.757876];
\draw[FivePartyWeb] (c1v2) arc[start angle=114.495770,end angle=0.000000,radius=1.028546];
\draw[FivePartyWeb] (c1v0) arc[start angle=43.795243,end angle=120.000000,radius=1.604389];
\FiveMarkVertex{c1v0}
\FiveMarkVertex{c1v1}
\FiveMarkVertex{c1v2}
\FiveMarkQ{c1q}{q}
\end{tikzpicture}}

\newcommand{\DrawFivePartyVI}{%
\begin{tikzpicture}[x=0.52cm,y=0.52cm]
\FivePartySetup
\coordinate (c1q) at (120:2.221260);
\coordinate (c1v0) at (1.825557,1.265439);
\coordinate (c1v1) at (1.664043,0.842230);
\coordinate (c2q) at (120:4.700000);
\draw[FivePartyWeb] (c1v0) arc[start angle=154.728923,end angle=163.493147,radius=2.964243];
\draw[FivePartyWeb] (c1v1) arc[start angle=103.493146,end angle=180.000000,radius=0.866138];
\draw[FivePartyWeb] (c1v1) arc[start angle=43.493147,end angle=0.000000,radius=1.223696];
\draw[FivePartyWeb] (c1v0) arc[start angle=94.728922,end angle=0.000000,radius=1.269762];
\draw[FivePartyWeb] (c1v0) arc[start angle=34.728922,end angle=120.000000,radius=2.221260];
\draw[FivePartyWeb] (b4) arc[start angle=0,end angle=120.000000,radius=4.700000];
\FiveMarkVertex{c1v0}
\FiveMarkVertex{c1v1}
\FiveMarkQ{c1q}{q_1}
\FiveMarkQ{c2q}{q_2}
\end{tikzpicture}}

\newcommand{\DrawFivePartyVII}{%
\begin{tikzpicture}[x=0.52cm,y=0.52cm]
\FivePartySetup
\coordinate (c1q) at (120:1.562690);
\coordinate (c1v0) at (1.167775,1.038413);
\coordinate (c1v1) at (2.076833,0.800609);
\coordinate (c2q) at (120:4.700000);
\draw[FivePartyWeb] (c1v0) arc[start angle=161.644251,end angle=180.000000,radius=3.297427];
\draw[FivePartyWeb] (c1v0) arc[start angle=101.644252,end angle=49.036367,radius=1.060233];
\draw[FivePartyWeb] (c1v1) arc[start angle=169.036369,end angle=180.000000,radius=4.209615];
\draw[FivePartyWeb] (c1v1) arc[start angle=109.036368,end angle=0.000000,radius=0.846926];
\draw[FivePartyWeb] (c1v0) arc[start angle=41.644252,end angle=120.000000,radius=1.562690];
\draw[FivePartyWeb] (b4) arc[start angle=0,end angle=120.000000,radius=4.700000];
\FiveMarkVertex{c1v0}
\FiveMarkVertex{c1v1}
\FiveMarkQ{c1q}{q_1}
\FiveMarkQ{c2q}{q_2}
\end{tikzpicture}}

\newcommand{\DrawFivePartyVIII}{%
\begin{tikzpicture}[x=0.52cm,y=0.52cm]
\FivePartySetup
\coordinate (c1q) at (120:1.414214);
\coordinate (c1v0) at (1.184699,0.772326);
\coordinate (c2q) at (120:3.878144);
\coordinate (c2v0) at (3.670278,1.252620);
\draw[FivePartyWeb] (c1v0) arc[start angle=153.101050,end angle=180.000000,radius=1.707107];
\draw[FivePartyWeb] (c1v0) arc[start angle=93.101049,end angle=0.000000,radius=0.773459];
\draw[FivePartyWeb] (c1v0) arc[start angle=33.101049,end angle=120.000000,radius=1.414214];
\draw[FivePartyWeb] (c2v0) arc[start angle=138.844129,end angle=180.000000,radius=1.903360];
\draw[FivePartyWeb] (c2v0) arc[start angle=78.844128,end angle=0.000000,radius=1.276745];
\draw[FivePartyWeb] (c2v0) arc[start angle=18.844127,end angle=120.000000,radius=3.878144];
\FiveMarkVertex{c1v0}
\FiveMarkQ{c1q}{q_1}
\FiveMarkVertex{c2v0}
\FiveMarkQ{c2q}{q_2}
\end{tikzpicture}}

\newcommand{\DrawFivePartyIX}{%
\begin{tikzpicture}[x=0.52cm,y=0.52cm]
\FivePartySetup
\coordinate (c1q) at (120:1.000000);
\coordinate (c2q) at (120:3.534851);
\coordinate (c2v0) at (3.134312,1.634398);
\coordinate (c2v1) at (2.827682,1.000750);
\draw[FivePartyWeb] (b1) arc[start angle=0,end angle=120.000000,radius=1.000000];
\draw[FivePartyWeb] (c2v0) arc[start angle=147.539949,end angle=160.814291,radius=3.045208];
\draw[FivePartyWeb] (c2v1) arc[start angle=100.814289,end angle=180.000000,radius=1.018844];
\draw[FivePartyWeb] (c2v1) arc[start angle=40.814290,end angle=0.000000,radius=1.531113];
\draw[FivePartyWeb] (c2v0) arc[start angle=87.539948,end angle=0.000000,radius=1.635906];
\draw[FivePartyWeb] (c2v0) arc[start angle=27.539948,end angle=120.000000,radius=3.534851];
\FiveMarkQ{c1q}{q_1}
\FiveMarkVertex{c2v0}
\FiveMarkVertex{c2v1}
\FiveMarkQ{c2q}{q_2}
\end{tikzpicture}}

\newcommand{\DrawFivePartyX}{%
\begin{tikzpicture}[x=0.52cm,y=0.52cm]
\FivePartySetup
\coordinate (c1q) at (120:1.000000);
\coordinate (c2q) at (120:2.742254);
\coordinate (c2v0) at (2.366767,1.385051);
\coordinate (c2v1) at (3.312650,1.019344);
\draw[FivePartyWeb] (b1) arc[start angle=0,end angle=120.000000,radius=1.000000];
\draw[FivePartyWeb] (c2v0) arc[start angle=150.336514,end angle=180.000000,radius=2.798620];
\draw[FivePartyWeb] (c2v0) arc[start angle=90.336512,end angle=47.387440,radius=1.385075];
\draw[FivePartyWeb] (c2v1) arc[start angle=167.387441,end angle=180.000000,radius=4.668243];
\draw[FivePartyWeb] (c2v1) arc[start angle=107.387440,end angle=0.000000,radius=1.068153];
\draw[FivePartyWeb] (c2v0) arc[start angle=30.336513,end angle=120.000000,radius=2.742254];
\FiveMarkQ{c1q}{q_1}
\FiveMarkVertex{c2v0}
\FiveMarkVertex{c2v1}
\FiveMarkQ{c2q}{q_2}
\end{tikzpicture}}

\newcommand{\DrawFivePartyXI}{%
\begin{tikzpicture}[x=0.52cm,y=0.52cm]
\FivePartySetup
\coordinate (c1q) at (120:1.414214);
\coordinate (c1v0) at (1.184699,0.772326);
\coordinate (c2q) at (120:3.200000);
\coordinate (c3q) at (120:4.700000);
\draw[FivePartyWeb] (c1v0) arc[start angle=153.101049,end angle=180.000000,radius=1.707107];
\draw[FivePartyWeb] (c1v0) arc[start angle=93.101049,end angle=0.000000,radius=0.773459];
\draw[FivePartyWeb] (c1v0) arc[start angle=33.101050,end angle=120.000000,radius=1.414214];
\draw[FivePartyWeb] (b3) arc[start angle=0,end angle=120.000000,radius=3.200000];
\draw[FivePartyWeb] (b4) arc[start angle=0,end angle=120.000000,radius=4.700000];
\FiveMarkVertex{c1v0}
\FiveMarkQ{c1q}{q_1}
\FiveMarkQ{c2q}{q_2}
\FiveMarkQ{c3q}{q_3}
\end{tikzpicture}}

\newcommand{\DrawFivePartyXII}{%
\begin{tikzpicture}[x=0.52cm,y=0.52cm]
\FivePartySetup
\coordinate (c1q) at (120:1.000000);
\coordinate (c2q) at (120:2.529822);
\coordinate (c2v0) at (2.330355,0.984604);
\coordinate (c3q) at (120:4.700000);
\draw[FivePartyWeb] (b1) arc[start angle=0,end angle=120.000000,radius=1.000000];
\draw[FivePartyWeb] (c2v0) arc[start angle=142.904666,end angle=180.000000,radius=1.632456];
\draw[FivePartyWeb] (c2v0) arc[start angle=82.904667,end angle=0.000000,radius=0.992202];
\draw[FivePartyWeb] (c2v0) arc[start angle=22.904666,end angle=120.000000,radius=2.529822];
\draw[FivePartyWeb] (b4) arc[start angle=0,end angle=120.000000,radius=4.700000];
\FiveMarkQ{c1q}{q_1}
\FiveMarkVertex{c2v0}
\FiveMarkQ{c2q}{q_2}
\FiveMarkQ{c3q}{q_3}
\end{tikzpicture}}

\newcommand{\DrawFivePartyXIII}{%
\begin{tikzpicture}[x=0.52cm,y=0.52cm]
\FivePartySetup
\coordinate (c1q) at (120:1.000000);
\coordinate (c2q) at (120:2.000000);
\coordinate (c3q) at (120:3.878144);
\coordinate (c3v0) at (3.670278,1.252620);
\draw[FivePartyWeb] (b1) arc[start angle=0,end angle=120.000000,radius=1.000000];
\draw[FivePartyWeb] (b2) arc[start angle=0,end angle=120.000000,radius=2.000000];
\draw[FivePartyWeb] (c3v0) arc[start angle=138.844127,end angle=180.000000,radius=1.903360];
\draw[FivePartyWeb] (c3v0) arc[start angle=78.844129,end angle=0.000000,radius=1.276745];
\draw[FivePartyWeb] (c3v0) arc[start angle=18.844127,end angle=120.000000,radius=3.878144];
\FiveMarkQ{c1q}{q_1}
\FiveMarkQ{c2q}{q_2}
\FiveMarkVertex{c3v0}
\FiveMarkQ{c3q}{q_3}
\end{tikzpicture}}

\newcommand{\DrawFivePartyXIV}{%
\begin{tikzpicture}[x=0.52cm,y=0.52cm]
\FivePartySetup
\coordinate (c1q) at (120:1.000000);
\coordinate (c2q) at (120:2.000000);
\coordinate (c3q) at (120:3.200000);
\coordinate (c4q) at (120:4.700000);
\draw[FivePartyWeb] (b1) arc[start angle=0,end angle=120.000000,radius=1.000000];
\draw[FivePartyWeb] (b2) arc[start angle=0,end angle=120.000000,radius=2.000000];
\draw[FivePartyWeb] (b3) arc[start angle=0,end angle=120.000000,radius=3.200000];
\draw[FivePartyWeb] (b4) arc[start angle=0,end angle=120.000000,radius=4.700000];
\FiveMarkQ{c1q}{q_1}
\FiveMarkQ{c2q}{q_2}
\FiveMarkQ{c3q}{q_3}
\FiveMarkQ{c4q}{q_4}
\end{tikzpicture}}

\begin{figure}[ht]
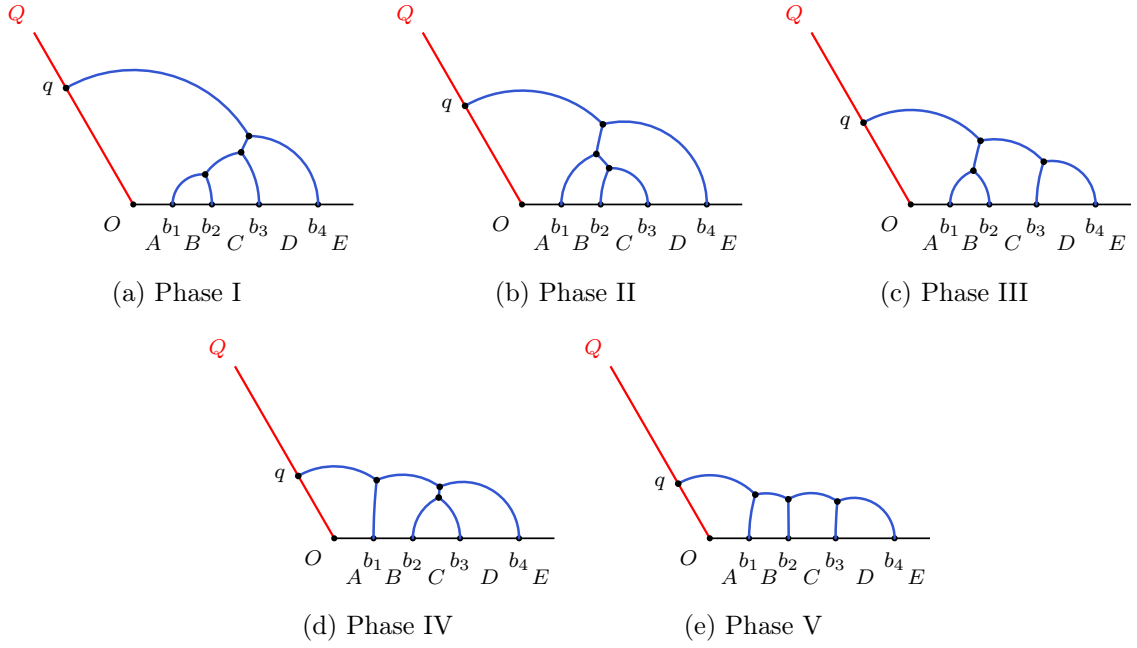

\centering
\begin{subfigure}[t]{0.32\textwidth}\centering\DrawFivePartyI\caption{Phase I}\end{subfigure}\hfill
\begin{subfigure}[t]{0.32\textwidth}\centering\DrawFivePartyII\caption{Phase II}\end{subfigure}\hfill
\begin{subfigure}[t]{0.32\textwidth}\centering\DrawFivePartyIII\caption{Phase III}\end{subfigure}

\medskip
\begin{subfigure}[t]{0.32\textwidth}\centering\DrawFivePartyIV\caption{Phase IV}\end{subfigure}
\begin{subfigure}[t]{0.32\textwidth}\centering\DrawFivePartyV\caption{Phase V}\end{subfigure}
\caption{The five connected five-party brane-web topologies with one endpoint on the EOW brane, $N_Q=1$, drawn as exact finite-tension representative geodesic networks.  Every segment is a circular geodesic arc centered on $z=0$; all three trivalent junctions satisfy the $2\pi/3$ condition and the brane-ending segment is orthogonal to $Q$.}
\label{fig:five-party-r1}
\end{figure}

\begin{figure}[ht]
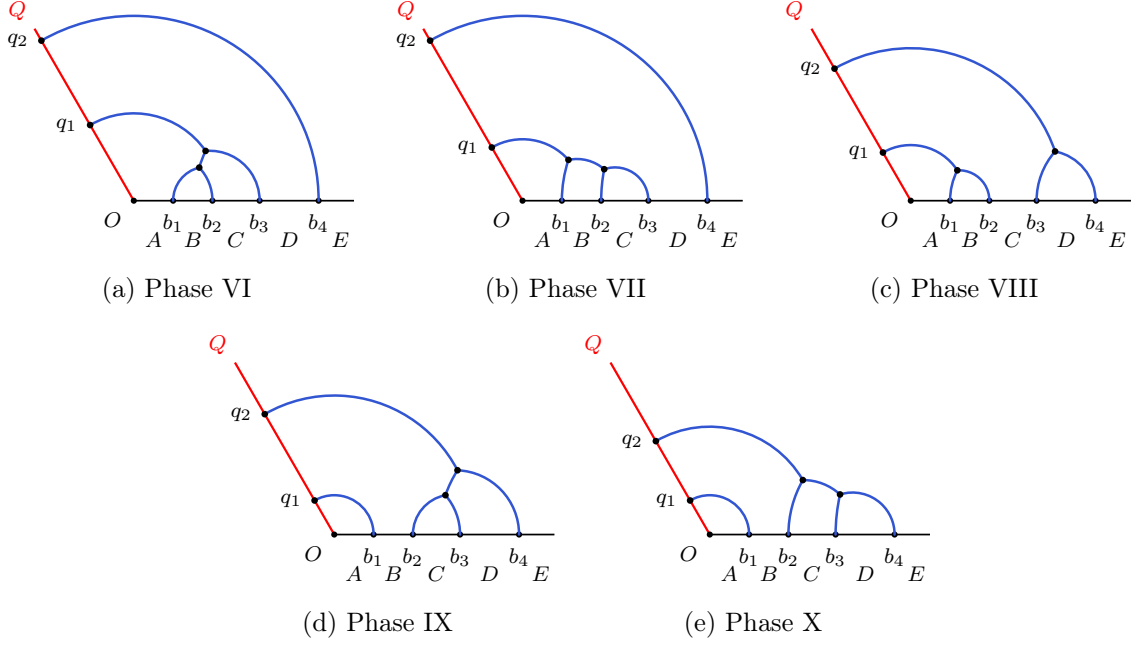

\centering
\begin{subfigure}[t]{0.32\textwidth}\centering\DrawFivePartyVI\caption{Phase VI}\end{subfigure}\hfill
\begin{subfigure}[t]{0.32\textwidth}\centering\DrawFivePartyVII\caption{Phase VII}\end{subfigure}\hfill
\begin{subfigure}[t]{0.32\textwidth}\centering\DrawFivePartyVIII\caption{Phase VIII}\end{subfigure}

\medskip
\begin{subfigure}[t]{0.32\textwidth}\centering\DrawFivePartyIX\caption{Phase IX}\end{subfigure}
\begin{subfigure}[t]{0.32\textwidth}\centering\DrawFivePartyX\caption{Phase X}\end{subfigure}
\caption{The five five-party brane-web topologies with two EOW-brane endpoints, $N_Q=2$, shown as exact finite-tension representative geodesic networks.  The three possible relative-homology assignments correspond to the ordered distributions $3+1$, $2+2$ and $1+3$ of the asymptotic endpoints among the two connected components. At every trivalent junction, the three incident segments meet pairwise at $2\pi/3$, and every brane-ending segment intersects $Q$ orthogonally.}
\label{fig:five-party-r2}
\end{figure}

\begin{figure}[ht]
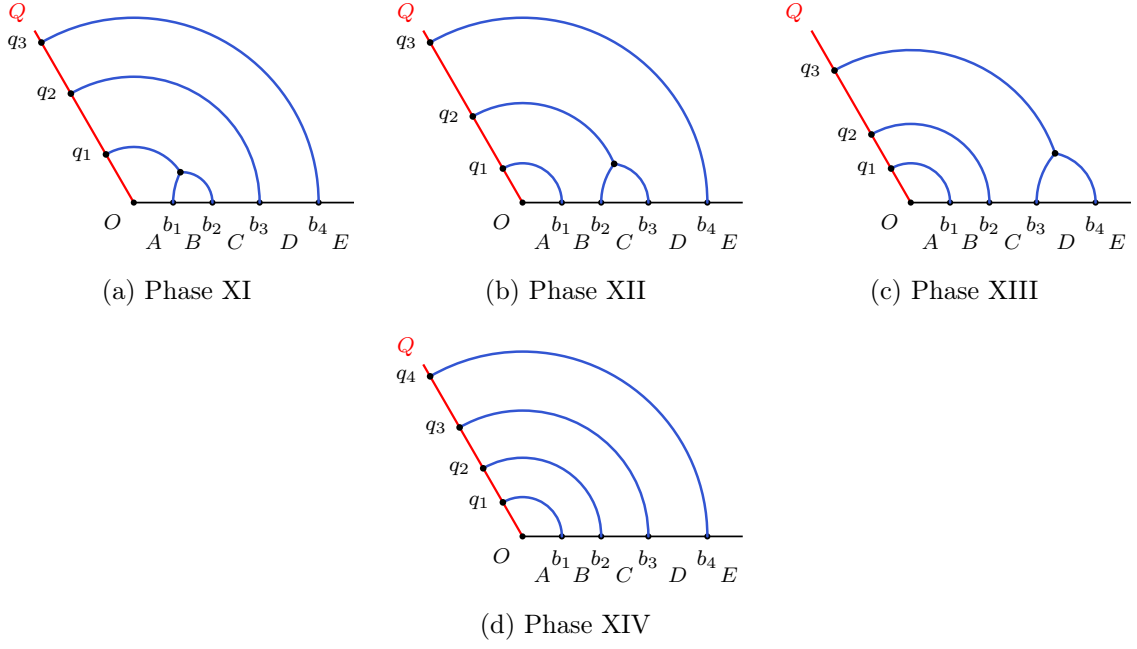

\centering
\begin{subfigure}[t]{0.32\textwidth}\centering\DrawFivePartyXI\caption{Phase XI}\end{subfigure}\hfill
\begin{subfigure}[t]{0.32\textwidth}\centering\DrawFivePartyXII\caption{Phase XII}\end{subfigure}\hfill
\begin{subfigure}[t]{0.32\textwidth}\centering\DrawFivePartyXIII\caption{Phase XIII}\end{subfigure}

\medskip
\begin{subfigure}[t]{0.32\textwidth}\centering\DrawFivePartyXIV\caption{Phase XIV}\end{subfigure}
\caption{Five-party topologies with three or four EOW-brane endpoints, again drawn as exact finite-tension geodesic representatives.  Phases XI--XIII have $N_Q=3$ and contain one $Y$ web with a trivalent junction at which the three incident segments meet pairwise at $2\pi/3$, together with two brane-ending geodesics.  Phase XIV has $N_Q=4$ and is completely factorized into four circular boundary-to-brane geodesics, each orthogonal to $Q$.}
\label{fig:five-party-r34}
\end{figure}
\FloatBarrier

\bibliographystyle{JHEP}
\bibliography{multi-entropy-Geng-citations}
\end{document}